\documentclass[aps,pra,reprint,amsfonts,amsmath,amssymb,longbibliography]{revtex4-2}
\usepackage{xcolor}

\usepackage{hyperref}
\usepackage{graphicx}
\usepackage{bm}
\usepackage{newtxtext}
\usepackage[varg]{newtxmath}
\usepackage[normalem]{ulem}
\usepackage{lipsum}
\usepackage{slashed}
\usepackage{mathrsfs}
\usepackage{blkarray}
\usepackage{mleftright}
\mleftright

\hypersetup{colorlinks=true,linkcolor=blue,citecolor=blue,urlcolor=blue}

\begin{document}
\title{Quantum geometric spectroscopy with Schr\"{o}dinger cat states}
\author{Shohei Imai}
\email{imai@ap.t.u-tokyo.ac.jp}
\affiliation{Department of Applied Physics, University of Tokyo, Hongo, Tokyo 113-0033, Japan}
\author{Marcelo F. Ciappina}
\email{marcelo.ciappina@gtiit.edu.cn}
\affiliation{Department of Physics, Guangdong Technion-Israel Institute of Technology, Shantou 515063, China}
\affiliation{Technion--Israel Institute of Technology, Haifa 32000, Israel}
\affiliation{Guangdong Provincial Key Laboratory of Materials and Technologies for Energy Conversion, Guangdong Technion-Israel Institute of Technology, Shantou 515063, China}
\date{\today}

\begin{abstract}
We propose a spectroscopic approach that reads out the quantum geometry of matter from the quantum decoherence of Schr\"{o}dinger-cat light, in the same way that optical absorption reads out the energy structure of matter.
For a cat-state probe comprising a superposition of two coherent states, the interbranch coherence of the outgoing light encodes the overlap between the matter states driven by the two branches.
We show that, when the electromagnetic fields of the two branches differ only slightly, cat-branch decoherence is quantified by dynamical quantum geometry, which characterizes infinitesimal displacements on the matter-state manifold.
For frequency-selective kitten probes of equilibrium matter, we further establish a partition law for the quantum geometric tensor between light and matter, enabling optical readout of matter quantum metric and Berry curvature through, for example, the brightening of a stellar zero in the outgoing Husimi function.
For nonequilibrium matter in a pure state, short-pulse cat probes also provide a time-resolved witness of multipartite entanglement through the transient quantum metric, which can be extracted from homodyne measurements of the outgoing light.
By turning cat-state decoherence, conventionally regarded as a loss of quantum information, into a carrier of information about matter, this approach establishes a spectroscopic principle that directly probes quantum-state geometry and fluctuations alongside conventional response-based spectroscopy.
\end{abstract}
\maketitle

\section{Introduction} \label{sec:introduction}
Spectroscopy turns seemingly ordinary changes in light into microscopic information about matter.
In absorption spectroscopy, for example, optical attenuation is itself a generic consequence of light--matter interactions, but its frequency dependence carries spectroscopic information by reflecting the excitation structure of matter.
This principle underlies the study of an exceptionally broad range of physical systems, from atomic and molecular energy levels~\cite{Bohr1913} and electronic and collective excitations in solids~\cite{Basov2011} to the optical responses of biomolecules~\cite{Barth2007} and tissues~\cite{Jacques2013}.

\begin{figure}[t]
\centering
\includegraphics[width=\columnwidth]{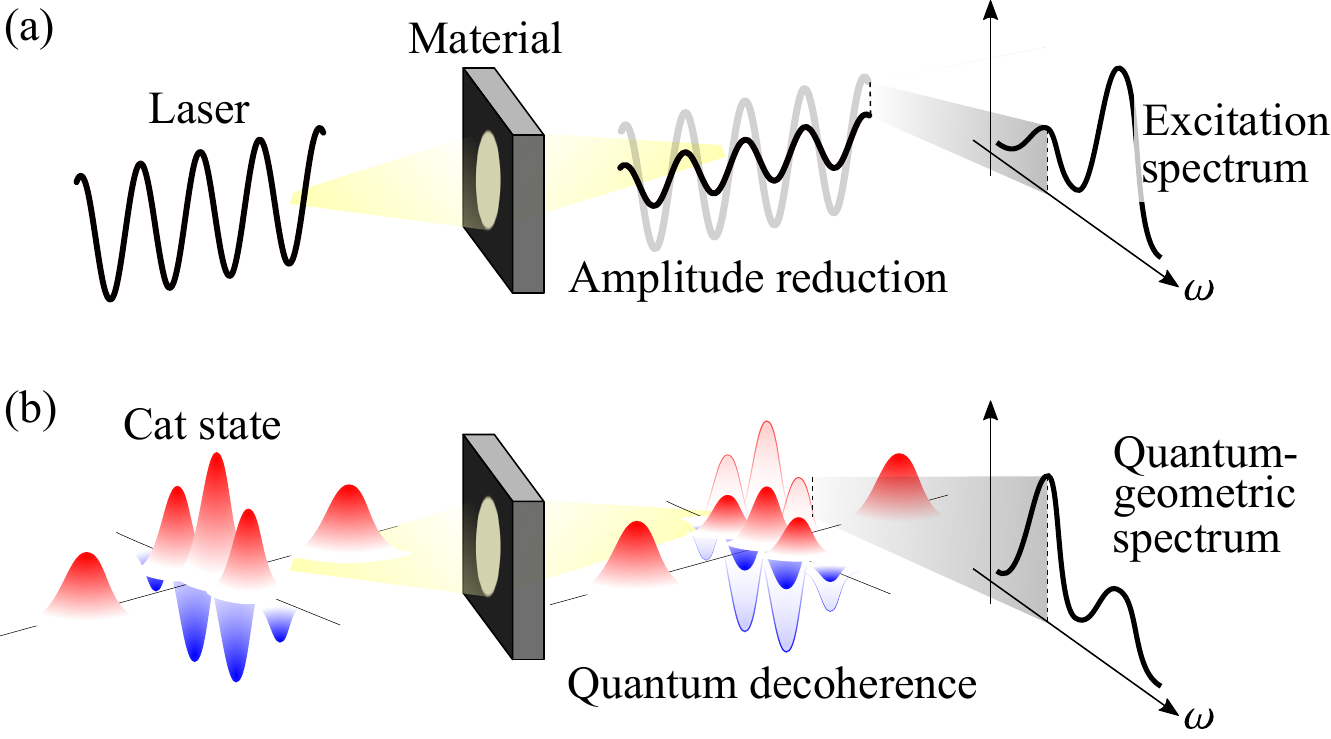}
\caption{
Concept of quantum geometric spectroscopy of matter through the decoherence of cat-state probe light.
(a)~Conventional spectroscopy. Optical absorption arising from light--matter interactions reflects the energy structure of matter.
(b)~Cat-state spectroscopy. Cat-branch decoherence arising from light--matter interactions reflects the quantum geometry of matter.
The quantum metric characterizes fluctuations of matter observables coupled to light, such as electric current and polarization, and can witness multipartite entanglement.
}
\label{fig:schematic}
\end{figure}

This observation suggests a natural quantum counterpart:
when the probe light is prepared in a nonclassical state, can the quantum decoherence that generically accompanies its interaction with matter also serve as a spectroscopic signal?
Here, we address this question using Schr\"{o}dinger cat states of light (Fig.~\ref{fig:schematic}).
A cat state is a quantum superposition of two coherent states with different amplitudes.
Each branch corresponds to a classical, laser-like field, whereas their interbranch coherence constitutes a quantum degree of freedom absent from classical light.
This interbranch coherence is readily lost through interactions with matter~\cite{Zurek2003}, a process conventionally viewed as the undesirable degradation of a quantum resource.
Here, we instead treat this ubiquitous branch decoherence as a carrier of information about matter and identify the physical quantities it encodes.

The framework can be outlined as follows.
The two coherent states $|\alpha_{\pm}\rangle$ forming the cat state drive matter with different electromagnetic (EM) fields, inducing distinct matter states $|\psi_{\pm}\rangle$.
Assuming that the light--matter dynamics within each branch admit a semiclassical description, the linearity of the Schr\"{o}dinger equation gives the schematic evolution
\begin{equation}
(|\alpha_+\rangle+|\alpha_-\rangle)|\psi(0)\rangle
\to
|\beta_+\rangle|\psi_+\rangle + |\beta_-\rangle|\psi_-\rangle,
\end{equation}
where $|\psi(0)\rangle$ is the initial matter state and $|\beta_{\pm}\rangle$ are the modulated coherent states.
After tracing out matter, the optical interbranch term transforms as 
\begin{equation}
|\alpha_+\rangle \langle \alpha_-| \to \langle \psi_-|\psi_+\rangle \, |\beta_+\rangle \langle \beta_-|,
\end{equation}
so that the interbranch coherence of the outgoing light encodes the overlap $\langle \psi_-|\psi_+\rangle$ between two many-body quantum histories driven by different EM fields.

This observation provides an intuitive picture of what cat-probe decoherence reveals.
The more distinguishable the matter states induced by the two optical fields become, the smaller their overlap and the stronger the cat-branch decoherence.
For an infinitesimal field difference between the branches, this reduction in overlap measures how sensitively the matter state changes under a slight variation of the optical field and is characterized by the dynamical quantum geometric tensor~\cite{Provost1980, Anandan1990, Braunstein1994, Kolodrubetz2017}.
Cat-branch decoherence therefore probes the geometry of matter quantum states.

For kitten states, whose two coherent-state components have similar amplitudes, a geometric interpretation applies not only to matter but also to the input light itself.
Certain kitten states can be understood as tangent vectors to the coherent-state manifold.
This property, combined with the unitarity of light--matter evolution and the branchwise separability condition specified below, leads to a partition law for the quantum geometric tensor (QGT) between light and matter in frequency-selective measurements of equilibrium matter.
Intuitively, the optical branches of the kitten probe move closer together as the matter branches move farther apart.
This partition law allows the quantum geometry of matter to be read out, for example, from the brightening of an initially dark point in the Husimi function---a stellar zero~\cite{Chabaud2020} that reflects the nonclassicality of the cat state.

Reading the quantum geometry of matter from cat-branch decoherence extends beyond stationary spectroscopy.
When a short-pulse cat probe interacts with nonequilibrium matter in a pure state, the decay of interbranch coherence is governed by a temporally filtered quantum metric.
This metric provides a lower bound on the transient quantum Fisher information (QFI) and thus serves as a time-resolved witness of multipartite entanglement generated within matter.
In this nonequilibrium regime, the cat-branch decoherence is read out from changes in the Weyl characteristic function---the phase-space Fourier transform of the Wigner function---or in the purity of the probe light.

The optical cat and kitten states required for this spectroscopy are rapidly becoming experimentally accessible.
In continuous-variable optical quantum information processing, generalized photon subtraction has enabled the generation of propagating Schr\"{o}dinger cat states at high rates and high purity, including homodyne characterization in picosecond temporal modes~\cite{Takase2021, Endo2026}.
Independently, attosecond optics have experimentally demonstrated that conditioning on high-harmonic generation (HHG) can herald kitten states that superpose distinct amplitudes of an intense driving field~\cite{Lewenstein2021}.
In the nearby-branch limit, these states naturally approach the tangent kitten probes used here.
These advances suggest that cat-state light is evolving from a quantum-information resource into a quantum probe for material spectroscopy.

We next place this work in context of existing spectroscopy.
The distinction from conventional response-based spectroscopy lies in the matter correlation function accessed by the probe.
In optical absorption, attenuation is governed by the dissipative part of a retarded response function, determined by the commutator of the measured matter observable~\cite{Kubo1957}.
By contrast, the quantum metric accessed here is related to symmetrized matter fluctuations, determined by the anticommutator of that observable, and manifests itself in cat-branch decoherence.
Just as attenuation reveals the excitation structure of matter, cat-branch decoherence reveals quantum-state geometry and fluctuations, providing a complementary spectroscopic principle.
Quantum-light spectroscopy~\cite{Dorfman2016} has exploited nonclassical photon statistics and entanglement to enhance spectroscopic signals, select excitation pathways, and improve temporal and spectral resolution~\cite{Fujihashi2023, Fujihashi2024}, with notable applications to photosynthetic pigment--protein complexes~\cite{Li2023d} and hemoglobin in red blood cells~\cite{Whittaker2017}.
Theoretical proposals have also used cavities to measure Bloch-electron quantum geometry even with coherent-state input light~\cite{Lysne2023}.
Here, we instead propose a versatile spectroscopic method that exploits changes in the nonclassicality of the probe light itself---interbranch coherence---as a signal, enabling its application to many-body and nonequilibrium systems, as well as cavity-free settings.
We do not claim, however, that the matter information obtained here is fundamentally inaccessible with classical light.
For example, equilibrium quantum metrics can be reconstructed from frequency integrals of the corresponding response functions~\cite{Hauke2016, Ozawa2018a}; moreover, complete coherent-state spectroscopy with infinite precision can, in principle, reconstruct general quantum-optical responses~\cite{Kira2011}.
The distinguishing feature of our approach is instead its operational architecture: two distinct many-body evolutions are coherently encoded within a single probe (in parallel, in the quantum sense), with their overlap read out directly as the interbranch decoherence of the outgoing light.
Other approaches that use probe decoherence for spectroscopy include noise spectroscopy~\cite{Alvarez2011} and Loschmidt echoes~\cite{Gorin2006a}, which employ time-reversal operations such as dynamical decoupling; our method accesses such echo-like signals without time reversal.
Ramsey interferometry, which reads out many-body overlaps from the coherence of an ancillary qubit~\cite{Quan2006, Knap2012, Mazzola2013}, is powerful in engineered quantum systems and quantum simulators, but its application to general solids or molecules requires a suitable ancilla and a Hamiltonian conditioned on its internal state.
Here, the quantized EM field that directly couples to matter also serves as the quantum ancilla, enabling a more broadly applicable form of interferometric spectroscopy.

Quantum geometry plays an important role across materials science.
In crystals, the quantum geometry of Bloch states in momentum space characterizes aspects of wave-function structure beyond the energy dispersion~\cite{Torma2023}.
The Berry curvature, associated with the imaginary part of the QGT, gives rise to topological invariants~\cite{Thouless1982}, while the quantum metric, its real part, characterizes many-body transport properties~\cite{Peotta2015}.
In chemistry, the geometry of electronic states parameterized by nuclear coordinates influences quantum interference between reaction pathways through geometric phases~\cite{Mead1979, Yuan2018}.
More generally, the quantum metric quantifies the sensitivity of quantum states to parameter variations, linking it through the QFI to quantum metrology~\cite{Braunstein1994} and multipartite entanglement witnessing~\cite{Hyllus2012, Toth2012}.
While experimental access to quantum geometry has often relied on band-structure analysis~\cite{Kang2025} or response-based measurements~\cite{Tan2019a}, our approach offers a broadly applicable quantum-geometric spectroscopy based on the picture that cat-branch coherence directly couples to the quantum geometry of matter parameterized by the optical EM field.

The remainder of this paper develops this framework in two complementary regimes.
Section~\ref{sec:GS_QGT} establishes quantum geometric spectroscopy of equilibrium matter using tangent kitten states and demonstrates the resulting geometric partition and tangent-excitation picture.
Section~\ref{sec:Neq_QFI} formulates time-resolved spectroscopy of nonequilibrium matter using cat-branch decoherence and shows how dynamical quantum metrics, QFI, and multipartite entanglement can be extracted from the outgoing quantum light.
Both protocols are demonstrated numerically using full light--matter dynamics in the Lipkin--Meshkov--Glick (LMG) model, an effective model of semiconductors.
Section~\ref{sec:conclusion} concludes the paper.

\section{Quantum geometric spectroscopy of equilibrium matter} \label{sec:GS_QGT}
We develop a spectroscopic method that uses tangent kitten probes to read out the quantum geometry of equilibrium matter from the outgoing light.
For nearby coherent-state branches, we treat the difference between their EM fields as a small parameter and describe the resulting dynamics in terms of tangent vectors on the quantum-state manifold.
Section~\ref{sec:QGT_conserv} first introduces the general framework of quantum geometry~\cite{Provost1980, Anandan1990, Braunstein1994, Kolodrubetz2017} and shows that the total QGT, parameterized by the initial coherent-state amplitude, is conserved under unitary light--matter evolution.
Section~\ref{sec:branchwise_separability_approximation} analyzes the evolution of each coherent-state branch through a weak-coupling expansion and shows that the conserved QGT is partitioned additively between light and matter under the branchwise separability approximation (BSA), in which intrabranch light--matter entanglement is negligible.
Section~\ref{sec:branchwise_separability_approximation:validity_of_BSA} establishes the conditions for BSA in frequency-selective stationary spectroscopy of ground states and thermal equilibrium states.
Section~\ref{sec:dynamical_quantum_geometry} summarizes the physical meaning of the QGT obtained in this protocol.
Section~\ref{sec:optical_readout} derives methods for reading out the matter QGT from the Husimi function and field fluctuations of the outgoing light.
Section~\ref{sec:GS_demo} numerically demonstrates this quantum geometric spectroscopy using the LMG model.

\subsection{Conservation of the QGT in light--matter systems}
\label{sec:QGT_conserv}
Consider a quantum state $|\chi\rangle$ that depends on a set of real parameters $\bm{\lambda} = \{\lambda_{\mu}\}$.
Its tangent state with respect to $\lambda_{\mu}$ is defined as
\begin{equation}
|D_{\mu}\chi\rangle = (1-|\chi\rangle\langle\chi|) \partial_{\mu}|\chi\rangle, \label{eq:def:tangent}
\end{equation}
and the corresponding
Fubini--Study QGT is defined as
\begin{equation}
\mathcal{Q}_{\mu\nu}[\chi] = \langle D_{\mu}\chi|D_{\nu}\chi\rangle = \langle \partial_{\mu} \chi| (1-|\chi\rangle\langle\chi|)|\partial_{\nu}\chi\rangle. \label{eq:def:QGT}
\end{equation}
Its real and imaginary parts give the quantum metric and Berry curvature, respectively, according to
\begin{equation}
\mathcal{Q}_{\mu\nu}=g_{\mu\nu} -\frac{i}{2}\Omega_{\mu\nu},
\end{equation}
which specifies our convention.
The quantum metric quantifies the local distinguishability of neighboring quantum states: a larger metric along a given parameter direction indicates greater sensitivity of the state to infinitesimal variation in that parameter.

The QGT is a kinematic invariant when the unitary evolution operator $\hat{U}^{\mathrm{tot}}(t)$ is independent of the parameters $\{\lambda_{\mu}\}$:
\begin{equation}
\mathcal{Q}^{\mathrm{tot}}_{\mu\nu}(t) = \mathcal{Q}^{\mathrm{tot}}_{\mu\nu}(0). \label{eq:QGT_conservation_law}
\end{equation}
This follows directly from $\partial_{\mu} (\hat{U}^{\mathrm{tot}}|\chi\rangle)=\hat{U}^{\mathrm{tot}} \partial_{\mu}|\chi\rangle$.

To connect this geometric framework to cat-state probes, we first consider the coherent-state manifold of the input light.
Let $\alpha_{\pm}$ denote the amplitudes of the two coherent states forming the cat state, and define their mean and difference as
\begin{equation}
\alpha_c = \frac{\alpha_+ + \alpha_-}{2},\ \alpha_q = \alpha_+ - \alpha_-.
\end{equation}
We consider the kitten state
\begin{equation}
|\mathrm{HK}\rangle = \frac{1}{\sqrt{1-|\langle \alpha_-|\alpha_+\rangle|^2}}(|\alpha_+\rangle - \langle \alpha_-|\alpha_+\rangle |\alpha_-\rangle ). \label{eq:def:HHG_kitten}
\end{equation}
This type of nonclassical light has recently been generated experimentally in gas-phase HHG by using conditioning~\cite{Lewenstein2021}.
The coherent state is defined using the displacement operator $\hat{\mathcal{D}}(\alpha)$ as
\begin{equation}
|\alpha\rangle = \hat{\mathcal{D}}(\alpha)|0\rangle,\ \hat{\mathcal{D}}(\alpha)=e^{\alpha \hat{a}^{\dagger} - \alpha^* \hat{a}}, \label{eq:def:coherent_state}
\end{equation}
where $\hat{a}$ is the bosonic annihilation operator and the complex amplitude $\alpha$ determines the corresponding classical EM field.
The resulting drive on matter scales as $g_{\mathrm{lm}}\alpha$, where $g_{\mathrm{lm}}$ is the electron--photon coupling constant introduced below.

The geometric meaning of $|\mathrm{HK}\rangle$ becomes clear as the two branches approach each other.
We parameterize the coherent-state family connecting $\alpha_-$ to $\alpha_+$ as
\begin{equation}
\alpha(\lambda) = \alpha_-+\lambda e^{i\varphi}, \qquad
0 \le \lambda \le |\alpha_q|, \label{eq:alpha_lambda}
\end{equation}
and, unless otherwise stated, set $\varphi=0$ by absorbing this phase into the temporal mode function.
In the limit of vanishing branch separation, $|\alpha_q| \to 0$, we have $\alpha_- \to \alpha_c$, and the HHG-conditioned kitten approaches a normalized tangent state to the coherent-state manifold:
\begin{equation}
\lim_{|\alpha_q|\to0}|\mathrm{HK}\rangle = \left. \frac{|D_{\lambda}\alpha\rangle}{\sqrt{g^{\mathrm{L}}_{\lambda\lambda}[\alpha]}} \right|_{\lambda=0}  = e^{i \varphi} |\alpha_c;1\rangle, \label{eq:kitten_tangent-limit}
\end{equation}
where $|\alpha;n\rangle=\hat{\mathcal{D}}(\alpha)|n\rangle$ denotes a displaced Fock state.
At $\alpha_c=0$, this construction also provides a geometric interpretation of the single-photon Fock state~\cite{Hackl2020, Coelho2016, Lenhard2015, Baragiola2012, Knill2001}.
For the coherent-state family parametrized by $\lambda$ [Eq.~\eqref{eq:alpha_lambda}], the corresponding quantum metric is
\begin{equation}
g^{\mathrm{L}}_{\lambda\lambda}[\alpha] = \langle D_{\lambda}\alpha|D_{\lambda}\alpha\rangle=|\partial_{\lambda}\alpha|^2 = 1, \label{eq:kitten_metric}
\end{equation}
where the final equality follows from our choice of $\lambda$.
This constant metric implies that the coherent-state manifold is flat, with the line element $ds^2=g^{\mathrm{L}}_{\lambda\lambda}d\lambda d\lambda$.
Consequently, the squared distance between the two coherent-state branches $\{|\alpha_+\rangle,|\alpha_-\rangle\}$ is measured by the tangent-vector norm and the difference in their amplitudes as $\Delta s^2 =\langle D_{\lambda}\alpha|D_{\lambda}\alpha\rangle |\alpha_q|^2$.

\begin{figure}[t]
\centering
\includegraphics[width=\columnwidth]{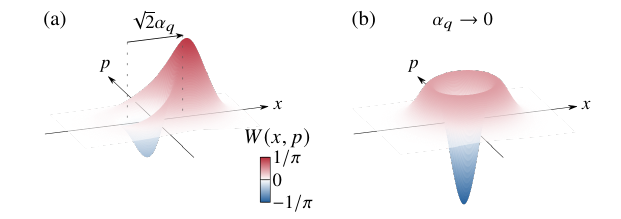}
\caption{
Wigner functions illustrating the tangent-state limit of a kitten state.
(a)~For a finite branch separation $\alpha_q \neq 0$, the distribution exhibits a positive peak near $(x+ip)/\sqrt{2}=\alpha_+$, associated with $|\alpha_+\rangle$, and Wigner negativity on the $\alpha_-$ side.
(b)~For a small branch separation, the distribution approaches that of a single-photon Fock state, with a negative minimum at the origin.
The phase-space coordinates $(x,p)$ correspond to the field quadratures, with the normalization $\int dx\,dp\,W(x,p)=1$.
Both panels use $\alpha_c=0$, with (a)~$\alpha_q=1.5$ and (b)~$\alpha_q=0.05$.
}
\label{fig:wigner}
\end{figure}

Figure~\ref{fig:wigner} visualizes the tangent-state limit of the kitten state~\eqref{eq:def:HHG_kitten} through its Wigner function.
Wigner-function negativity is a signature of nonclassicality.
For a finite branch separation, the negative region arises from quantum interference between the two coherent-state components [Fig.~\ref{fig:wigner}(a)].
As the branch separation vanishes, the Wigner function continuously approaches that of a single-photon Fock state [Fig.~\ref{fig:wigner}(b)], consistent with the tangent-state interpretation in Eq.~\eqref{eq:kitten_tangent-limit}.

For a multimode coherent state $|\bm{\alpha}\rangle=\otimes_j |\alpha_j\rangle$, the QGT is given by
\begin{equation}
\mathcal{Q}_{\mu\nu}^{\mathrm{L}}[\bm{\alpha}] = \langle D_{\mu}\bm{\alpha}|D_{\nu}\bm{\alpha}\rangle = \sum_{j}\partial_{\mu}\alpha^*_j\, \partial_{\nu}\alpha_j.
\end{equation}
In particular, parameterizing the branch separation in each mode $\mu$ by $\lambda_{\mu}$, with $\alpha_{\mu}=\alpha_{-,\mu} + \lambda_{\mu} \exp(i\varphi_{\mu})$, gives
\begin{equation}
\mathcal{Q}_{\mu\nu}^{\mathrm{L}} = \delta_{\mu\nu},
\end{equation}
where $\delta_{\mu\nu}$ is the Kronecker delta.

Crucially, $\bm{\lambda}$ parametrizes the initial coherent-state family of the light but does not enter into the total light--matter Hamiltonian.
For example, the Jaynes--Cummings model~\cite{Jaynes1963}, $\hat{H}_{\mathrm{JC}} = \Delta \hat{\sigma}^z/2 + \omega \hat{a}^{\dagger} \hat{a} + g_{\mathrm{lm}} \hat{a} \hat{\sigma}^+ + g_{\mathrm{lm}}\hat{a}^{\dagger} \hat{\sigma}^-$, contains the field operators $\hat{a}$ and $\hat{a}^{\dagger}$ but is independent of the initial coherent-state amplitude $\alpha$.
Consequently, $\partial_{\mu} \hat{U}^{\mathrm{tot}}=0$, and the total QGT is conserved under unitary light--matter evolution [Eq.~\eqref{eq:QGT_conservation_law}].
Moreover, because the initial matter state is also independent of $\bm{\lambda}$, the total QGT is initially carried entirely by light, $\mathcal{Q}_{\mu\nu}^{\mathrm{tot}}(0)=\mathcal{Q}_{\mu\nu}^{\mathrm{L}}(0)$.
By contrast, the branch-dependent effective matter evolution introduced below depends on the input coherent-state amplitude. This dependence arises from resolving the same exact joint dynamics into individual input branches and does not contradict unitary conservation of the total QGT.

\subsection{Partition of the QGT under the branchwise separability approximation}
\label{sec:branchwise_separability_approximation}
Although the total QGT is conserved, the sensitivity to the initial coherent-state amplitude can spread from light to matter and light--matter correlations.
To examine this redistribution, we analyze the evolution of each coherent-state branch in the weak-coupling limit, which is generally relevant to interactions with propagating probe light.
In this regime, the dynamics can be separated into a semiclassical light--matter trajectory and corrections arising from intrabranch light--matter entanglement.
Neglecting the latter defines the branchwise separability approximation (BSA).
When the BSA holds at the state level, the total QGT can be additively partitioned between light and matter.

\subsubsection{Branchwise dynamics}
\label{sec:branchwise_separability_approximation:branchwise_dynamics}
We first consider the evolution of the coupled light--matter system starting from a single coherent-state branch.
To first order in the electron--photon coupling constant $g_{\mathrm{lm}}$, a general light--matter Hamiltonian takes the form
\begin{equation}
\hat{H}(t) = \omega\hat{a}^{\dagger} \hat{a} + \hat{H}_{\mathrm{M}}(t) - g_{\mathrm{lm}} \left[ f(t) \hat{a} + f^*(t) \hat{a}^{\dagger} \right] \hat{O}(t) + \mathcal{O}(g_{\mathrm{lm}}^2). \label{eq:linearized_hamiltonian}
\end{equation}
Here, light is described by a single bosonic mode with central frequency $\omega$ and temporal mode function $f(t)$.
The operator $\hat{H}_{\mathrm{M}}(t)$ is the matter Hamiltonian, and $\hat{O}(t)$ is the Hermitian matter operator conjugate to the EM field.
When the matter Hamiltonian depends on the EM field $\hat{A}$, this conjugate operator can be defined as $\hat{O}(t)=-\partial\hat{H}_{\mathrm{M}}[\hat{A}]/\partial A|_{\hat{A}=A_c}$.
In particular, when $A$ is the vector potential, $\hat{O}$ is the electric current operator.

We write the evolution from the coherent-state branch $|\alpha\rangle$ in a moving coherent frame as
\begin{equation}
\hat{U}^{\mathrm{tot}}(t) |\alpha\rangle |\psi(0)\rangle=
e^{i\theta_\alpha(t)}
\hat{\mathcal{D}}[\beta_\alpha(t)]
|\widetilde{\Psi}_\alpha(t)\rangle.
\end{equation}
Here, $\hat{U}^{\mathrm{tot}}$ is the full evolution operator generated by $\hat{H}(t)$, and $|\psi(0)\rangle$ is the initial matter state.
The branch-dependent overall phase $\theta_{\alpha}$ includes both dynamical and geometric contributions; however, since this does not affect the QGT, we omit it in subsequent expressions unless relative phases between branches are relevant.
The amplitude $\beta_{\alpha}$ is the moving coherent-state center of each branch.
The classical EM field, represented by $\beta_{\alpha}(t)$, and the matter state it drives are defined by the coupled equations
\begin{align}
&i\partial_t \beta_{\alpha} = \omega\beta_{\alpha} - g_{\mathrm{lm}} f^*(t) \langle\hat{O}(t)\rangle_{\psi_\alpha}, \\
&i\partial_t|\psi_\alpha(t)\rangle = \hat{H}_{\alpha}(t)|\psi_\alpha(t)\rangle,\\
&\hat{H}_\alpha(t) = \hat{H}_{\mathrm{M}}(t) - g_{\mathrm{lm}} \left[ f(t) \beta_{\alpha}(t) + f^*(t) \beta_{\alpha}^*(t) \right] \hat{O}(t), \label{eq:H_alpha}
\end{align}
with initial conditions $\beta_\alpha(0)=\alpha,\, |\psi_\alpha(0)\rangle=|\psi(0)\rangle$.
The coupled equations define the semiclassical light--matter dynamics of each branch, which serve as the reference trajectory in the moving frame.

The moving-frame Hamiltonian, including the residual coupling between optical and matter fluctuations, takes the form
\begin{equation}
\widetilde{H}_{\alpha}(t) = \omega \hat{a}^{\dagger} \hat{a} + \hat{H}_{\alpha}(t) - g_{\mathrm{lm}} \left[ f(t) \hat{a} + f^*(t) \hat{a}^{\dagger} \right] \delta \hat{O}_{\alpha}(t), \label{eq:branch_dependent_Hamiltonian}
\end{equation}
where the matter fluctuation operator is defined as
\begin{equation}
\delta\hat{O}_{\alpha}(t)= \hat{O}(t) -\langle \psi_{\alpha}(t)|\hat{O}(t)|\psi_{\alpha}(t)\rangle.
\end{equation}
Writing $\hat{U}_{\alpha}(t,s)$ for the matter propagator generated by $\hat{H}_{\alpha}(t)$, a first-order Dyson expansion in the residual interaction gives, up to a branch-dependent overall phase,
\begin{align}
&\hat{U}^{\mathrm{tot}}(t) |\alpha\rangle |\psi(0)\rangle \nonumber \\
&= |\beta_{\alpha}(t)\rangle|\psi_{\alpha}(t)\rangle + g_{\mathrm{lm}} |\beta_{\alpha}(t);1\rangle |\eta_{\alpha}(t)\rangle + \mathcal{O}(g_{\mathrm{lm}}^2), \label{eq:branch_time_evolution}
\end{align}
where
\begin{equation}
|\eta_{\alpha}(t)\rangle = i\int_{0}^{t}ds\, f^*(s)e^{-i\omega(t-s)} \hat{U}_{\alpha}(t,s) \delta\hat{O}_{\alpha}(s) |\psi_{\alpha}(s)\rangle. \label{eq:eta_alpha}
\end{equation}
Equation~\eqref{eq:branch_time_evolution} plays a central role in this work.
The first term is a separable light–matter reference trajectory, describing semiclassical dynamics driven by a branch-dependent classical EM field.
The second term describes branch deformation beyond coherent displacement and generates intrabranch light--matter entanglement.

Two properties of Eq.~\eqref{eq:branch_time_evolution} are useful below.
First, the correction vector $|\eta_{\alpha}(t)\rangle$ is orthogonal to the classically driven state, $\langle \psi_{\alpha}(t)| \eta_{\alpha}(t)\rangle =0$, because the matter fluctuation operator has a vanishing expectation value, $\langle \psi_{\alpha}(s)|\delta \hat{O}_{\alpha}(s)|\psi_{\alpha}(s)\rangle=0$.
Consequently, the first moment of the optical field coincides with the moving coherent-state center up to second-order corrections, $\langle\hat{a}\rangle_{\alpha} = \beta_{\alpha} + \mathcal{O}(g_{\mathrm{lm}}^2)$.
Second, the coupled equations for the moving center and matter give
\begin{equation}
\beta_{q}(t)=\alpha_q e^{-i\omega t} + \mathcal{O}(g_{\mathrm{lm}}^2\alpha_q), \label{eq:beta_vs_alpha}
\end{equation}
where we use $\langle \hat{O}(s)\rangle_{+} - \langle \hat{O}(s)\rangle_{-}= \mathcal{O}(g_{\mathrm{lm}}\alpha_q)$.
This relation allows the self-consistent moving center $\beta_{\alpha}$ in the semiclassical matter Hamiltonian $\hat{H}_\alpha(t)$ [Eq.~\eqref{eq:H_alpha}] to be replaced by the freely evolving input amplitude $\alpha \exp(-i\omega t)$ when evaluating the leading-order matter QGT (see Sec.~\ref{sec:branchwise_separability_approximation:quantum_geometric_partition}).

We define the BSA by neglecting the branch-deformation term in Eq.~\eqref{eq:branch_time_evolution}, when its contribution is negligible at the accuracy required for the observable of interest.
The contribution from each branch can be evaluated using a separable light--matter state:
\begin{equation}
\hat{U}^{\mathrm{tot}}(t) |\alpha\rangle |\psi(0)\rangle \overset{\mathrm{BSA}}{\longmapsto} |\beta_{\alpha}(t)\rangle|\psi_{\alpha}(t)\rangle. \label{eq:BSA}
\end{equation}
For equilibrium spectroscopy, Sec.~\ref{sec:branchwise_separability_approximation:validity_of_BSA} examines the condition under which the BSA holds at the state level.
For nonequilibrium spectroscopy, Sec.~\ref{sec:Neq_QFI} identifies optical observables for which the leading effects of branch deformation are suppressed.

\subsubsection{Quantum geometric partition law under BSA}
\label{sec:branchwise_separability_approximation:quantum_geometric_partition}
We assume that the BSA holds for each coherent-state branch, allowing it to be described by the product~\eqref{eq:BSA}.
Taking the projected derivative $D_{\mu}$ of this product state with respect to the coherent-state amplitude parameter $\lambda_{\mu}$ [Eq.~\eqref{eq:alpha_lambda}] gives
\begin{equation}
D_{\mu} (|\beta_{\alpha}\rangle|\psi_{\alpha}\rangle) = |D_{\mu}\beta_{\alpha}\rangle|\psi_{\alpha}\rangle + |\beta_{\alpha}\rangle|D_{\mu}\psi_{\alpha}\rangle. \label{eq:product_rule_beta_psi}
\end{equation}
Since $\langle \beta_{\alpha}|D_{\mu} \beta_{\alpha}\rangle = \langle\psi_{\alpha}|D_{\mu}\psi_{\alpha}\rangle =0$, the two tangent components are orthogonal, and the total QGT decomposes additively as
\begin{equation}
\mathcal{Q}_{\mu\nu}^{\mathrm{tot}}(t) = \mathcal{Q}_{\mu\nu}^{\mathrm{L}} + \mathcal{Q}_{\mu\nu}^{\mathrm{M}}, \label{eq:QGT_partition_law}
\end{equation}
with
\begin{equation}
\mathcal{Q}_{\mu\nu}^{\mathrm{L}} = \partial_{\mu}\beta^*\, \partial_{\nu}\beta,\qquad \mathcal{Q}_{\mu\nu}^{\mathrm{M}}=\langle D_{\mu}\psi|D_{\nu}\psi\rangle.
\end{equation}
Combining this additivity with conservation of the total QGT [Eq.~\eqref{eq:QGT_conservation_law}] yields
\begin{equation}
\mathcal{Q}_{\mu\nu}^{\mathrm{M}} = \mathcal{Q}_{\mu\nu}^{\mathrm{L}}(0) - \mathcal{Q}_{\mu\nu}^{\mathrm{L}}(t), \label{eq:QGT_decrease}
\end{equation}
identifying the matter QGT with the reduction in the optical QGT.
Thus, light--matter interactions redistribute the quantum geometry of the input coherent-state manifold between light and matter.
In this sense, matter excitation by small-separation kitten light can be interpreted as geometric tangent excitation.
This mechanism provides an intuitive picture analogous to energy partition in optical absorption spectroscopy.

This quantum-geometric partition law admits a more general formulation based on the state-level BSA~\eqref{eq:BSA}.
For simplicity, we focus on the metric part.
Consider Eq.~\eqref{eq:BSA} for the two coherent-state branches $\{|\alpha_+\rangle, |\alpha_-\rangle\}$ of a cat probe.
Taking the negative logarithm of the squared magnitude of the overlap between the two branches, we obtain, up to $|\alpha_q|^2$,
\begin{equation}
|\alpha_q|^2 = |\beta_q|^2 +|\alpha_q|^2 g^{\mathrm{M}}_{\lambda\lambda}, \label{eq:geometric_partition_law_finite_distance}
\end{equation}
where we used $-\ln|\langle \psi_-|\psi_+\rangle|^2 = |\alpha_q|^2 g^{\mathrm{M}}_{\lambda\lambda}+ \mathcal{O}(|\alpha_q|^4)$~\cite{Provost1980}.
This has a simple geometric interpretation.
The two initial branches, separated by a squared distance $|\alpha_q|^2$ on the coherent-state manifold, evolve under light--matter interaction into states whose optical components lose part of the initial squared separation, retaining  $|\beta_q|^2$ on the coherent-state manifold, while their matter components acquire a squared separation $|\alpha_q|^2 g^{\mathrm{M}}_{\lambda\lambda}$ on the matter-state manifold.
Equation~\eqref{eq:geometric_partition_law_finite_distance} reduces to Eq.~\eqref{eq:QGT_decrease} as the differential form of this relation by dividing both sides by $|\alpha_q|^2$ and taking $\alpha_q \to 0$.

The physical content of the matter QGT probed by kitten light and its relation to many-body correlation functions are discussed in Sec.~\ref{sec:dynamical_quantum_geometry}.
Here, we only specify its order in the electron--photon coupling constant $g_{\mathrm{lm}}$.
Since the matter tangent vector satisfies $|D_{\mu}\psi\rangle = \mathcal{O}(g_{\mathrm{lm}})$, the matter QGT of interest is of order $\mathcal{O}(g_{\mathrm{lm}}^2)$.
To evaluate this leading-order coefficient, the self-consistent moving center $\beta_{\alpha}$ in the semiclassical matter Hamiltonian [Eq.~\eqref{eq:H_alpha}] can be replaced by the freely evolving input amplitude $\alpha \exp(-i\omega t)$.
Since the input-amplitude-dependent correction to the moving center is $\mathcal{O}(g_{\mathrm{lm}}^2\alpha)$ and the matter state responds linearly to changes in the driving field $g_{\mathrm{lm}}\beta_\alpha$, this correction affects the matter tangent vector only at $\mathcal{O}(g_{\mathrm{lm}}^3)$ and the matter QGT at $\mathcal{O}(g_{\mathrm{lm}}^4)$.
Unless otherwise stated, we therefore evaluate the classically driven matter dynamics in Eq.~\eqref{eq:H_alpha} using the prescribed EM field specified by $\alpha$.
The $\mathcal{O}(g_{\mathrm{lm}}^2)$ change in the optical QGT, however, originates from this moving-center correction, which must therefore be retained on the optical side.

\subsection{Validity of BSA in equilibrium matter}
\label{sec:branchwise_separability_approximation:validity_of_BSA}
We examine when the BSA~\eqref{eq:BSA} holds at the state level with sufficient accuracy to evaluate the matter QGT.
Although the branch-deformation term in Eq.~\eqref{eq:branch_time_evolution} is first order in $g_{\mathrm{lm}}$, it can contribute to the QGT at $\mathcal{O}(g_{\mathrm{lm}}^2)$, the same order as the matter QGT itself.
Thus, the weak-coupling limit $g_{\mathrm{lm}} \to 0$ alone is insufficient to justify the QGT partition law~\eqref{eq:QGT_partition_law}; an additional mechanism is needed to suppress the branch-deforming channel relative to the tangent-transfer channel.
Below, we establish the conditions for this separation in frequency-selective spectroscopy of equilibrium matter.

We consider matter initially prepared in its ground state $|0\rangle_{\mathrm{M}}$ and examine the spectral representation of the branchwise evolution in Eq.~\eqref{eq:branch_time_evolution}.
We work in the interaction picture with respect to the free Hamiltonian $\omega\hat{a}^\dagger\hat{a}+\hat{H}_{\mathrm{M}}$ and evaluate the state after the probe pulse has passed.
Although the kitten probe approaches the displaced single-photon state $|\alpha_c;1\rangle$ [Eq.~\eqref{eq:kitten_tangent-limit}] in the tangent limit, we set $\alpha_c=0$ to eliminate the common classical drive.

To first order in $g_{\mathrm{lm}}$, the classically driven matter state and the branch-deformation amplitude in Eq.~\eqref{eq:branch_time_evolution} are given by
\begin{align}
|\psi_{\alpha,\mathrm{I}}\rangle &=|0\rangle_{\mathrm{M}} +g_{\mathrm{lm}} \left(\alpha|\xi_{0,\mathrm{I}}\rangle+\alpha^*|\eta_{0,\mathrm{I}}\rangle\right)+\mathcal{O}(g_{\mathrm{lm}}^2), \label{eq:tangent_transfer_channel} \\
|\eta_{\alpha,\mathrm{I}}\rangle &=|\eta_{0,\mathrm{I}}\rangle+\mathcal{O}(g_{\mathrm{lm}}), \label{eq:branch_deforming_channel}
\end{align}
where the matter excitation amplitudes have the spectral representations
\begin{align}
|\xi_{0,\mathrm{I}}\rangle &=i\sum_{n>0}\langle n|\delta \hat{O}|0\rangle \tilde{f}(\Delta_n-\omega)|n\rangle_{\mathrm{M}},\\
|\eta_{0,\mathrm{I}}\rangle &=i\sum_{n>0}\langle n|\delta \hat{O}|0\rangle \tilde{f}^*(-\Delta_n-\omega)|n\rangle_{\mathrm{M}}.
\end{align}
Here, $\hat{H}_{\mathrm{M}}|n\rangle_{\mathrm{M}}=E_n|n\rangle_{\mathrm{M}}$ and $\Delta_n=E_n-E_0>0$ define the matter eigenstates and excitation energies, while $\tilde{f}(\nu)=\int dt\,f(t)e^{i\nu t}$ denotes the Fourier transform of the probe envelope.

Crucially, $|\xi_{0,\mathrm{I}}\rangle$ can contain resonant contributions at $\Delta_n\simeq\omega$, whereas $|\eta_{0,\mathrm{I}}\rangle$ involves off-resonant spectral components at $-(\Delta_n+\omega)$.
A sufficiently long, narrowband probe can therefore suppress the branch-deforming channel [Eq.~\eqref{eq:branch_deforming_channel}] while retaining the resonant tangent-transfer channel [Eq.~\eqref{eq:tangent_transfer_channel}].
Specifically, for relevant transitions with excitation energies $\Delta_n$ near the probe frequency $\omega$, this requires
\begin{equation}
\frac{|\tilde{f}(-\Delta_n -\omega)|}{|\tilde{f}(\Delta_n - \omega)|} \ll 1. \label{eq:BSA_condition}
\end{equation}
Under this condition, branch deformation is sufficiently suppressed to justify the BSA~\eqref{eq:BSA} at the state level and the QGT partition law~\eqref{eq:QGT_partition_law} to leading order in $g_{\mathrm{lm}}^2$.

For a thermal state with the $|m\rangle_{\mathrm{M}}$ populations $p_m \propto \exp(-\beta_{\mathrm{th}}E_m)$ at inverse temperature $\beta_{\mathrm{th}}$, a narrowband probe also selects resonant emission from thermally occupied upper levels.
For a transition with $E_n-E_m \simeq \omega$, detailed balance gives an emission-to-absorption weight ratio of $p_n/p_m =\exp[-\beta_{\mathrm{th}}(E_n-E_m)] \simeq \exp(-\beta_{\mathrm{th}}\omega)$.
Thus, in addition to the spectral selectivity discussed above, the condition
\begin{equation}
\beta_{\mathrm{th}}\omega \gg 1
\label{eq:finiteT_BSA_condition}
\end{equation}
suppresses resonant branch deformation.
Under these conditions, the state-level BSA and the QGT partition law extend to thermal equilibrium to leading order in $g_{\mathrm{lm}}^2$ (Appendix~\ref{sec:app:mix}).
For a concrete estimate, room temperature gives $k_{\mathrm{B}}T/h \approx 6.3~\mathrm{THz}$, so a mid-infrared or higher-frequency probe can suppress the resonant branch-deformation weight to below approximately $1\%$ of the absorption weight.

This frequency-selection mechanism requires the matter state to remain stationary during the probe and a sufficiently narrowband, long pulse to be available.
It cannot justify BSA for rapidly evolving nonequilibrium matter states, whose dynamics must instead be resolved using short probe pulses.
Quantum-metric measurements in that regime are discussed in Sec.~\ref{sec:Neq_QFI}.

\subsection{Dynamical quantum geometry probed by cat light}
\label{sec:dynamical_quantum_geometry}
To clarify the physical meaning of the matter QGT accessed by kitten probes, we relate it to two-time many-body correlation functions filtered by the probe waveform.

To account for polarization and multiple optical modes, we express the EM-field difference between the two input coherent-state branches as
\begin{equation}
\bm{A}_{q}(t;\bm{\lambda})=\sum_{\mu} \lambda_{\mu} \bm{F}_{\mu}(t),
\end{equation}
where $\bm{F}_{\mu}(t)$ denotes the tangent field associated with the parameter $\lambda_{\mu}$.
For a probe mode with central frequency $\omega_{\mu}$, temporal envelope $f_{\mu}(t)$, and Jones vector $\bm{s}_{\mu}$, this field takes the form
\begin{equation}
\bm{F}_{\mu}(t) = g_{\mathrm{lm},\mu}[\bm{s}_{\mu}f_{\mu}(t) e^{-i\omega_{\mu}t} + \bm{s}_{\mu}^* f_{\mu}^*(t) e^{i\omega_{\mu}t}].
\end{equation}

Let $\hat{U}_{\bm{\lambda}}$ denote the matter evolution operator driven by the parameter-dependent classical EM field.
The corresponding local generator is
\begin{equation}
\hat{h}_{\mu} = \left. i \hat{U}_{\bm{\lambda}}^{\dagger} \partial_{\mu} \hat{U}_{\bm{\lambda}}\right|_{\bm{\lambda}=0} = - \int dt \bm{F}_{\mu}(t) \cdot \hat{\bm{O}}_{\mathrm{I}}(t), \label{eq:local_generator}
\end{equation}
where $\hat{\bm{O}}_{\mathrm{I}}(t)$ is the matter operator in the interaction picture defined by the reference evolution at $\bm{\lambda}=0$.

For a pure initial matter state $|\psi(0)\rangle$, the QGT is given by the covariance of the local generators:
\begin{equation}
\mathcal{Q}^{\mathrm{M}}_{\mu\nu} 
= \langle \Delta \hat{h}_{\mu} \Delta \hat{h}_{\nu}\rangle,
\qquad
\Delta \hat{h}_{\mu} = \hat{h}_{\mu} - \langle \hat{h}_{\mu} \rangle.
\label{eq:covariance_QGT}
\end{equation}
This follows from $\partial_\mu|\psi_{\bm{\lambda}}\rangle|_{\bm{\lambda}=0} = -i \hat{U}_0 \hat{h}_{\mu}|\psi(0)\rangle$ after projecting out the component parallel to the state.

Substituting Eq.~\eqref{eq:local_generator} yields
\begin{equation}
\mathcal{Q}^{\mathrm{M}}_{\mu\nu} 
= \int dt\,dt' \,\bm{F}_{\mu}^{\mathsf{T}}(t) \bm{C}^{>}(t,t') \bm{F}_{\nu}(t'),
\label{eq:QGT_correlation_time}
\end{equation}
where the greater correlation function is defined as
\begin{equation}
C^{>}_{ab}(t,t') = \langle \delta \hat{O}_{a,\mathrm{I}}(t)\, \delta \hat{O}_{b,\mathrm{I}}(t') \rangle.
\label{eq:def:greater_correlator}
\end{equation}
Here, $a,b$ label the spatial components of $\bm{F}_\mu$ and $\hat{\bm{O}}$ (e.g., $a,b=x,y$ for two transverse polarizations).
Thus, the matter QGT is determined by two-time fluctuations filtered by the probe waveform.
In particular, for a single-mode probe with fixed polarization, the relevant quantum-geometric quantity is the quantum metric, which is determined by the symmetrized matter fluctuations (the Keldysh component):
\begin{equation}
C^{\mathrm{sym}}_{ab}(t,t') = \frac{1}{2} \langle \{ \delta \hat{O}_{a,\mathrm{I}}(t), \delta \hat{O}_{b,\mathrm{I}}(t') \} \rangle = \frac{i}{2} G^K_{ab}(t,t'), \label{eq:def:symmetric_fluctuation}
\end{equation}
through
\begin{equation}
g^{\mathrm{M}}_{\mu\nu} = \sum_{a,b} \int dt\,dt'\, F_{\mu,a}(t) F_{\nu,b}(t') C^{\mathrm{sym}}_{ab}(t,t'). \label{eq:quantum_metric_real-time}
\end{equation}

For the stationary equilibrium states considered in Sec.~\ref{sec:branchwise_separability_approximation:validity_of_BSA}, the correlation function depends only on the time difference, $C^{>}_{ab}(t,t')=C^{>}_{ab}(t-t')$.
In the frequency domain, the QGT becomes
\begin{align}
&\mathcal{Q}^{\mathrm{M}}_{\mu\nu}[F]
= \frac{1}{2\pi} \int d\omega \,
\tilde{\bm{F}}_{\mu}^{\dagger}(\omega)
\bm{C}^{>}(\omega)
\tilde{\bm{F}}_{\nu}(\omega)
\nonumber\\
&= \sum_{n}\sum_{a,b}
\tilde{F}_{\mu,a}^*(\Delta_{n})
\tilde{F}_{\nu,b}(\Delta_{n}) \langle 0|\delta \hat{O}_{a}|n\rangle
\langle n|\delta \hat{O}_{b}|0\rangle,
\label{eq:frequency_filtered_QGT}
\end{align}
where $\tilde{\bm{F}}(\omega)$ and $\bm{C}^{>}(\omega)$ denote the Fourier transforms of the tangent field and stationary correlation function, respectively, and the second equality assumes an initial ground state.

For a long, narrowband probe, the frequency-selection condition~\eqref{eq:BSA_condition} yields $\tilde{\bm{F}}_{\mu}(\Delta_n) \approx g_{\mathrm{lm},\mu} \bm{s}_{\mu}\tilde{f}_{\mu}(\Delta_n-\omega_{\mu})$.
The frequency-filtered quantum metric then reduces to
\begin{equation}
g^{\mathrm{M}}_{\mu\mu}(\omega) = g_{\mathrm{lm},\mu}^2 \sum_{n>0} |\langle n|\bm{s}_{\mu}\cdot \delta \hat{\bm{O}}|0\rangle|^2 |\tilde{f}_{\mu}(\Delta_n -\omega_{\mu})|^2. \label{eq:frequencty_filtered_metric}
\end{equation}
This quantity should be distinguished from the static ground-state quantum metric~\cite{CamposVenuti2007} and the Bloch quantum metric~\cite{Torma2023}.
Rather, it is a dynamical quantum metric: the pullback of the Fubini--Study metric on the matter-state manifold to the probe-parameter space $\{\lambda_{\mu}\}$ through the time-dependent source $\bm{A}_q(t;\bm{\lambda})$.
Varying the probe frequency or polarization therefore enables selective measurements of different components of many-body fluctuations.

In thermal equilibrium, the symmetrized fluctuations are related to the dissipative response through the fluctuation--dissipation theorem.
Defining the retarded response function as
\begin{equation}
\chi^{R}_{ab}(t,t') = i \Theta(t-t') \langle [\hat{O}_{a,\mathrm{I}}(t), \hat{O}_{b,\mathrm{I}}(t')]\rangle,
\end{equation}
the fluctuation--dissipation theorem~\cite{Callen1951, Kubo1966} gives
\begin{equation}
C^{\mathrm{sym}}_{ab}(\omega) = \coth \left( \frac{\beta_{\mathrm{th}} \omega}{2} \right)\, \chi''_{ab}(\omega).
\label{eq:FDT}
\end{equation}
Here, $\chi''_{ab}=(\chi^R_{ab} - [\chi^R_{ba}]^*)/(2i)$ denotes the dissipative part of the retarded response.
Thus, in thermal equilibrium, the quantum metric can also be reconstructed from conventional linear-response spectroscopy, provided that the temperature and probe spectrum are known.
Our approach instead directly accesses the symmetrized matter fluctuations and the associated quantum geometry through changes in the outgoing kitten probe, without reconstructing a response function.

Furthermore, this constraint linking fluctuations and response clarifies two distinct forms of loss in a cat probe: absorption and interbranch decoherence.
Equation~\eqref{eq:geometric_partition_law_finite_distance}, derived from interbranch decoherence, is expressed in terms of the reduction in branch separation, $|\alpha_q|^2 \to |\beta_q|^2$.
This indicates that the absorption-induced kittenization and the interbranch decoherence are not independent processes.
This interdependence follows from the unitary light--matter evolution and the state-level BSA~\eqref{eq:BSA}.
Consequently, when the state-level BSA holds, the quantum geometry of matter inferred from cat-branch decoherence can also be reconstructed from the response coefficients obtained using classical-light probes.
By contrast, nonequilibrium matter systems in which the fluctuation--dissipation theorem is violated~\cite{Harada2005} provide promising settings in which cat-branch decoherence can probe matter fluctuations that cannot be determined from the response alone.
Such nonequilibrium systems are analyzed in Sec.~\ref{sec:Neq_QFI}.

\subsection{Optical readout of quantum geometry}
\label{sec:optical_readout}
We present concrete optical protocols for reading out the quantum geometry of matter.
The QGT partition law [Eq.~\eqref{eq:QGT_decrease}] relates the matter QGT to changes in the optical tangent geometry of the coherent-state family, which can be accessed using kitten probes in the small-separation limit.
For a single mode, we first discuss readout of the matter quantum metric through the Husimi $Q$ function (Sec.~\ref{sec:Husimi}) and low-order field moments (Sec.~\ref{sec:field_fluctuation}).
We then show how two orthogonal polarization modes enable Berry-curvature readout (Sec.~\ref{sec:Berry_curvature}).

Tracing out matter from the normalized tangent state constructed from Eq.~\eqref{eq:product_rule_beta_psi} yields the outgoing optical state
\begin{equation}
\hat{\rho}_{\mathrm L}^{\mathrm{out}}(t)
= \frac{g^{\mathrm{L}}(t)}{g^{\mathrm{L}}(0)} |\beta_c;1\rangle\langle\beta_c;1|
+\frac{g^{\mathrm{M}}(t)}{g^{\mathrm{L}}(0)}|\beta_c\rangle\langle\beta_c|, \label{eq:reduced_tangent_light}
\end{equation}
where $g^{\mathrm{L}}(0) = 1$ is the quantum metric of the coherent-state manifold for the chosen parametrization~\eqref{eq:kitten_metric}.
The two coefficients represent the fractions of the initial quantum metric retained in light and transferred to matter, respectively.
We retain a general input center $\alpha_c$ here, although $\alpha_c=0$ is preferable for equilibrium spectroscopy to eliminate the common classical drive.

\subsubsection{Brightening of the Husimi dark point}
\label{sec:Husimi}
\begin{figure}[t]
\centering
\includegraphics[width=\columnwidth]{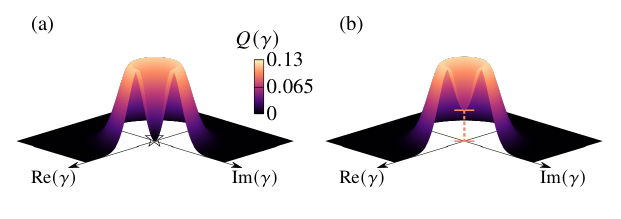}
\caption{
Brightening of the Husimi dark point of a tangent kitten probe induced by light--matter interaction.
(a)~Incident kitten state $|\alpha_c;1\rangle$.
The stellar zero at the coherent-state center is marked by an open star.
(b)~Outgoing light state after light--matter interaction [Eq.~\eqref{eq:reduced_tangent_light}].
The brightening of the initial stellar zero is illustrated using a representative mixing fraction $g^{\mathrm{M}}(t)/g^{\mathrm{L}}(0)=0.15$.
Both panels use $\alpha_c=\beta_c=0$ and cut away the first quadrant to reveal the central region.
}
\label{fig:husimi}
\end{figure}

We consider the Husimi $Q$ function
\begin{equation}
Q(\gamma) = \frac{1}{\pi} \langle\gamma|\hat{\rho}_{\mathrm{L}}|\gamma\rangle. \label{eq:def:Husimi}
\end{equation}
In the small-separation limit, the Husimi function of the incident kitten state $|\alpha_c;1\rangle$ has a zero (dark point) at the coherent-state center $\gamma = \langle \hat{a} \rangle = \alpha_c$.
Such zeros, known as stellar zeros, characterize the nonclassicality of optical states through the stellar representation~\cite{Chabaud2020}.
Figure~\ref{fig:husimi}(a) shows the Husimi function of this tangent kitten state $|\alpha_c;1\rangle$.

At the moving coherent-state center $\gamma = \langle \hat{a} \rangle = \beta_c$, the Husimi function of the outgoing light [Eq.~\eqref{eq:reduced_tangent_light}] satisfies
\begin{equation}
g^{\mathrm{M}} = g^{\mathrm{L}}(0)\, \pi Q(\langle \hat{a} \rangle). \label{eq:Husimi_metric_readout}
\end{equation}
As part of the optical quantum metric is transferred to matter through the light--matter interaction, the outgoing light undergoes quantum decoherence and becomes mixed, brightening the stellar zero of the incident kitten state.
Figure~\ref{fig:husimi}(b) illustrates the corresponding Husimi function of the outgoing light.
The Husimi function acquires a finite value at the point where it initially vanished, and the magnitude of this brightening is determined by the matter quantum metric.

Experimentally, several methods have been used to measure the Husimi function of quantum light~\cite{Banaszek1996, Muller2016, Luders2021}.
The moving center $\beta_c = \langle \hat{a}\rangle$ can first be determined by balanced homodyne detection~\cite{Lvovsky2009, Smithey1993, Yoon2026}.
Applying a displacement $\hat{\mathcal{D}}(-\beta_c)$ then allows the Husimi function at this point to be measured through the vacuum probability
\begin{equation}
P_0 = \langle0|
\hat{\mathcal{D}}(-\beta_c) \hat{\rho}_{\mathrm{L}}^{\mathrm{out}} \hat{\mathcal{D}}^{\dagger}(-\beta_c)
|0\rangle = \pi Q(\beta_c). \label{eq:vacuum_probability_Husimi}
\end{equation}
A particularly simple implementation is obtained for ground-state spectroscopy with $\alpha_c=\beta_c=0$, where the probe reduces to a single-photon Fock state.
In this case, the matter quantum metric is simply the no-click probability, $g^{\mathrm M}=P_0$, which can be estimated from repeated single-photon measurements.
This gives the conventional single-photon absorption signal a quantum-geometric interpretation in terms of the dynamical quantum metric of matter.
Alternatively, heterodyne detection provides direct samples of the Husimi $Q$ distribution~\cite{Muller2016, Luders2021}.

\subsubsection{Reduction of field fluctuations}
\label{sec:field_fluctuation}
For the same reduced state [Eq.~\eqref{eq:reduced_tangent_light}], consider the field fluctuations about the moving coherent center.
Defining $\delta \hat{a} = \hat{a} - \beta_c$ gives
\begin{equation}
n_{\mathrm{fl}} = \langle (\hat{a}-\beta_c)^{\dagger}(\hat{a}-\beta_c)\rangle = \frac{g^{\mathrm{L}}(t)}{g^{\mathrm{L}}(0)}.
\end{equation}
The QGT partition law then yields
\begin{equation}
g^{\mathrm{M}} =  (1 - \langle \delta \hat{a}^{\dagger} \delta \hat{a}\rangle )g^{\mathrm{L}}(0).
\end{equation}
The reduction in the weight of the displaced single-photon component thus directly yields the matter quantum metric.

These field fluctuations can also be measured by balanced homodyne detection~\cite{Lvovsky2009, Smithey1993, Yoon2026}.
Writing $\hat{a}=(\hat{X}+i\hat{P})/\sqrt{2}$ and defining the quadrature variances as $V_{X}=\langle (\delta \hat{X})^2\rangle$ and $V_{P}=\langle (\delta \hat{P})^2\rangle$, we obtain
\begin{equation}
n_{\mathrm{fl}} = \frac{V_X+V_P -1}{2}.
\end{equation}
Thus, the first moments determine the moving center $\beta_c$, while the second central moments determine $g^{\mathrm{M}}$.

\subsubsection{Berry-curvature measurements with two-polarization kitten light}
\label{sec:Berry_curvature}
Consider a kitten probe spanning two orthogonal polarization modes, $x$ and $y$.
Using the normalized Jones vector $\bm{s}=(s_x,s_y)^{\mathsf T}$, with $\bm{s}^{\dagger} \bm{s}=1$, to describe its polarization, we define the corresponding photon annihilation operator as $\hat{a}_{\bm{s}} = s_x^* \hat{a}_x + s_y^* \hat{a}_y$.

The two-mode Husimi $Q$ function is
\begin{equation}
Q^{(2)}_{\bm{s}}(\bm{\gamma}) = \frac{1}{\pi^2} \langle \bm{\gamma}|\hat{\rho}_{\mathrm{L}}|\bm{\gamma}\rangle.
\end{equation}
Generalizing Eq.~\eqref{eq:reduced_tangent_light} to two polarization modes, the Husimi function at the moving coherent center $\bm{\beta}_c=\langle\hat{\bm{a}}\rangle_{\bm{s}}$ satisfies
\begin{equation}
\pi^2 Q^{(2)}_{\bm{s}}(\langle \hat{\bm{a}}\rangle_{\bm{s}}) = \frac{ \bm{s}^{\dagger}\mathcal{Q}^{\mathrm{M}} \bm{s}}{g^{\mathrm{L}}(0)}, 
\label{eq:polarization_projected_QGT}
\end{equation}
where $\langle\hat{\bm{a}}\rangle_{\bm{s}}=(\langle\hat{a}_x\rangle_{\bm{s}},\langle\hat{a}_y\rangle_{\bm{s}})^{\mathsf T}$.
This relation follows because, in the frequency-selective regime considered here, the matter tangent state depends linearly on the probe polarization, $|D_{\bm{s}}\psi\rangle \simeq \sum_\mu s_\mu|D_\mu\psi\rangle$, since the off-resonant (counter-rotating) contribution proportional to $s_\mu^*$ is suppressed.
Consequently, $g_{\bm{s}}^{\mathrm{M}}=\langle D_{\bm{s}}\psi|D_{\bm{s}}\psi\rangle \simeq \bm{s}^{\dagger}\mathcal{Q}^{\mathrm{M}}\bm{s}$.

In particular, left- and right-circular polarizations give access to the Berry curvature, a nontrivial off-diagonal component of the QGT.
Their Jones vectors are defined by
\begin{equation}
\bm{s}_{\pm} = \frac{\bm{e}_x \pm i\bm{e}_y}{\sqrt{2}}
\end{equation}
and the brightened stellar zeros yield
\begin{equation}
\pi^2 Q^{(2)}_{\pm}(\langle \hat{\bm{a}}\rangle_{\pm}) = \frac{1}{2} \left( g_{xx}^{\mathrm{M}} + g_{yy}^{\mathrm{M}} \pm \Omega_{xy}^{\mathrm{M}} \right).
\end{equation}
Taking the difference between the results for the two circularly polarized kitten probes therefore gives
\begin{equation}
\Omega_{xy}^{\mathrm{M}} = \pi^2 [Q^{(2)}_+(\langle \hat{\bm{a}}\rangle_{+}) - Q^{(2)}_-(\langle \hat{\bm{a}}\rangle_{-})].
\label{eq:Berry_circular_dichroism}
\end{equation}
The Berry curvature can thus be read out directly from the difference in stellar-zero brightening between left- and right-circularly polarized tangent kittens.
This is a quantum-geometric counterpart of circular dichroism.
More generally, combining tangent kittens with linear, diagonal, and circular polarizations~\cite{Gaharwar2026} allows the full matter QGT to be reconstructed from phase-space signals of the outgoing light.

\subsection{Numerical demonstration}
\label{sec:GS_demo}
We numerically test the quantum geometric spectroscopy developed above using the LMG model, a collective many-body system.
This model also describes semiconductors whose conduction and valence bands are treated as an effective two-level system~\cite{Lipkin1965a}.
The LMG Hamiltonian for $N$ particles is
\begin{equation}
\hat{H}_{\mathrm{LMG}} = E_g \hat{J}_z - \frac{V}{N} \hat{J}_x^2. \label{eq:def:LMG_model}
\end{equation}
We set $\hbar=1$, work in the total-spin sector $J=N/2$, and use the energy splitting $E_g=1$ as the energy unit.
The parameter $V$ is the effective interaction strength. In the thermodynamic limit, a quantum phase transition occurs at $V=E_g$, separating the normal phase at $V<E_g$ from the ordered phase at $V >E_g$~\cite{Dusuel2004}.

We initialize matter in the ground state of the LMG Hamiltonian, $|\psi(0)\rangle=|0\rangle_{\mathrm{M}}$, with excitation energies $\Delta_n=E_n-E_0$.

We consider the electric-dipole light--matter interaction
\begin{equation}
\hat{H}_{\mathrm{lm}} = - g_{\mathrm{lm}}\left[\mathcal{E}(t) \hat{a}  + \mathcal{E}^*(t)  \hat{a}^{\dagger} \right] \, d_0\hat{J}_x.
\end{equation}
Here, $d_0$ is the electric dipole moment, and we set $d_0=1$ unless otherwise stated.
We choose the probe vector-potential envelope $\mathcal{A}(t) = \exp[-(t-t_c)^2/(2\sigma^2)]/(\sqrt{2\pi}\sigma)$ and the corresponding electric-field envelope $\mathcal{E}(t) = i\omega \mathcal{A}(t) - \partial_t \mathcal{A}(t)$.
The optical field is a harmonic oscillator of frequency $\omega$, with Hamiltonian $\hat{H}_{\mathrm{L}}=\omega \hat{a}^{\dagger} \hat{a}$.

As discussed above, spectroscopy of ground-state quantum geometry requires a sufficiently narrowband, long probe pulse to select the tangent-transfer channel resonantly while suppressing the branch-deforming channel [Eq.~\eqref{eq:BSA_condition}].
For the Gaussian envelope above, $|\tilde{\mathcal{E}}(-\Delta_n-\omega)|/|\tilde{\mathcal{E}}(\Delta_n-\omega)| = \exp[-2\sigma^2\Delta_n\omega]$, so the state-level BSA is justified for resonant transitions $\Delta_n\simeq\omega$ when $\sigma\omega\gg1$.

The initial optical state is an HHG-conditioned kitten [Eq.~\eqref{eq:def:HHG_kitten}] with a finite but sufficiently small branch separation $|\alpha_q|$.
Starting from the product initial state $|\mathrm{HK}\rangle_{\mathrm{L}}|0\rangle_{\mathrm{M}}$, we compute the full light--matter dynamics by numerically solving the Schr\"{o}dinger equation with $\hat{H}_{\mathrm{L}}+\hat{H}_{\mathrm{lm}}+\hat{H}_{\mathrm{LMG}}$ using the fourth-order Runge--Kutta method.
The numerical calculations are performed with a time step of $dt=0.001$ and a photon Hilbert-space dimension of $120$.

We compare the optical estimate obtained from this full calculation with the frequency-filtered quantum metric calculated from the eigenstates of the isolated matter system.
Equation~\eqref{eq:frequencty_filtered_metric} gives the frequency-filtered quantum metric as
\begin{equation}
g^{\mathrm{M}}(\omega) = g_{\mathrm{lm}}^2 \sum_{n>0} |\langle n|d_0\hat{J}_x|0\rangle|^2 |\tilde{\mathcal{E}}(\Delta_n -\omega)|^2.
\label{eq:frequencty_filtered_metric_LMG}
\end{equation}
For the Gaussian vector-potential envelope, $\tilde{\mathcal{E}}(\Delta_n - \omega) = i \Delta_n e^{-\sigma^2(\Delta_n - \omega)^2/2} e^{i(\Delta_n - \omega) t_c}$, so $g^{\mathrm M}(\omega)$ is a frequency-resolved dynamical metric that selectively probes matter excitations coupled by $\hat J_x$ through the central frequency $\omega$.

Because the numerical calculation uses a finite $|\alpha_q|$, we reduce finite-separation corrections to the exact expression in Eq.~\eqref{eq:Husimi_metric_readout} by using the baseline-subtracted Husimi estimator
\begin{equation}
g^{\mathrm{M}}_{\mathrm{H}}(\omega) = \pi [Q^{\mathrm{out}}(\langle \hat{a}\rangle_{\mathrm{out}}) - Q^{\mathrm{in}}(\langle \hat{a}\rangle_{\mathrm{in}})].
\label{eq:Husimi_metric_finite_alphaq}
\end{equation}
In the tangent limit discussed above, we expect $\lim_{|\alpha_q|\to0} g^{\mathrm{M}}_{\mathrm{H}} = g^{\mathrm{M}}$.

\begin{figure}[t]
\centering
\includegraphics[width=\columnwidth]{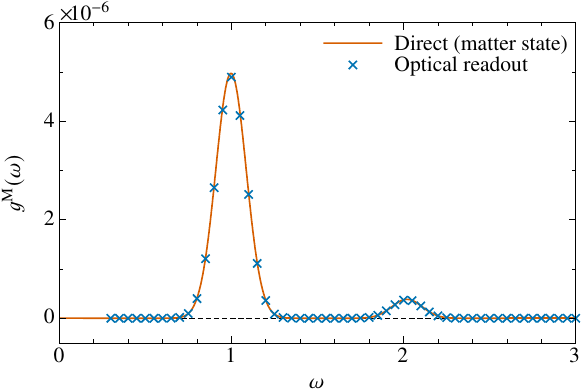}
\caption{
Quantum geometric spectrum probed by tangent kitten light.
Orange curve: frequency-filtered matter quantum metric $g^{\mathrm{M}}(\omega)$ directly evaluated from the spectral representation in Eq.~\eqref{eq:frequencty_filtered_metric_LMG}.
Blue crosses: optical estimate $g^{\mathrm{M}}_{\mathrm{H}}(\omega)$ reconstructed from the Husimi function after the full light--matter evolution using Eq.~\eqref{eq:Husimi_metric_finite_alphaq}.
Parameters are $N=32$, $V = 1.5$, $\alpha_q = 0.1$, $\alpha_c = 0$, $\sigma = 8$, and $g_{\mathrm{lm}}=10^{-3}$.
}
\label{fig:gs}
\end{figure}
Figure~\ref{fig:gs} uses $N=32$, $V=1.5$, $\alpha_q=0.1$, $\alpha_c=0$, $\sigma=8$, and $g_{\mathrm{lm}}=10^{-3}$ to compare the spectral representation of the matter quantum metric, $g^{\mathrm{M}}$ [Eq.~\eqref{eq:frequencty_filtered_metric_LMG}], with $g^{\mathrm{M}}_{\mathrm{H}}$ obtained from the full light--matter evolution [Eq.~\eqref{eq:Husimi_metric_finite_alphaq}], as a function of the optical frequency $\omega$.
For these parameters, the dominant finite-frequency excitations coupled by $\hat{J}_x$ occur at
$\Delta \simeq 1.0$ and $\Delta \simeq 2.0$.
The frequency-filtered metric therefore exhibits two resonant peaks.
A nearly degenerate low-energy state associated with the symmetry-broken regime is also present. Its spectral weight is strongly suppressed, however, because the electric field is the time derivative of the vector potential, giving $\tilde{\mathcal{E}}(\Delta_n - \omega) \propto \Delta_n$.
We therefore focus on the finite-frequency resonances for which the long-pulse BSA condition is satisfied.

As shown in Fig.~\ref{fig:gs}, the optical estimate $g^{\mathrm{M}}_{\mathrm{H}}$ (Blue crosses) quantitatively reproduces both the positions and amplitudes of the two resonant peaks in $g^{\mathrm M}(\omega)$ calculated directly from the matter states (Orange curve).
This agreement is consistent with the mechanism discussed in Sec.~\ref{sec:branchwise_separability_approximation:validity_of_BSA}.
It provides a numerical proof of principle that long-pulse frequency filtering suppresses the branch-deforming channel in ground-state spectroscopy, thereby validating the state-level BSA~\eqref{eq:BSA} and enabling the quantum geometry of matter to be read out through the QGT partition law~\eqref{eq:QGT_partition_law}.

\section{Time-resolved quantum-metric spectroscopy and entanglement witnessing} \label{sec:Neq_QFI}
We develop here a time-resolved quantum-metric spectroscopy of nonequilibrium matter using short cat-state probe pulses.
Unlike the equilibrium protocol of Sec.~\ref{sec:GS_QGT}, short-pulse spectroscopy cannot rely on frequency filtering to suppress branch deformation and thereby justify the QGT partition law.
Instead, we identify optical observables for which branch-deformation contributions are subleading relative to the matter-metric signal.

Section~\ref{sec:Neq_QFI:theory} considers general cat-state probes without taking the tangent limit.
We show that the overlap between the branch-dependent matter states encodes the transient quantum metric when the EM-field difference experienced by matter is sufficiently small, $g_{\mathrm{lm}}|\alpha_q|\ll1$, even if the optical branches are well separated.
For pure matter states, this metric also provides a QFI-based witness of multipartite entanglement.
Section~\ref{sec:Neq_QFI:readout} develops optical readout protocols using a branch-matched Weyl characteristic function or the purity of the outgoing light.
Finally, Section~\ref{sec:Neq_QFI:demo} numerically demonstrates time-resolved QFI witnessing in the quench dynamics of the LMG model.
Throughout this section, we set the coherent-state center $\alpha_c=0$ to eliminate the common classical drive.

\subsection{Nonequilibrium quantum metric and multipartite entanglement from cat-branch decoherence}
\label{sec:Neq_QFI:theory}

\subsubsection{Matter-tangent regime}
\label{sec:Neq_QFI:theory:matter_tangent}
We consider a cat probe formed by a quantum superposition of two coherent states $|\alpha_{\pm}\rangle$.
For weak electron--photon coupling $g_{\mathrm{lm}} \ll 1$, Eq.~\eqref{eq:branch_time_evolution} gives the joint evolution
\begin{align}
&(|\alpha_+\rangle + |\alpha_-\rangle)|\psi(0)\rangle \nonumber \\
\to&  |\beta_+\rangle|\psi_+\rangle + |\beta_-\rangle |\psi_-\rangle + g_{\mathrm{lm}} (|\beta_+;1\rangle|\eta_+\rangle + |\beta_-;1\rangle|\eta_-\rangle) \nonumber \\
&+ \mathcal{O}(g_{\mathrm{lm}}^2). \label{eq:cat_induced_total_system_dynamics}
\end{align}
Here, we omit the branch phases and normalization for simplicity.
The first two terms describe semiclassical light--matter dynamics in each branch, as product states of the classically driven matter states $|\psi_{\pm}\rangle$ and the outgoing coherent states $|\beta_{\pm}\rangle$.
The remaining terms describe branch deformation arising at first order in $g_{\mathrm{lm}}$.

The matter-state overlap $\langle\psi_-|\psi_+\rangle$, encoded in the interbranch coherence of the outgoing light, contains information about the quantum geometry of matter.
This overlap is a generalized Loschmidt amplitude~\cite{Gorin2006a, Quan2006} and can also be viewed as a Schwinger--Keldysh generating functional (a Feynman--Vernon influence functional)~\cite{Kamenev2009a, Feynman1963}, whose functional derivatives generate general nonequilibrium dynamics.
To derive quantum geometry from this overlap, we parameterize the EM-field amplitude experienced by matter, rather than the input coherent-state amplitude $\alpha$ used in Sec.~\ref{sec:GS_QGT}, as
\begin{equation}
g_{\mathrm{lm}}\alpha = g_{\mathrm{lm}}\alpha_- + \lambda' e^{i\varphi}, \label{eq:galpha_lambda'}
\end{equation}
where $0 \leq \lambda' \leq g_{\mathrm{lm}}|\alpha_q|$.
With $\alpha_c=0$, the midpoint between the two branches corresponds to $g_{\mathrm{lm}}\alpha=0$, or $\lambda'=g_{\mathrm{lm}}|\alpha_q|/2$.
For $g_{\mathrm{lm}}|\alpha_q|\ll1$, expanding the overlap about this midpoint yields the standard fidelity relation associated with the Fubini--Study metric~\cite{Provost1980}:
\begin{equation}
-\ln |\langle \psi_-|\psi_+\rangle|^2 = (g_{\mathrm{lm}}|\alpha_q|)^2 g^{\mathrm{M}}_{\lambda' \lambda'} + \mathcal{O}((g_{\mathrm{lm}}|\alpha_q|)^4).
\end{equation}
Thus, cat-branch decoherence directly encodes the quantum metric of the matter-state family with respect to the applied EM field.

The metric $g^{\mathrm{M}}_{\lambda\lambda}$ [Eq.~\eqref{eq:quantum_metric_real-time}] discussed in Sec.~\ref{sec:GS_QGT} is related to $g^{\mathrm{M}}_{\lambda'\lambda'}$ by the parameter rescaling in Eqs.~\eqref{eq:alpha_lambda} and~\eqref{eq:galpha_lambda'}:
\begin{equation}
g^{\mathrm{M}}_{\lambda\lambda}
=g_{\mathrm{lm}}^2g^{\mathrm{M}}_{\lambda'\lambda'},
\label{eq:quantum_metric_transformation}
\end{equation}
showing that both metrics characterize the same matter-state family in different parameter coordinates.
Unlike in Sec.~\ref{sec:GS_QGT}, however, no restriction is placed on the optical branch separation $|\alpha_q|$ itself; the condition applies only to the field difference experienced by matter, $g_{\mathrm{lm}}|\alpha_q|$, which must remain small. The tangent expansion is therefore applied to the matter-state family without requiring the optical tangent limit.

For time-resolved spectroscopy, we consider a finite pulse centered at $t_c$.
The corresponding tangent field conjugate to $\lambda'$ is $F'_{t_c}(t) = \partial A_q(t;\lambda')/\partial \lambda' = f_{t_c}(t) \exp(-i\omega t) + \mathrm{c.c.}$.
The general expression in Eq.~\eqref{eq:quantum_metric_real-time} gives the quantum metric measured by this pulse as
\begin{equation}
g^{\mathrm{M}}_{\lambda'\lambda'}[F';t_c] = \int dtdt'\, F'_{t_c}(t) F'_{t_c}(t') C^{\mathrm{sym}}(t,t'),
\label{eq:noneq_metric_time}
\end{equation}
where $C^{\mathrm{sym}}(t,t')$ denotes the symmetrized fluctuations of the matter operator $\hat{O}$ [Eq.~\eqref{eq:def:symmetric_fluctuation}].
Out of equilibrium, these fluctuations are not determined by the retarded response function through the fluctuation--dissipation theorem; the two carry independent information.
The transient quantum metric therefore contains information beyond the retarded response alone, motivating its direct readout through cat-branch decoherence.

\subsubsection{From quantum metric to time-resolved QFI}
\label{sec:Neq_QFI:theory:QFI}
We relate the transient quantum metric in Eq.~\eqref{eq:noneq_metric_time} to multipartite entanglement in pure matter states.
The QFI generated by a Hermitian operator $\hat{O}$ is
\begin{equation}
\mathcal{F}_{\mathrm{Q}}[\hat{O}](t) = 4\, \mathrm{Var}[\hat{O}(t)].
\end{equation}
For $\hat{O}=\sum_j\hat{o}_j$, where each local operator $\hat{o}_j$ has an eigenvalue range of at most $1$, the QFI density witnesses multipartite entanglement depth: $\mathcal{F}_{\mathrm{Q}}/N>k$ guarantees entanglement among at least $k+1$ particles in an $N$-particle system~\cite{Hyllus2012,Toth2012}.

The quantum metric measured by a finite pulse [Eq.~\eqref{eq:noneq_metric_time}] and the instantaneous QFI obey
\begin{align}
\frac{4g^{\mathrm{M}}_{\lambda'\lambda'}[F';t_c]}{\mathcal{I}_F^2} 
&\leq \left[ \int dt \, w_{t_c}(t)\sqrt{\mathcal{F}_{\mathrm{Q}}[\hat{O}](t)}
\right]^2 \nonumber \\
&\leq \int dt \, w_{t_c}(t) \mathcal{F}_{\mathrm{Q}}[\hat{O}](t). \label{eq:quantum_metric_QFI_inequality}
\end{align}
Here, $\mathcal{I}_F = \int dt |F'_{t_c}(t)|$ is the pulse area and $w_{t_c}(t) = |F'_{t_c}(t)|/\mathcal{I}_F$ is the normalized absolute-value pulse waveform.
To derive this inequality, we define the fluctuation vector
\begin{equation}
|v(t)\rangle = \delta \hat{O}_{\mathrm{I}}(t)|\psi(0)\rangle, \label{eq:|v>}
\end{equation}
which gives
\begin{equation}
g^{\mathrm{M}}_{\lambda'\lambda'}[F';t_c] = \left\| \int dt\,F'_{t_c}(t)|v(t)\rangle \right\|^2, \quad \mathcal{F}_{\mathrm{Q}}[\hat{O}](t)=4\| |v(t)\rangle \|^2.
\end{equation}
The triangle inequality gives
\begin{equation}
\sqrt{g^{\mathrm{M}}_{\lambda'\lambda'}[F';t_c]} 
\le \frac{1}{2} \int dt\,|F'_{t_c}(t)| \sqrt{\mathcal{F}_{\mathrm{Q}}[\hat{O}](t)},
\end{equation}
and applying the Cauchy--Schwarz inequality with weight $w_{t_c}$ yields inequality~\eqref{eq:quantum_metric_QFI_inequality}.
Thus, the quantity defined from the finite-pulse quantum metric~\eqref{eq:noneq_metric_time},
\begin{equation}
\mathcal{F}_{\mathrm{Q}}^{\mathrm{lb}}[\hat{O}](t_c) \equiv \frac{4g^{\mathrm{M}}_{\lambda'\lambda'}[F';t_c]}{\mathcal{I}_F^2}
\label{eq:QFI_lower_bound_metric}
\end{equation}
provides a lower bound on the QFI averaged over the probe time window.

We next examine the conditions under which inequality~\eqref{eq:quantum_metric_QFI_inequality} is saturated.
The first inequality is saturated when the signed fluctuation vectors $\mathrm{sgn}[F'_{t_c}(t)]|v(t)\rangle$ remain aligned in both direction and phase throughout the pulse window, requiring compensation of the carrier-phase evolution and sign changes of the probe waveform.
Raman-type interactions may help achieve this condition by reducing the carrier detuning or producing a slowly varying, approximately unipolar coupling in an appropriate rotating frame~\cite{Brion2007, Hammerer2010}.
The second inequality is saturated when the norm of $|v(t)\rangle$, and hence the QFI, remains constant over the pulse window.
In this case, the QFI averaged over the probe window approaches the instantaneous QFI at the pulse center:
\begin{equation}
\int dt\, w_{t_c}(t) \mathcal{F}_{\mathrm{Q}}[\hat{O}](t) \simeq \mathcal{F}_{\mathrm{Q}}[\hat{O}](t_c).
\end{equation}
When the first inequality is also saturated, the metric-based QFI lower bound in Eq.~\eqref{eq:QFI_lower_bound_metric} approaches the instantaneous QFI.
Thus, multipartite entanglement generated transiently within matter can be bounded from below through the branch decoherence of a short cat-state probe, enabling time-resolved optical tracking of this nonclassical many-body property.

\subsection{Optical readout of cat-branch decoherence}
\label{sec:Neq_QFI:readout}
We develop two optical protocols for extracting the magnitude of the matter-state overlap $\langle\psi_-|\psi_+\rangle$ from the outgoing light.
The first uses the Weyl characteristic function (Sec.~\ref{sec:Neq_QFI:Weyl}), while the second uses optical purity (Sec.~\ref{sec:Neq_QFI:purity}).
Because intrabranch light--matter entanglement [the branch-deformation terms in Eq.~\eqref{eq:cat_induced_total_system_dynamics}] can affect optical observables at the same order in $g_{\mathrm{lm}}$ as the matter-metric signal, we identify observables for which these corrections are subleading.
Unless otherwise stated, we describe optical readout in a rotating frame defined by the phase-locked local oscillator (LO).

\subsubsection{Branch-matched Weyl characteristic function}
\label{sec:Neq_QFI:Weyl}
We consider a general cat-state probe
\begin{equation}
|\mathrm{cat}\rangle =\frac{1}{\sqrt{\mathcal{N}}} (c_+ |\alpha_+\rangle + c_- |\alpha_-\rangle), \label{eq:general_cat_state}
\end{equation}
with normalization $\mathcal{N} = |c_+|^2 + |c_-|^2+2\mathrm{Re}[c_-^* c_+\langle \alpha_-|\alpha_+\rangle]$.
The evolution of the coupled light--matter system for this input can be obtained by applying Eq.~\eqref{eq:branch_time_evolution} to each branch.

\begin{figure}[t]
\centering
\includegraphics[width=\columnwidth]{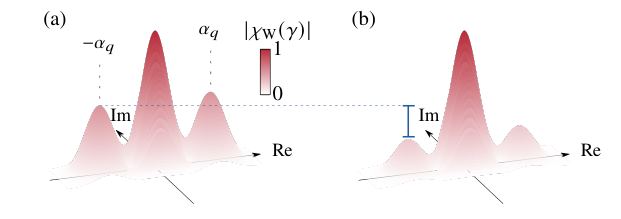}
\caption{
Effect of interbranch decoherence on the Weyl characteristic function of a symmetric cat state.
(a)~Incident pure cat state.
Peaks associated with interference between the coherent-state branches appear near $\gamma=\pm \alpha_q$.
(b)~Mixed state with phenomenologically reduced interbranch coherence [Eq.~\eqref{eq:decohered_cat}].
Quantum decoherence suppresses the peaks arising from interbranch interference.
Both panels use $\alpha_q=4$, with a coherence coefficient $c=0.5$ in (b).
}
\label{fig:weyl}
\end{figure}

We use the optical Weyl characteristic function, defined as the expectation value of the displacement operator $\hat{\mathcal{D}}(\gamma)$:
\begin{equation}
\chi_{\mathrm{W}}(\gamma) = \mathrm{Tr}_{\mathrm{L}} [ \hat{\rho}_{\mathrm{L}} \hat{\mathcal{D}}(\gamma)]. \label{eq:def:Weyl_characteristic_function}
\end{equation}
For a cat state such as Eq.~\eqref{eq:general_cat_state}, the diagonal coherent-state contributions decay with the Gaussian envelope $\exp(-|\gamma|^2/2)$, whereas interference between the two branches produces additional peaks around the branch separations $\gamma \simeq \pm \alpha_q$.
Figure~\ref{fig:weyl}(a) illustrates this feature for a symmetric cat state ($\alpha_+=-\alpha_-$, $c_+=c_-$).
We next illustrate phenomenologically how quantum decoherence affects these interbranch interference peaks.
To this end, we introduce an interbranch coherence coefficient $c$ and consider the mixed state
\begin{equation}
\hat{\rho}_{\mathrm{L}}= \frac{1}{\mathcal{N'}}\left( |\alpha\rangle \langle \alpha| + |{-}\alpha\rangle \langle {-}\alpha| + c|{-}\alpha\rangle \langle \alpha| + c^{*}|\alpha\rangle \langle{-}\alpha|\right).
\label{eq:decohered_cat}
\end{equation}
Decoherence can be modeled by taking $|c|<1$. Figure~\ref{fig:weyl}(b) shows the Weyl characteristic function for $c=0.5$.
Compared with the incident pure state [Fig.~\ref{fig:weyl}(a)], the peaks near $\gamma=\pm\alpha_q$ arising from interbranch interference are suppressed.

We show how changes in the Weyl interference peaks provide a readout of the matter quantum metric.
Their usefulness can be quantified using displaced Fock states $|\beta;n\rangle$.
We focus on the Weyl characteristic function $\chi_{\mathrm{W}}(-\beta_q)$, evaluated at a displacement matched to the outgoing branch separation $\beta_q$.
From the Weyl algebra, we have $\langle \beta_-;n|\hat{\mathcal{D}}(-\beta_q)|\beta_+;m\rangle = \exp(i \varphi_{\mathrm{W}})\delta_{nm}$.
Here, $\varphi_{\mathrm{W}}$ is a phase arising from coherent displacement and is independent of $m$ and $n$.
In particular, the matrix elements with $n=0$ and $m=1$ (or $n=1$ and $m=0$) vanish.
Therefore, cross terms between the leading semiclassical component and the first-order branch-deformation component in Eq.~\eqref{eq:branch_time_evolution} do not contribute to interbranch interference in the Weyl function.
Meanwhile, contributions from the same branch are suppressed by $\exp(-|\beta_q|^2/2)$, while those from the oppositely ordered interbranch term are even more strongly suppressed by $\exp(-2|\beta_q|^2)$.
Thus, for $|\beta_q|\gg1$, we obtain
\begin{align}
\chi_{\mathrm{W}}^{\mathrm{out}}(-\beta_q) =& \frac{c_-^* c_+}{\mathcal{N}} e^{i \Phi_{-+}} \left[ \langle \psi_-|\psi_+\rangle + \mathcal{O}(g_{\mathrm{lm}}^2) \right] \nonumber \\
&+ \mathcal{O}\left(\mathrm{poly}(|\beta_q|) e^{-|\beta_q|^2/2} \right).
\label{eq:Weyl_overlap_readout}
\end{align}
The phase $\Phi_{-+}$ includes the branch-dependent phase $\theta_+-\theta_-$ from Eq.~\eqref{eq:branch_time_evolution} and $\varphi_{\mathrm{W}}$.
Equation~\eqref{eq:Weyl_overlap_readout} is a central result of this work: an appropriately branch-matched Weyl characteristic function isolates the overlap between the matter branches to leading order, even when light and matter become entangled within each branch.
As the matter quantum metric governs the loss of the overlap $|\langle\psi_-|\psi_+\rangle|$, it appears as a reduction in the height of the Weyl interference peak [Fig.~\ref{fig:weyl}(b)].
The same readout applies to the equilibrium states considered in Sec.~\ref{sec:GS_QGT} and to both transient and stationary nonequilibrium states.
For an initially mixed matter state $\hat{\rho}_{\mathrm{M}}$, the corresponding overlap is $\mathrm{Tr}[\hat{U}_+\hat{\rho}_{\mathrm{M}}\hat{U}_-^{\dagger}]$.

Quantitative metric readout requires $g_{\mathrm{lm}}|\alpha_q| \ll 1$ for the matter-tangent expansion, while the residual branch-deformation corrections in Eq.~\eqref{eq:Weyl_overlap_readout} must remain subleading.
We therefore consider the regime
\begin{equation}
1\ll|\alpha_q|\ll g_{\mathrm{lm}}^{-1}.
\label{eq:large_cat_small_matter_source}
\end{equation}
The target matter-metric signal is $\mathcal{O}[(g_{\mathrm{lm}}|\alpha_q|)^2]$, whereas the residual branch-deformation contribution is $\mathcal{O}(g_{\mathrm{lm}}^2)$.
Their ratio therefore scales as $\mathcal{O}(|\alpha_q|^{-2})$, so that $|\alpha_q|\gg1$ parametrically separates the effects of microscopic quantum fluctuations of light from the matter metric signal.
The large-separation condition also exponentially suppresses the residual diagonal and oppositely matched interference contributions to the branch-matched Weyl function.

We estimate the quantum metric and QFI by normalizing the outgoing Weyl signal to its input value.
In the regime of Eq.~\eqref{eq:large_cat_small_matter_source}, the input Weyl function is
\begin{equation}
\chi_{\mathrm{W}}^{\mathrm{in}}(-\alpha_q) = \frac{c_-^* c_+}{\mathcal{N}} e^{i \Phi_{-+}^{\mathrm{in}}}  + \mathcal{O}\left(\mathrm{poly}(|\alpha_q|) e^{-|\alpha_q|^2/2} \right),
\label{eq:Weyl_input_reference}
\end{equation}
so that the ratio to this input reference yields the many-body-history overlap:
\begin{equation}
|\langle \psi_-|\psi_+\rangle| = \left| \frac{\chi_{\mathrm{W}}^{\mathrm{out}}(-\beta_q)}{\chi_{\mathrm{W}}^{\mathrm{in}}(-\alpha_q)} \right| + \mathcal{O}\left(g_{\mathrm{lm}}^2,\, \mathrm{poly}(|\alpha_q|)e^{-|\alpha_q|^2/2} \right).
\end{equation}

Moreover, the experimentally unknown output branch separation $\beta_q$ can be replaced by the known input value $\alpha_q$ without changing the accuracy of the matter-metric estimate.
Since $\beta_q = \alpha_q + \mathcal{O}(g_{\mathrm{lm}}^2 \alpha_q)$ [Eq.~\eqref{eq:beta_vs_alpha}], this mismatch corrects the absolute value of the Weyl function at $\mathcal{O}(|\beta_q - \alpha_q|^2)$ in the first term of Eq.~\eqref{eq:Weyl_overlap_readout} and at $\mathcal{O}(g_{\mathrm{lm}}|\beta_q -\alpha_q|)$ in the second term.
Both are included in the existing $\mathcal{O}(g_{\mathrm{lm}}^2)$ residual and do not affect the metric-estimation accuracy.
Substituting this result into Eq.~\eqref{eq:QFI_lower_bound_metric} yields an optical estimate of the lower bound on the QFI averaged over the probe time window~\eqref{eq:quantum_metric_QFI_inequality}:
\begin{equation}
\mathcal{F}_{\mathrm{Q},\mathrm{W}}[\hat{O}](t_c) \equiv \frac{-4}{\mathcal{I}_{F}^2(g_{\mathrm{lm}}|\alpha_q|)^2} \ln \left| \frac{\chi_{\mathrm{W}}^{\mathrm{out}}(-\alpha_q)}{\chi_{\mathrm{W}}^{\mathrm{in}}(-\alpha_q)} \right|^2 \approx \mathcal{F}_{\mathrm{Q}}^{\mathrm{lb}}[\hat{O}](t_c). \label{eq:QFI_from_Weyl}
\end{equation}

The Weyl characteristic function at a selected displacement can be obtained directly from a quadrature distribution measured by balanced homodyne detection~\cite{Lvovsky2009,Kiesel2008}.
Consider the quadratures
\begin{equation}
\hat{x}_\phi= \frac{ \hat{a} e^{-i\phi} + \hat{a}^{\dagger} e^{i\phi}}{\sqrt{2}},
\qquad
\hat{p}_{\phi}=\frac{\hat{a}e^{-i\phi}- \hat{a}^{\dagger} e^{i\phi}}{\sqrt{2}i},
\end{equation}
and let $W_{\phi}(x,p)$ denote the Wigner function in the corresponding phase space.
We use the normalization $\int dxdpW_{\phi}(x,p)=1$.
Choosing the LO phase $\phi_q = \arg(\alpha_q) + \pi/2$ gives the displacement operator $\hat{\mathcal{D}}(-\alpha_q)=\exp(i\sqrt{2} |\alpha_q| \hat{x}_\phi)$, and hence the Weyl characteristic function
\begin{equation}
\chi_{\mathrm{W}}(-\alpha_q) = \int dxdp\, W_{\phi}(x,p) e^{i\sqrt{2}|\alpha_q| x},
\label{eq:Weyl_homodyne}
\end{equation}
where $P_\phi(x)=\int dp W_{\phi}(x,p)$ is the probability distribution of the quadrature $\hat{x}_{\phi}$ measured by single-axis homodyne detection.
Thus, a single LO phase $\phi_q$ suffices to determine $\chi_{\mathrm{W}}(-\alpha_q)$ from the Fourier component of the measured quadrature distribution at wave number $\sqrt{2}|\alpha_q|$.
No full reconstruction of the Wigner function is required.

In practice, the input Weyl signal $\chi_{\mathrm{W}}^{\mathrm{in}}$ can be replaced by a sample-free reference measured at the output of the same optical path.
This can mitigate common-mode effects of optical loss and phase noise, while changes in branch separation or visibility require separate calibration.

\subsubsection{Purity-based readout}
\label{sec:Neq_QFI:purity}
Because the magnitude of the branch overlap is sufficient to determine the matter quantum metric, the purity of the outgoing light provides an alternative readout.
Defining the purity as $\mathcal{P}_{\mathrm{L}}=\mathrm{Tr}[(\hat{\rho}_{\mathrm{L}}^{\mathrm{out}})^2]$, we obtain, for $|\alpha_q| \gg 1$,
\begin{align}
\mathcal{P}_{\mathrm{L}} =& \frac{|c_+|^4+|c_-|^4 + 2|c_+c_-|^2 |\langle \psi_-|\psi_+\rangle|^2}{(|c_+|^2+|c_-|^2)^2} \nonumber \\
&+ \mathcal{O}(g_{\mathrm{lm}}^2) + \mathcal{O}\left[{\mathrm{poly}}(|\alpha_q|) e^{-|\alpha_q|^2/2}\right].
\label{eq:purity_general_cat}
\end{align}
Here again, the moving-center condition $\langle\psi_\pm|\eta_\pm\rangle=0$ and the orthogonality of displaced Fock states eliminate nonexponential $\mathcal{O}(g_{\mathrm{lm}})$ corrections.

In particular, consider a balanced cat with $|c_+|=|c_-|$.
For an ideal optical setup, the metric is estimated as
\begin{equation}
g_{\lambda'\lambda'}^{\mathrm{M}}[F';t_c] \simeq -\frac{1}{(g_{\mathrm{lm}}|\alpha_q|)^2} \ln(2\mathcal{P}_{\mathrm{L}} - 1) \simeq 2\frac{1-\mathcal{P}_{\mathrm{L}}}{(g_{\mathrm{lm}}|\alpha_q|)^2},
\label{eq:metric_from_purity_ideal}
\end{equation}
where the final approximation is the leading-order expansion for $g_{\mathrm{lm}}|\alpha_q|\ll1$.

Optical purity can be measured by e.g., interfering two copies of the outgoing light on a 50:50 beam splitter~\cite{Ekert2002, Griffet2023}.
Two successive output pulses prepared under identical conditions can be overlapped by delaying one pulse by a repetition period.
The beam splitter transforms the input modes $\hat{a}_1$ and $\hat{a}_2$ into the sum and difference modes $\hat{a}_{\pm}=(\hat{a}_1\pm\hat{a}_2)/\sqrt{2}$.
Exchanging the inputs leaves the sum mode unchanged and reverses the sign of the difference mode.
Thus, the photon-number parity expectation value $\langle(-1)^{\hat{n}_-}\rangle$ of the difference mode equals the expectation value of the SWAP operator $S_{12}$, defined by $S_{12}|u\rangle_1|v\rangle_2=|v\rangle_1|u\rangle_2$.
For two independently prepared copies of $\hat{\rho}_{\mathrm{L}}$, this expectation value gives the optical purity $\mathrm{Tr}(\hat{\rho}_{\mathrm{L}}^2)$ without full state tomography.

As in the Weyl-based protocol, a sample-free reference can help account for extrinsic optical decoherence.
Using the purity contrast $2\mathcal{P}_{\mathrm{L}}-1$, we obtain the QFI estimate
\begin{equation}
\mathcal{F}_{\mathrm{Q}, \mathrm{P}}[\hat{O}](t_c) \simeq -\frac{4}{\mathcal{I}_{F}^2\, (g_{\mathrm{lm}}|\alpha_q|)^2}\, \ln\left| \frac{2\mathcal{P}_{\mathrm{L}}^{\mathrm{sample}}-1}{2\mathcal{P}_{\mathrm{L}}^{\mathrm{ref}}-1} \right|.
\end{equation}
Additional corrections are required, however, when extrinsic decoherence changes the branch separation.

More broadly, light--matter entanglement generically reduces the purity of initially pure probe light, suggesting the possibility of purity-based spectroscopy with optical states beyond cat states.
Which properties of matter can be extracted from such measurements, however, depends on the probe state and light--matter interaction and remains an open direction for future research.

\subsection{Numerical demonstration}
\label{sec:Neq_QFI:demo}
We analyze quench dynamics in the LMG model to demonstrate, as a proof of principle, that the time-dependent quantum metric of nonequilibrium matter can be read out through the interbranch decoherence of cat-probe light.
Using the relation between the pulse-filtered metric and QFI derived in Sec.~\ref{sec:Neq_QFI:theory:QFI}, we show that multipartite entanglement generated during the evolution can be tracked using only the outgoing light.

We use the matter Hamiltonian $\hat{H}_{\mathrm{LMG}}$ [Eq.~\eqref{eq:def:LMG_model}] and prepare its ground state at interaction strength $V=0$, namely $|\psi(0)\rangle = |J,-J\rangle$.
At $t=0$, we suddenly change the interaction strength to $V =  2 E_g$ and evolve the system with the postquench Hamiltonian.
In the thermodynamic limit, the postquench Hamiltonian lies in the symmetry-broken regime $V>E_g$, characterized by an order parameter associated with $\hat{J}_x$.
Although the quench preserves $\langle\hat{J}_x\rangle_t=0$, it can strongly enhance the fluctuations of $\hat{J}_x$, producing strong state distinguishability and a large quantum metric.
This evolution is characterized by the QFI of the pure state $|\psi(t)\rangle$,
\begin{equation}
\mathcal{F}_{\mathrm{Q}}(t)=
4\left[
\langle \hat J_x^2\rangle_t
-\langle\hat J_x\rangle_t^2
\right], \label{eq:QFI_J_x}
\end{equation}
and, for a system of $N$ spin-$1/2$ particles with $\hat{J}_x=\sum_{j=1}^{N} \hat{\sigma}_j^x/2$, the QFI-density criterion $\mathcal{F}_{\mathrm{Q}}/N>k$ is sufficient to guarantee an entanglement depth of at least $k+1$~\cite{Hyllus2012,Toth2012}.

To measure this transient QFI,
we use the balanced cat in Eq.~\eqref{eq:general_cat_state}, with $c_+=c_-$ and $\alpha_c=0$, as the input probe and couple it to the matter operator $\hat{J}_x$.
The total Hamiltonian consists of the free-photon term $\hat{H}_{\mathrm{L}} = \omega\hat{a}^{\dagger} \hat{a}$, the postquench Hamiltonian $\hat{H}_{\mathrm{LMG}}$, and an effective light--matter interaction tailored to measure the QFI of $\hat{J}_x$:
\begin{equation}
\hat{H}_{\mathrm{lm}} = -g_{\mathrm{lm}}  (f_{t_c}(t) \hat{a}  + f_{t_c}^*(t)\hat{a}^{\dagger} ) d_0\hat{J}_x. \label{eq:gamma_coupling}
\end{equation}
The temporal mode function is a Gaussian centered at $t_c$ with width $\sigma$,
\begin{equation}
f_{t_c}(t)=
\frac{1}{\sqrt{2\pi}\sigma}
\exp\left[
-\frac{(t-t_c)^2}{2\sigma^2}
\right] e^{i \omega t_c}.
\end{equation}
The phase factor $\exp(i\omega t_c)$ ensures that the same optical quadrature probes matter as the probe delay $t_c$ is varied.
We estimate the matter QFI from the change in the optical Weyl characteristic function using Eq.~\eqref{eq:QFI_from_Weyl}.
For a sufficiently short pulse, $\sigma\omega \ll 1$, the effective probe field is approximately unipolar, giving $\mathcal{I}_F = \int dt |F'_{t_c}(t)| \approx 2$.
By choosing $\sigma$ sufficiently short compared with both the carrier timescale $\omega^{-1}$ and the relevant matter-evolution timescales $V^{-1},E_g^{-1}$, the fluctuation vectors~\eqref{eq:|v>} vary little over the pulse window, allowing the metric-based QFI lower bound to approach the instantaneous QFI [as the equality holds in inequality~\eqref{eq:quantum_metric_QFI_inequality}].

We obtain the outgoing Weyl characteristic function by numerically solving the full light--matter Schr\"{o}dinger equation governed by $\hat{H}_{\mathrm{L}}+\hat{H}_{\mathrm{lm}}+\hat{H}_{\mathrm{LMG}}$ [Eq.~\eqref{eq:gamma_coupling}].
The numerical calculations are performed using the fourth-order Runge--Kutta method with a time step of $dt=0.001$ and a photon Hilbert-space dimension of $120$.

\begin{figure}[t]
\centering
\includegraphics[width=\columnwidth]{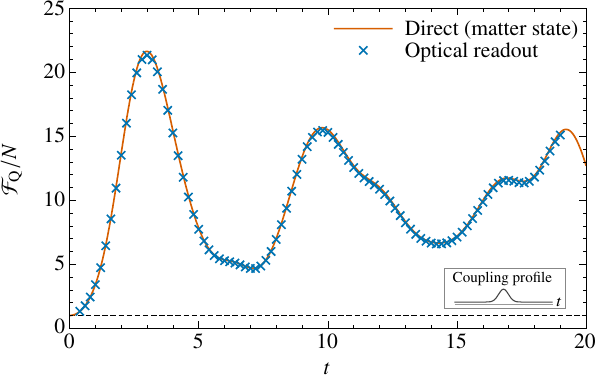}
\caption{
Time-resolved QFI spectroscopy with cat-state probe light,
demonstrated for quench dynamics in the LMG model.
Orange curve: QFI density $\mathcal{F}_{\mathrm{Q}}[\hat{J}_x]/N$ calculated directly from the matter state.
Blue crosses: QFI density estimated using Eq.~\eqref{eq:QFI_from_Weyl} from the change in the Weyl characteristic function $\chi_{\mathrm{W}}(-\alpha_q)$ of the cat-state probe.
Parameters: $N=32$, $E_g=1$, $V=0 \to 2E_g$, $g_{\mathrm{lm}}=10^{-3}$, $\alpha_q=4$, $\alpha_c=0$, $\omega = 1$, $\sigma=0.1$, and $d_0=1$.
}
\label{fig:te}
\end{figure}
Figure~\ref{fig:te} compares the time evolution of the QFI $\mathcal{F}_{\mathrm{Q}}[\hat{J}_x](t)$ calculated directly from the matter state [Eq.~\eqref{eq:QFI_J_x}; Orange curve] with the QFI inferred from a cat-state probe pulse centered at $t=t_c$, $\mathcal{F}_{\mathrm{Q},\mathrm{W}}(t_c)$ [Eq.~\eqref{eq:QFI_from_Weyl}; Blue crosses], using the change in its Weyl characteristic function.
The two agree well throughout the evolution.
This demonstrates that the appropriate branch-matched Weyl characteristic function admits an accurate BSA description and quantitatively reads out the matter quantum metric and QFI.
In particular, by the time of the first peak ($t_c \simeq 3$), the QFI grows from its initial value $\mathcal{F}_{\mathrm{Q}}(0)=N$ to a value of order $N^2$.
The LMG quench thus transiently generates macroscopic order-parameter fluctuations, multipartite entanglement, and QFI exhibiting Heisenberg scaling.
Cat-branch decoherence therefore enables time-resolved optical measurements of the dynamics that generate these macroscopic quantum properties in matter.

\section{Conclusions} \label{sec:conclusion}
We have developed a spectroscopic method that reads out the quantum geometry of matter through the interbranch decoherence of cat-state light.
For equilibrium matter, the light--matter QGT partition law enables frequency-selective measurements of the quantum metric and Berry curvature, for example through the brightening of a stellar zero in the outgoing Husimi function.
For pure nonequilibrium matter, changes in the optical Weyl characteristic function or purity reveal the transient quantum metric and, under the specified conditions, provide a time-resolved witness of multipartite entanglement.
This framework turns the degradation of a nonclassical optical superposition into a carrier of information about the sensitivity of matter to the optical field---how readily its quantum state changes on the state manifold.

Under BSA and in the tangent limit used for equilibrium spectroscopy, the outgoing light is a mixture of a displaced vacuum and a displaced single-photon state, with the matter quantum metric encoded in their relative weights.
Readout is therefore not restricted to the Husimi function and field fluctuations considered here; other optical observables, such as the Wigner function, can be used whenever they determine the same mixture weights.
Even when they access the same matter quantity, however, measurement schemes differ in statistical precision, making both ease of implementation and precision per input photon important criteria for choosing a readout.
In particular, Husimi-dark-point brightening can be read out directly as the vacuum probability $P_0$ [Eq.~\eqref{eq:vacuum_probability_Husimi}] through a binary vacuum/nonvacuum measurement.
In the ideal tangent limit without a common displacement, this binary measurement has Fisher information $I_{P_0}=1/[P_0(1-P_0)]$ and saturates the optical state's QFI for estimating $P_0$.
Because this readout has the same statistical structure as optimal single-photon loss estimation, its estimation variance can be reduced to $P_0$ times that obtained with a coherent-state probe with the same mean photon number~\cite{Adesso2009, Whittaker2017}.
This improved precision per photon reduces the exposure required for a given measurement accuracy.
Applications to biomolecular spectroscopy, including hemoglobin~\cite{Whittaker2017}, are therefore expected to benefit when optical heating or photodamage limits the allowable sample exposure~\cite{Casacio2021}.

The framework can also be extended to scattering into multiple optical modes.
For example, consider input light split into reflected ($\mathrm{R}$) and transmitted ($\mathrm{T}$) modes.
For stationary spectroscopy of equilibrium states, the state-level BSA gives $|\alpha(\lambda)\rangle_{\mathrm{in}} |\psi(0)\rangle_{\mathrm{M}} \to |\beta_{\mathrm{T}}(\lambda)\rangle _{\mathrm{T}} |\beta_{\mathrm{R}}(\lambda)\rangle_{\mathrm{R}} |\psi_{\lambda}\rangle_{\mathrm{M}}$, and the QGT partition law becomes $\mathcal{Q}^{\mathrm{M}} = \mathcal{Q}^{\mathrm{in}} -  \mathcal{Q}^{\mathrm{T}}-  \mathcal{Q}^{\mathrm{R}}$.
The matter QGT can therefore be read out by measuring the QGT changes in both transmission and reflection.
The stellar-zero-brightening relation [Eq.~\eqref{eq:Husimi_metric_readout}] likewise generalizes to $g^{\mathrm{M}}=g^{\mathrm{L}}(0)[\pi Q_{\mathrm{T}}(\langle \hat{a}\rangle_{\mathrm{T}}) + \pi Q_{\mathrm{R}}(\langle \hat{a}\rangle_{\mathrm{R}}) -1]$.
For time-resolved quantum metric spectroscopy of nonequilibrium matter, a suitably branch-matched multimode Weyl function provides an analogous readout.

Recent advances in cat-light-driven nonlinear optics~\cite{Lamprou2025} and nonperturbative light--matter dynamics~\cite{Imai2026, Imai2026b} motivate extending this approach beyond a local quantum-geometric description valid for small interbranch field differences.
At finite field differences or under nonlinear coupling, interbranch coherence can encode higher-order many-body correlations.
Identifying which correlations can be extracted quantitatively from such signals remains an important direction for future research.

Another natural extension is to other forms of nonclassical light, such as squeezed light~\cite{Casacio2021,Rasputnyi2024, Eto2025, Imai2026a}.
The underlying idea---using changes in optical nonclassicality to read out material properties spectroscopically---can also apply to quantum resources such as squeezing and entanglement.
Establishing the correspondence between optical nonclassicality and matter correlations for different probe states and light--matter interactions will clarify the broader applicability of this principle.

Even within these broader generalizations, cat states remain fundamental quantum probes.
The matter-branch overlap $\langle \psi_-|\psi_+\rangle$ encoded in the cat branches corresponds to the Schwinger--Keldysh generating functional for the matter operator coupled to light.
Functional derivatives with respect to external sources on the forward and backward time branches systematically generate nonequilibrium many-body correlation functions~\cite{Kamenev2009a}.
Expanding the logarithm of $\langle\psi_-|\psi_+\rangle$ in the classical source $A_c\propto \alpha_+ + \alpha_-$ and quantum source $A_q \propto \alpha_+ - \alpha_-$ gives, at low order, the retarded response from the $A_c A_q$ term and symmetrized fluctuations from the $A_q^2$ term.
Of these two essential quantities in the generating functional for general quantum many-body dynamics, conventional linear-response spectroscopy accesses the retarded response, while the cat-branch decoherence accesses symmetrized fluctuations.
By physically implementing a Keldysh quantum source $A_q$ through their interbranch field difference, cat-state probes provide a fluctuation-sensitive channel complementary to conventional response-based spectroscopy.

\section*{Acknowledgments}
We thank Koichiro Tanaka, Yujiro Eto, Kento Uchida, Akihito Ishizaki, Yuta Fujihashi, and Mamoru Endo for fruitful discussions.
S.I. was supported by JSPS KAKENHI (Grant Nos. JP25K17343 and JP26K21749).
M.F.C.~acknowledges support by the Quantum Science and Technology-National Science and Technology Major Project (Grant No. 2025ZD0301000),  the National Key Research and Development Program of China (Grant No. 2023YFA1407100), the Guangdong Province Science, Technology Major Project (Future functional materials under extreme conditions - 2021B0301030005) and the National Natural Science Foundation of China (Grant No. 12574092).

\appendix
\section{Mixed initial states and the thermal criterion for BSA} \label{sec:app:mix}
We derive the condition~\eqref{eq:finiteT_BSA_condition} under which the BSA holds for matter in thermal equilibrium.
Consider a general mixed state, independent of the initial coherent-state amplitude parameter $\lambda$,
\begin{equation}
\hat{\rho}(0) = \sum_m p_m |m\rangle \langle m|, \label{eq:def:initial_mixed_state}
\end{equation}
which can be purified by introducing an auxiliary system $\mathrm{R}$ as
\begin{equation}
|\psi(0)\rangle = \sum_m \sqrt{p_m} |m\rangle_{\mathrm{M}}|m\rangle_{\mathrm{R}}. \label{eq:purification}
\end{equation}
Leaving $\mathrm{R}$ unevolved and denoting the branch-dependent matter evolution generated by $\hat{H}_{\alpha}(t)$ [Eq.~\eqref{eq:H_alpha}] by $\hat{U}_{\alpha}$, we define the driven state family as
\begin{equation}
|\psi_{\alpha}\rangle = (\hat{U}_{\alpha} \otimes \hat{I}_{\mathrm{R}})|\psi(0)\rangle.
\end{equation}
Using the Fubini--Study QGT of this purified state family as the matter geometry, we extend the BSA [Eq.~\eqref{eq:BSA}] and the associated QGT partition law [Eq.~\eqref{eq:QGT_partition_law}] to mixed initial states.

The auxiliary system $\mathrm{R}$ is introduced only to purify the mixed state and need not be realized experimentally.
The resulting geometry corresponds to a fixed purification and generally differs from the Bures metric~\cite{Bures1969} or mixed-state QFI~\cite{Braunstein1994}.

We now specialize the fixed purification in Eq.~\eqref{eq:purification} to a Gibbs state and derive the thermal criterion in Eq.~\eqref{eq:finiteT_BSA_condition}.
The initial matter density matrix is
\begin{equation}
\hat{\rho}_{\mathrm{th}} =
\frac{e^{-\beta_{\mathrm{th}}\hat{H}_{\mathrm{M}}}}{Z},
\qquad
p_m=\frac{e^{-\beta_{\mathrm{th}}E_m}}{Z}.
\end{equation}
The thermally weighted matter excitation amplitudes can then be written as
\begin{align}
&\sum_m \sqrt{p_m}|\xi_{m,\mathrm{I}}\rangle|m\rangle_{\mathrm{R}} \nonumber \\
&=i\sum_{m,n}\langle n|\delta\hat{O}|m\rangle\sqrt{p_m} \tilde{f}(\Delta_{nm}-\omega)|n\rangle_{\mathrm{M}}|m\rangle_{\mathrm{R}}, \label{eq:thermal_partial_tangent_transfer_channel} \\
&\sum_m \sqrt{p_m}|\eta_{m,\mathrm{I}}\rangle|m\rangle_{\mathrm{R}} \nonumber \\
&=i\sum_{m,n}(\langle n|\delta\hat{O}|m\rangle)^*\sqrt{p_n} \tilde{f}^*(\Delta_{nm}-\omega)|m\rangle_{\mathrm{M}}|n\rangle_{\mathrm{R}}.
\label{eq:thermal_branch_deforming_channel}
\end{align}
Here, $\Delta_{nm}=E_n-E_m$, and the second expression follows by interchanging $m$ and $n$ and using the Hermiticity of $\hat{O}$.

For a sufficiently narrowband probe, the spectral factor selects transitions with $\Delta_{nm}\simeq\omega$.
The first amplitude~\eqref{eq:thermal_partial_tangent_transfer_channel} describes photon absorption from $|m\rangle_{\mathrm{M}}$ to $|n\rangle_{\mathrm{M}}$, whereas the second one~\eqref{eq:thermal_branch_deforming_channel} describes the reverse thermal emission process.
The corresponding squared coefficients differ by the thermal occupation ratio $p_n/p_m = \exp(-\beta_{\mathrm{th}}\Delta_{nm}) \simeq \exp(-\beta_{\mathrm{th}}\omega)$.
This gives the thermal criterion in Eq.~\eqref{eq:finiteT_BSA_condition}, under which the branch-deformation contribution is exponentially suppressed relative to the tangent-transfer contribution.
Together with the spectral selectivity of Eq.~\eqref{eq:BSA_condition}, it justifies the partition law at leading order in $g_{\mathrm{lm}}^2$.

\bibliography{reference}

\begin{thebibliography}{72}%
\makeatletter
\providecommand \@ifxundefined [1]{%
 \@ifx{#1\undefined}
}%
\providecommand \@ifnum [1]{%
 \ifnum #1\expandafter \@firstoftwo
 \else \expandafter \@secondoftwo
 \fi
}%
\providecommand \@ifx [1]{%
 \ifx #1\expandafter \@firstoftwo
 \else \expandafter \@secondoftwo
 \fi
}%
\providecommand \natexlab [1]{#1}%
\providecommand \enquote  [1]{``#1''}%
\providecommand \bibnamefont  [1]{#1}%
\providecommand \bibfnamefont [1]{#1}%
\providecommand \citenamefont [1]{#1}%
\providecommand \href@noop [0]{\@secondoftwo}%
\providecommand \href [0]{\begingroup \@sanitize@url \@href}%
\providecommand \@href[1]{\@@startlink{#1}\@@href}%
\providecommand \@@href[1]{\endgroup#1\@@endlink}%
\providecommand \@sanitize@url [0]{\catcode `\\12\catcode `\$12\catcode `\&12\catcode `\#12\catcode `\^12\catcode `\_12\catcode `\%12\relax}%
\providecommand \@@startlink[1]{}%
\providecommand \@@endlink[0]{}%
\providecommand \url  [0]{\begingroup\@sanitize@url \@url }%
\providecommand \@url [1]{\endgroup\@href {#1}{\urlprefix }}%
\providecommand \urlprefix  [0]{URL }%
\providecommand \Eprint [0]{\href }%
\providecommand \doibase [0]{https://doi.org/}%
\providecommand \selectlanguage [0]{\@gobble}%
\providecommand \bibinfo  [0]{\@secondoftwo}%
\providecommand \bibfield  [0]{\@secondoftwo}%
\providecommand \translation [1]{[#1]}%
\providecommand \BibitemOpen [0]{}%
\providecommand \bibitemStop [0]{}%
\providecommand \bibitemNoStop [0]{.\EOS\space}%
\providecommand \EOS [0]{\spacefactor3000\relax}%
\providecommand \BibitemShut  [1]{\csname bibitem#1\endcsname}%
\let\auto@bib@innerbib\@empty
\bibitem [{\citenamefont {Bohr}(1913)}]{Bohr1913}%
  \BibitemOpen
  \bibfield  {author} {\bibinfo {author} {\bibfnamefont {N.}~\bibnamefont {Bohr}},\ }\bibfield  {title} {\bibinfo {title} {{I. On the constitution of atoms and molecules}},\ }\href {https://doi.org/10.1080/14786441308634955} {\bibfield  {journal} {\bibinfo  {journal} {London, Edinburgh, Dublin Philos. Mag. J. Sci.}\ }\textbf {\bibinfo {volume} {26}},\ \bibinfo {pages} {1} (\bibinfo {year} {1913})}\BibitemShut {NoStop}%
\bibitem [{\citenamefont {Basov}\ \emph {et~al.}(2011)\citenamefont {Basov}, \citenamefont {Averitt}, \citenamefont {van~der Marel}, \citenamefont {Dressel},\ and\ \citenamefont {Haule}}]{Basov2011}%
  \BibitemOpen
  \bibfield  {author} {\bibinfo {author} {\bibfnamefont {D.~N.}\ \bibnamefont {Basov}}, \bibinfo {author} {\bibfnamefont {R.~D.}\ \bibnamefont {Averitt}}, \bibinfo {author} {\bibfnamefont {D.}~\bibnamefont {van~der Marel}}, \bibinfo {author} {\bibfnamefont {M.}~\bibnamefont {Dressel}},\ and\ \bibinfo {author} {\bibfnamefont {K.}~\bibnamefont {Haule}},\ }\bibfield  {title} {\bibinfo {title} {{Electrodynamics of correlated electron materials}},\ }\href {https://doi.org/10.1103/RevModPhys.83.471} {\bibfield  {journal} {\bibinfo  {journal} {Rev. Mod. Phys.}\ }\textbf {\bibinfo {volume} {83}},\ \bibinfo {pages} {471} (\bibinfo {year} {2011})}\BibitemShut {NoStop}%
\bibitem [{\citenamefont {Barth}(2007)}]{Barth2007}%
  \BibitemOpen
  \bibfield  {author} {\bibinfo {author} {\bibfnamefont {A.}~\bibnamefont {Barth}},\ }\bibfield  {title} {\bibinfo {title} {{Infrared spectroscopy of proteins}},\ }\href {https://doi.org/10.1016/j.bbabio.2007.06.004} {\bibfield  {journal} {\bibinfo  {journal} {Biochim. Biophys. Acta - Bioenerg.}\ }\textbf {\bibinfo {volume} {1767}},\ \bibinfo {pages} {1073} (\bibinfo {year} {2007})}\BibitemShut {NoStop}%
\bibitem [{\citenamefont {Jacques}(2013)}]{Jacques2013}%
  \BibitemOpen
  \bibfield  {author} {\bibinfo {author} {\bibfnamefont {S.~L.}\ \bibnamefont {Jacques}},\ }\bibfield  {title} {\bibinfo {title} {{Optical properties of biological tissues: a review}},\ }\href {https://doi.org/10.1088/0031-9155/58/11/R37} {\bibfield  {journal} {\bibinfo  {journal} {Phys. Med. Biol.}\ }\textbf {\bibinfo {volume} {58}},\ \bibinfo {pages} {R37} (\bibinfo {year} {2013})}\BibitemShut {NoStop}%
\bibitem [{\citenamefont {Zurek}(2003)}]{Zurek2003}%
  \BibitemOpen
  \bibfield  {author} {\bibinfo {author} {\bibfnamefont {W.~H.}\ \bibnamefont {Zurek}},\ }\bibfield  {title} {\bibinfo {title} {{Decoherence, einselection, and the quantum origins of the classical}},\ }\href {https://doi.org/10.1103/RevModPhys.75.715} {\bibfield  {journal} {\bibinfo  {journal} {Rev. Mod. Phys.}\ }\textbf {\bibinfo {volume} {75}},\ \bibinfo {pages} {715} (\bibinfo {year} {2003})}\BibitemShut {NoStop}%
\bibitem [{\citenamefont {Provost}\ and\ \citenamefont {Vallee}(1980)}]{Provost1980}%
  \BibitemOpen
  \bibfield  {author} {\bibinfo {author} {\bibfnamefont {J.~P.}\ \bibnamefont {Provost}}\ and\ \bibinfo {author} {\bibfnamefont {G.}~\bibnamefont {Vallee}},\ }\bibfield  {title} {\bibinfo {title} {{Riemannian structure on manifolds of quantum states}},\ }\href {https://doi.org/10.1007/BF02193559} {\bibfield  {journal} {\bibinfo  {journal} {Commun. Math. Phys.}\ }\textbf {\bibinfo {volume} {76}},\ \bibinfo {pages} {289} (\bibinfo {year} {1980})}\BibitemShut {NoStop}%
\bibitem [{\citenamefont {Anandan}\ and\ \citenamefont {Aharonov}(1990)}]{Anandan1990}%
  \BibitemOpen
  \bibfield  {author} {\bibinfo {author} {\bibfnamefont {J.}~\bibnamefont {Anandan}}\ and\ \bibinfo {author} {\bibfnamefont {Y.}~\bibnamefont {Aharonov}},\ }\bibfield  {title} {\bibinfo {title} {{Geometry of quantum evolution}},\ }\href {https://doi.org/10.1103/PhysRevLett.65.1697} {\bibfield  {journal} {\bibinfo  {journal} {Phys. Rev. Lett.}\ }\textbf {\bibinfo {volume} {65}},\ \bibinfo {pages} {1697} (\bibinfo {year} {1990})}\BibitemShut {NoStop}%
\bibitem [{\citenamefont {Braunstein}\ and\ \citenamefont {Caves}(1994)}]{Braunstein1994}%
  \BibitemOpen
  \bibfield  {author} {\bibinfo {author} {\bibfnamefont {S.~L.}\ \bibnamefont {Braunstein}}\ and\ \bibinfo {author} {\bibfnamefont {C.~M.}\ \bibnamefont {Caves}},\ }\bibfield  {title} {\bibinfo {title} {{Statistical distance and the geometry of quantum states}},\ }\href {https://doi.org/10.1103/PhysRevLett.72.3439} {\bibfield  {journal} {\bibinfo  {journal} {Phys. Rev. Lett.}\ }\textbf {\bibinfo {volume} {72}},\ \bibinfo {pages} {3439} (\bibinfo {year} {1994})}\BibitemShut {NoStop}%
\bibitem [{\citenamefont {Kolodrubetz}\ \emph {et~al.}(2017)\citenamefont {Kolodrubetz}, \citenamefont {Sels}, \citenamefont {Mehta},\ and\ \citenamefont {Polkovnikov}}]{Kolodrubetz2017}%
  \BibitemOpen
  \bibfield  {author} {\bibinfo {author} {\bibfnamefont {M.}~\bibnamefont {Kolodrubetz}}, \bibinfo {author} {\bibfnamefont {D.}~\bibnamefont {Sels}}, \bibinfo {author} {\bibfnamefont {P.}~\bibnamefont {Mehta}},\ and\ \bibinfo {author} {\bibfnamefont {A.}~\bibnamefont {Polkovnikov}},\ }\bibfield  {title} {\bibinfo {title} {{Geometry and non-adiabatic response in quantum and classical systems}},\ }\href {https://doi.org/10.1016/j.physrep.2017.07.001} {\bibfield  {journal} {\bibinfo  {journal} {Phys. Rep.}\ }\textbf {\bibinfo {volume} {697}},\ \bibinfo {pages} {1} (\bibinfo {year} {2017})}\BibitemShut {NoStop}%
\bibitem [{\citenamefont {Chabaud}\ \emph {et~al.}(2020)\citenamefont {Chabaud}, \citenamefont {Markham},\ and\ \citenamefont {Grosshans}}]{Chabaud2020}%
  \BibitemOpen
  \bibfield  {author} {\bibinfo {author} {\bibfnamefont {U.}~\bibnamefont {Chabaud}}, \bibinfo {author} {\bibfnamefont {D.}~\bibnamefont {Markham}},\ and\ \bibinfo {author} {\bibfnamefont {F.}~\bibnamefont {Grosshans}},\ }\bibfield  {title} {\bibinfo {title} {{Stellar Representation of Non-Gaussian Quantum States}},\ }\href {https://doi.org/10.1103/PhysRevLett.124.063605} {\bibfield  {journal} {\bibinfo  {journal} {Phys. Rev. Lett.}\ }\textbf {\bibinfo {volume} {124}},\ \bibinfo {pages} {063605} (\bibinfo {year} {2020})}\BibitemShut {NoStop}%
\bibitem [{\citenamefont {Takase}\ \emph {et~al.}(2021)\citenamefont {Takase}, \citenamefont {Yoshikawa}, \citenamefont {Asavanant}, \citenamefont {Endo},\ and\ \citenamefont {Furusawa}}]{Takase2021}%
  \BibitemOpen
  \bibfield  {author} {\bibinfo {author} {\bibfnamefont {K.}~\bibnamefont {Takase}}, \bibinfo {author} {\bibfnamefont {J.-i.}\ \bibnamefont {Yoshikawa}}, \bibinfo {author} {\bibfnamefont {W.}~\bibnamefont {Asavanant}}, \bibinfo {author} {\bibfnamefont {M.}~\bibnamefont {Endo}},\ and\ \bibinfo {author} {\bibfnamefont {A.}~\bibnamefont {Furusawa}},\ }\bibfield  {title} {\bibinfo {title} {{Generation of optical Schr{\"{o}}dinger cat states by generalized photon subtraction}},\ }\href {https://doi.org/10.1103/PhysRevA.103.013710} {\bibfield  {journal} {\bibinfo  {journal} {Phys. Rev. A}\ }\textbf {\bibinfo {volume} {103}},\ \bibinfo {pages} {013710} (\bibinfo {year} {2021})}\BibitemShut {NoStop}%
\bibitem [{\citenamefont {Endo}\ \emph {et~al.}()\citenamefont {Endo}, \citenamefont {Takase}, \citenamefont {Nomura}, \citenamefont {Sonoyama}, \citenamefont {Takahashi}, \citenamefont {Takasu}, \citenamefont {Fukuda}, \citenamefont {Kashiwazaki}, \citenamefont {Inoue}, \citenamefont {Umeki}, \citenamefont {van Loock}, \citenamefont {Marek}, \citenamefont {Filip}, \citenamefont {Asavanant},\ and\ \citenamefont {Furusawa}}]{Endo2026}%
  \BibitemOpen
  \bibfield  {author} {\bibinfo {author} {\bibfnamefont {M.}~\bibnamefont {Endo}}, \bibinfo {author} {\bibfnamefont {K.}~\bibnamefont {Takase}}, \bibinfo {author} {\bibfnamefont {T.}~\bibnamefont {Nomura}}, \bibinfo {author} {\bibfnamefont {T.}~\bibnamefont {Sonoyama}}, \bibinfo {author} {\bibfnamefont {K.}~\bibnamefont {Takahashi}}, \bibinfo {author} {\bibfnamefont {S.}~\bibnamefont {Takasu}}, \bibinfo {author} {\bibfnamefont {D.}~\bibnamefont {Fukuda}}, \bibinfo {author} {\bibfnamefont {T.}~\bibnamefont {Kashiwazaki}}, \bibinfo {author} {\bibfnamefont {A.}~\bibnamefont {Inoue}}, \bibinfo {author} {\bibfnamefont {T.}~\bibnamefont {Umeki}}, \bibinfo {author} {\bibfnamefont {P.}~\bibnamefont {van Loock}}, \bibinfo {author} {\bibfnamefont {P.}~\bibnamefont {Marek}}, \bibinfo {author} {\bibfnamefont {R.}~\bibnamefont {Filip}}, \bibinfo {author} {\bibfnamefont {W.}~\bibnamefont {Asavanant}},\ and\ \bibinfo {author} {\bibfnamefont {A.}~\bibnamefont {Furusawa}},\ }\bibfield  {title} {\bibinfo {title} {{Picosecond
  Schr{\"{o}}dinger cat states for ultrafast optical quantum processing}},\ }\Eprint {https://arxiv.org/abs/2606.24002} {arXiv:2606.24002} \BibitemShut {NoStop}%
\bibitem [{\citenamefont {Lewenstein}\ \emph {et~al.}(2021)\citenamefont {Lewenstein}, \citenamefont {Ciappina}, \citenamefont {Pisanty}, \citenamefont {Rivera-Dean}, \citenamefont {Stammer}, \citenamefont {Lamprou},\ and\ \citenamefont {Tzallas}}]{Lewenstein2021}%
  \BibitemOpen
  \bibfield  {author} {\bibinfo {author} {\bibfnamefont {M.}~\bibnamefont {Lewenstein}}, \bibinfo {author} {\bibfnamefont {M.~F.}\ \bibnamefont {Ciappina}}, \bibinfo {author} {\bibfnamefont {E.}~\bibnamefont {Pisanty}}, \bibinfo {author} {\bibfnamefont {J.}~\bibnamefont {Rivera-Dean}}, \bibinfo {author} {\bibfnamefont {P.}~\bibnamefont {Stammer}}, \bibinfo {author} {\bibfnamefont {T.}~\bibnamefont {Lamprou}},\ and\ \bibinfo {author} {\bibfnamefont {P.}~\bibnamefont {Tzallas}},\ }\bibfield  {title} {\bibinfo {title} {{Generation of optical Schr{\"{o}}dinger cat states in intense laser^^e2^^80^^93matter interactions}},\ }\href {https://doi.org/10.1038/s41567-021-01317-w} {\bibfield  {journal} {\bibinfo  {journal} {Nat. Phys.}\ }\textbf {\bibinfo {volume} {17}},\ \bibinfo {pages} {1104} (\bibinfo {year} {2021})}\BibitemShut {NoStop}%
\bibitem [{\citenamefont {Kubo}(1957)}]{Kubo1957}%
  \BibitemOpen
  \bibfield  {author} {\bibinfo {author} {\bibfnamefont {R.}~\bibnamefont {Kubo}},\ }\bibfield  {title} {\bibinfo {title} {{Statistical-Mechanical Theory of Irreversible Processes. I. General Theory and Simple Applications to Magnetic and Conduction Problems}},\ }\href {https://doi.org/10.1143/JPSJ.12.570} {\bibfield  {journal} {\bibinfo  {journal} {J. Phys. Soc. Japan}\ }\textbf {\bibinfo {volume} {12}},\ \bibinfo {pages} {570} (\bibinfo {year} {1957})}\BibitemShut {NoStop}%
\bibitem [{\citenamefont {Dorfman}\ \emph {et~al.}(2016)\citenamefont {Dorfman}, \citenamefont {Schlawin},\ and\ \citenamefont {Mukamel}}]{Dorfman2016}%
  \BibitemOpen
  \bibfield  {author} {\bibinfo {author} {\bibfnamefont {K.~E.}\ \bibnamefont {Dorfman}}, \bibinfo {author} {\bibfnamefont {F.}~\bibnamefont {Schlawin}},\ and\ \bibinfo {author} {\bibfnamefont {S.}~\bibnamefont {Mukamel}},\ }\bibfield  {title} {\bibinfo {title} {{Nonlinear optical signals and spectroscopy with quantum light}},\ }\href {https://doi.org/10.1103/RevModPhys.88.045008} {\bibfield  {journal} {\bibinfo  {journal} {Rev. Mod. Phys.}\ }\textbf {\bibinfo {volume} {88}},\ \bibinfo {pages} {045008} (\bibinfo {year} {2016})}\BibitemShut {NoStop}%
\bibitem [{\citenamefont {Fujihashi}\ \emph {et~al.}(2023)\citenamefont {Fujihashi}, \citenamefont {Miwa}, \citenamefont {Higashi},\ and\ \citenamefont {Ishizaki}}]{Fujihashi2023}%
  \BibitemOpen
  \bibfield  {author} {\bibinfo {author} {\bibfnamefont {Y.}~\bibnamefont {Fujihashi}}, \bibinfo {author} {\bibfnamefont {K.}~\bibnamefont {Miwa}}, \bibinfo {author} {\bibfnamefont {M.}~\bibnamefont {Higashi}},\ and\ \bibinfo {author} {\bibfnamefont {A.}~\bibnamefont {Ishizaki}},\ }\bibfield  {title} {\bibinfo {title} {{Probing exciton dynamics with spectral selectivity through the use of quantum entangled photons}},\ }\href {https://doi.org/10.1063/5.0169768} {\bibfield  {journal} {\bibinfo  {journal} {J. Chem. Phys.}\ }\textbf {\bibinfo {volume} {159}},\ \bibinfo {pages} {114201} (\bibinfo {year} {2023})}\BibitemShut {NoStop}%
\bibitem [{\citenamefont {Fujihashi}\ \emph {et~al.}(2024)\citenamefont {Fujihashi}, \citenamefont {Ishizaki},\ and\ \citenamefont {Shimizu}}]{Fujihashi2024}%
  \BibitemOpen
  \bibfield  {author} {\bibinfo {author} {\bibfnamefont {Y.}~\bibnamefont {Fujihashi}}, \bibinfo {author} {\bibfnamefont {A.}~\bibnamefont {Ishizaki}},\ and\ \bibinfo {author} {\bibfnamefont {R.}~\bibnamefont {Shimizu}},\ }\bibfield  {title} {\bibinfo {title} {{Pathway selectivity in time-resolved spectroscopy using two-photon coincidence counting with quantum entangled photons}},\ }\href {https://doi.org/10.1063/5.0189134} {\bibfield  {journal} {\bibinfo  {journal} {J. Chem. Phys.}\ }\textbf {\bibinfo {volume} {160}},\ \bibinfo {pages} {104201} (\bibinfo {year} {2024})}\BibitemShut {NoStop}%
\bibitem [{\citenamefont {Li}\ \emph {et~al.}(2023)\citenamefont {Li}, \citenamefont {Orcutt}, \citenamefont {Cook}, \citenamefont {Sabines-Chesterking}, \citenamefont {Tong}, \citenamefont {Schlau-Cohen}, \citenamefont {Zhang}, \citenamefont {Fleming},\ and\ \citenamefont {Whaley}}]{Li2023d}%
  \BibitemOpen
  \bibfield  {author} {\bibinfo {author} {\bibfnamefont {Q.}~\bibnamefont {Li}}, \bibinfo {author} {\bibfnamefont {K.}~\bibnamefont {Orcutt}}, \bibinfo {author} {\bibfnamefont {R.~L.}\ \bibnamefont {Cook}}, \bibinfo {author} {\bibfnamefont {J.}~\bibnamefont {Sabines-Chesterking}}, \bibinfo {author} {\bibfnamefont {A.~L.}\ \bibnamefont {Tong}}, \bibinfo {author} {\bibfnamefont {G.~S.}\ \bibnamefont {Schlau-Cohen}}, \bibinfo {author} {\bibfnamefont {X.}~\bibnamefont {Zhang}}, \bibinfo {author} {\bibfnamefont {G.~R.}\ \bibnamefont {Fleming}},\ and\ \bibinfo {author} {\bibfnamefont {K.~B.}\ \bibnamefont {Whaley}},\ }\bibfield  {title} {\bibinfo {title} {{Single-photon absorption and emission from a natural photosynthetic complex}},\ }\href {https://doi.org/10.1038/s41586-023-06121-5} {\bibfield  {journal} {\bibinfo  {journal} {Nature}\ }\textbf {\bibinfo {volume} {619}},\ \bibinfo {pages} {300} (\bibinfo {year} {2023})}\BibitemShut {NoStop}%
\bibitem [{\citenamefont {Whittaker}\ \emph {et~al.}(2017)\citenamefont {Whittaker}, \citenamefont {Erven}, \citenamefont {Neville}, \citenamefont {Berry}, \citenamefont {O'Brien}, \citenamefont {Cable},\ and\ \citenamefont {Matthews}}]{Whittaker2017}%
  \BibitemOpen
  \bibfield  {author} {\bibinfo {author} {\bibfnamefont {R.}~\bibnamefont {Whittaker}}, \bibinfo {author} {\bibfnamefont {C.}~\bibnamefont {Erven}}, \bibinfo {author} {\bibfnamefont {A.}~\bibnamefont {Neville}}, \bibinfo {author} {\bibfnamefont {M.}~\bibnamefont {Berry}}, \bibinfo {author} {\bibfnamefont {J.~L.}\ \bibnamefont {O'Brien}}, \bibinfo {author} {\bibfnamefont {H.}~\bibnamefont {Cable}},\ and\ \bibinfo {author} {\bibfnamefont {J.~C.~F.}\ \bibnamefont {Matthews}},\ }\bibfield  {title} {\bibinfo {title} {{Absorption spectroscopy at the ultimate quantum limit from single-photon states}},\ }\href {https://doi.org/10.1088/1367-2630/aa5512} {\bibfield  {journal} {\bibinfo  {journal} {New J. Phys.}\ }\textbf {\bibinfo {volume} {19}},\ \bibinfo {pages} {023013} (\bibinfo {year} {2017})}\BibitemShut {NoStop}%
\bibitem [{\citenamefont {Lysne}\ \emph {et~al.}(2023)\citenamefont {Lysne}, \citenamefont {Sch{\"{u}}ler},\ and\ \citenamefont {Werner}}]{Lysne2023}%
  \BibitemOpen
  \bibfield  {author} {\bibinfo {author} {\bibfnamefont {M.}~\bibnamefont {Lysne}}, \bibinfo {author} {\bibfnamefont {M.}~\bibnamefont {Sch{\"{u}}ler}},\ and\ \bibinfo {author} {\bibfnamefont {P.}~\bibnamefont {Werner}},\ }\bibfield  {title} {\bibinfo {title} {{Quantum Optics Measurement Scheme for Quantum Geometry and Topological Invariants}},\ }\href {https://doi.org/10.1103/PhysRevLett.131.156901} {\bibfield  {journal} {\bibinfo  {journal} {Phys. Rev. Lett.}\ }\textbf {\bibinfo {volume} {131}},\ \bibinfo {pages} {156901} (\bibinfo {year} {2023})}\BibitemShut {NoStop}%
\bibitem [{\citenamefont {Hauke}\ \emph {et~al.}(2016)\citenamefont {Hauke}, \citenamefont {Heyl}, \citenamefont {Tagliacozzo},\ and\ \citenamefont {Zoller}}]{Hauke2016}%
  \BibitemOpen
  \bibfield  {author} {\bibinfo {author} {\bibfnamefont {P.}~\bibnamefont {Hauke}}, \bibinfo {author} {\bibfnamefont {M.}~\bibnamefont {Heyl}}, \bibinfo {author} {\bibfnamefont {L.}~\bibnamefont {Tagliacozzo}},\ and\ \bibinfo {author} {\bibfnamefont {P.}~\bibnamefont {Zoller}},\ }\bibfield  {title} {\bibinfo {title} {{Measuring multipartite entanglement through dynamic susceptibilities}},\ }\href {https://doi.org/10.1038/nphys3700} {\bibfield  {journal} {\bibinfo  {journal} {Nat. Phys.}\ }\textbf {\bibinfo {volume} {12}},\ \bibinfo {pages} {778} (\bibinfo {year} {2016})}\BibitemShut {NoStop}%
\bibitem [{\citenamefont {Ozawa}\ and\ \citenamefont {Goldman}(2018)}]{Ozawa2018a}%
  \BibitemOpen
  \bibfield  {author} {\bibinfo {author} {\bibfnamefont {T.}~\bibnamefont {Ozawa}}\ and\ \bibinfo {author} {\bibfnamefont {N.}~\bibnamefont {Goldman}},\ }\bibfield  {title} {\bibinfo {title} {{Extracting the quantum metric tensor through periodic driving}},\ }\href {https://doi.org/10.1103/PhysRevB.97.201117} {\bibfield  {journal} {\bibinfo  {journal} {Phys. Rev. B}\ }\textbf {\bibinfo {volume} {97}},\ \bibinfo {pages} {201117} (\bibinfo {year} {2018})}\BibitemShut {NoStop}%
\bibitem [{\citenamefont {Kira}\ \emph {et~al.}(2011)\citenamefont {Kira}, \citenamefont {Koch}, \citenamefont {Smith}, \citenamefont {Hunter},\ and\ \citenamefont {Cundiff}}]{Kira2011}%
  \BibitemOpen
  \bibfield  {author} {\bibinfo {author} {\bibfnamefont {M.}~\bibnamefont {Kira}}, \bibinfo {author} {\bibfnamefont {S.~W.}\ \bibnamefont {Koch}}, \bibinfo {author} {\bibfnamefont {R.~P.}\ \bibnamefont {Smith}}, \bibinfo {author} {\bibfnamefont {A.~E.}\ \bibnamefont {Hunter}},\ and\ \bibinfo {author} {\bibfnamefont {S.~T.}\ \bibnamefont {Cundiff}},\ }\bibfield  {title} {\bibinfo {title} {{Quantum spectroscopy with Schr{\"{o}}dinger-cat states}},\ }\href {https://doi.org/10.1038/nphys2091} {\bibfield  {journal} {\bibinfo  {journal} {Nat. Phys.}\ }\textbf {\bibinfo {volume} {7}},\ \bibinfo {pages} {799} (\bibinfo {year} {2011})}\BibitemShut {NoStop}%
\bibitem [{\citenamefont {{\'{A}}lvarez}\ and\ \citenamefont {Suter}(2011)}]{Alvarez2011}%
  \BibitemOpen
  \bibfield  {author} {\bibinfo {author} {\bibfnamefont {G.~A.}\ \bibnamefont {{\'{A}}lvarez}}\ and\ \bibinfo {author} {\bibfnamefont {D.}~\bibnamefont {Suter}},\ }\bibfield  {title} {\bibinfo {title} {{Measuring the Spectrum of Colored Noise by Dynamical Decoupling}},\ }\href {https://doi.org/10.1103/PhysRevLett.107.230501} {\bibfield  {journal} {\bibinfo  {journal} {Phys. Rev. Lett.}\ }\textbf {\bibinfo {volume} {107}},\ \bibinfo {pages} {230501} (\bibinfo {year} {2011})}\BibitemShut {NoStop}%
\bibitem [{\citenamefont {Gorin}\ \emph {et~al.}(2006)\citenamefont {Gorin}, \citenamefont {Prosen}, \citenamefont {Seligman},\ and\ \citenamefont {{\v{Z}}nidari{\v{c}}}}]{Gorin2006a}%
  \BibitemOpen
  \bibfield  {author} {\bibinfo {author} {\bibfnamefont {T.}~\bibnamefont {Gorin}}, \bibinfo {author} {\bibfnamefont {T.}~\bibnamefont {Prosen}}, \bibinfo {author} {\bibfnamefont {T.~H.}\ \bibnamefont {Seligman}},\ and\ \bibinfo {author} {\bibfnamefont {M.}~\bibnamefont {{\v{Z}}nidari{\v{c}}}},\ }\bibfield  {title} {\bibinfo {title} {{Dynamics of Loschmidt echoes and fidelity decay}},\ }\href {https://doi.org/10.1016/j.physrep.2006.09.003} {\bibfield  {journal} {\bibinfo  {journal} {Phys. Rep.}\ }\textbf {\bibinfo {volume} {435}},\ \bibinfo {pages} {33} (\bibinfo {year} {2006})}\BibitemShut {NoStop}%
\bibitem [{\citenamefont {Quan}\ \emph {et~al.}(2006)\citenamefont {Quan}, \citenamefont {Song}, \citenamefont {Liu}, \citenamefont {Zanardi},\ and\ \citenamefont {Sun}}]{Quan2006}%
  \BibitemOpen
  \bibfield  {author} {\bibinfo {author} {\bibfnamefont {H.~T.}\ \bibnamefont {Quan}}, \bibinfo {author} {\bibfnamefont {Z.}~\bibnamefont {Song}}, \bibinfo {author} {\bibfnamefont {X.~F.}\ \bibnamefont {Liu}}, \bibinfo {author} {\bibfnamefont {P.}~\bibnamefont {Zanardi}},\ and\ \bibinfo {author} {\bibfnamefont {C.~P.}\ \bibnamefont {Sun}},\ }\bibfield  {title} {\bibinfo {title} {{Decay of Loschmidt Echo Enhanced by Quantum Criticality}},\ }\href {https://doi.org/10.1103/PhysRevLett.96.140604} {\bibfield  {journal} {\bibinfo  {journal} {Phys. Rev. Lett.}\ }\textbf {\bibinfo {volume} {96}},\ \bibinfo {pages} {140604} (\bibinfo {year} {2006})}\BibitemShut {NoStop}%
\bibitem [{\citenamefont {Knap}\ \emph {et~al.}(2012)\citenamefont {Knap}, \citenamefont {Shashi}, \citenamefont {Nishida}, \citenamefont {Imambekov}, \citenamefont {Abanin},\ and\ \citenamefont {Demler}}]{Knap2012}%
  \BibitemOpen
  \bibfield  {author} {\bibinfo {author} {\bibfnamefont {M.}~\bibnamefont {Knap}}, \bibinfo {author} {\bibfnamefont {A.}~\bibnamefont {Shashi}}, \bibinfo {author} {\bibfnamefont {Y.}~\bibnamefont {Nishida}}, \bibinfo {author} {\bibfnamefont {A.}~\bibnamefont {Imambekov}}, \bibinfo {author} {\bibfnamefont {D.~A.}\ \bibnamefont {Abanin}},\ and\ \bibinfo {author} {\bibfnamefont {E.}~\bibnamefont {Demler}},\ }\bibfield  {title} {\bibinfo {title} {{Time-Dependent Impurity in Ultracold Fermions: Orthogonality Catastrophe and Beyond}},\ }\href {https://doi.org/10.1103/PhysRevX.2.041020} {\bibfield  {journal} {\bibinfo  {journal} {Phys. Rev. X}\ }\textbf {\bibinfo {volume} {2}},\ \bibinfo {pages} {041020} (\bibinfo {year} {2012})}\BibitemShut {NoStop}%
\bibitem [{\citenamefont {Mazzola}\ \emph {et~al.}(2013)\citenamefont {Mazzola}, \citenamefont {{De Chiara}},\ and\ \citenamefont {Paternostro}}]{Mazzola2013}%
  \BibitemOpen
  \bibfield  {author} {\bibinfo {author} {\bibfnamefont {L.}~\bibnamefont {Mazzola}}, \bibinfo {author} {\bibfnamefont {G.}~\bibnamefont {{De Chiara}}},\ and\ \bibinfo {author} {\bibfnamefont {M.}~\bibnamefont {Paternostro}},\ }\bibfield  {title} {\bibinfo {title} {{Measuring the Characteristic Function of the Work Distribution}},\ }\href {https://doi.org/10.1103/PhysRevLett.110.230602} {\bibfield  {journal} {\bibinfo  {journal} {Phys. Rev. Lett.}\ }\textbf {\bibinfo {volume} {110}},\ \bibinfo {pages} {230602} (\bibinfo {year} {2013})}\BibitemShut {NoStop}%
\bibitem [{\citenamefont {T{\"{o}}rm{\"{a}}}(2023)}]{Torma2023}%
  \BibitemOpen
  \bibfield  {author} {\bibinfo {author} {\bibfnamefont {P.}~\bibnamefont {T{\"{o}}rm{\"{a}}}},\ }\bibfield  {title} {\bibinfo {title} {{Essay: Where Can Quantum Geometry Lead Us?}},\ }\href {https://doi.org/10.1103/PhysRevLett.131.240001} {\bibfield  {journal} {\bibinfo  {journal} {Phys. Rev. Lett.}\ }\textbf {\bibinfo {volume} {131}},\ \bibinfo {pages} {240001} (\bibinfo {year} {2023})}\BibitemShut {NoStop}%
\bibitem [{\citenamefont {Thouless}\ \emph {et~al.}(1982)\citenamefont {Thouless}, \citenamefont {Kohmoto}, \citenamefont {Nightingale},\ and\ \citenamefont {den Nijs}}]{Thouless1982}%
  \BibitemOpen
  \bibfield  {author} {\bibinfo {author} {\bibfnamefont {D.~J.}\ \bibnamefont {Thouless}}, \bibinfo {author} {\bibfnamefont {M.}~\bibnamefont {Kohmoto}}, \bibinfo {author} {\bibfnamefont {M.~P.}\ \bibnamefont {Nightingale}},\ and\ \bibinfo {author} {\bibfnamefont {M.}~\bibnamefont {den Nijs}},\ }\bibfield  {title} {\bibinfo {title} {{Quantized Hall Conductance in a Two-Dimensional Periodic Potential}},\ }\href {https://doi.org/10.1103/PhysRevLett.49.405} {\bibfield  {journal} {\bibinfo  {journal} {Phys. Rev. Lett.}\ }\textbf {\bibinfo {volume} {49}},\ \bibinfo {pages} {405} (\bibinfo {year} {1982})}\BibitemShut {NoStop}%
\bibitem [{\citenamefont {Peotta}\ and\ \citenamefont {T{\"{o}}rm{\"{a}}}(2015)}]{Peotta2015}%
  \BibitemOpen
  \bibfield  {author} {\bibinfo {author} {\bibfnamefont {S.}~\bibnamefont {Peotta}}\ and\ \bibinfo {author} {\bibfnamefont {P.}~\bibnamefont {T{\"{o}}rm{\"{a}}}},\ }\bibfield  {title} {\bibinfo {title} {{Superfluidity in topologically nontrivial flat bands}},\ }\href {https://doi.org/10.1038/ncomms9944} {\bibfield  {journal} {\bibinfo  {journal} {Nat. Commun.}\ }\textbf {\bibinfo {volume} {6}},\ \bibinfo {pages} {8944} (\bibinfo {year} {2015})}\BibitemShut {NoStop}%
\bibitem [{\citenamefont {Mead}\ and\ \citenamefont {Truhlar}(1979)}]{Mead1979}%
  \BibitemOpen
  \bibfield  {author} {\bibinfo {author} {\bibfnamefont {C.~A.}\ \bibnamefont {Mead}}\ and\ \bibinfo {author} {\bibfnamefont {D.~G.}\ \bibnamefont {Truhlar}},\ }\bibfield  {title} {\bibinfo {title} {{On the determination of Born^^e2^^80^^93Oppenheimer nuclear motion wave functions including complications due to conical intersections and identical nuclei}},\ }\href {https://doi.org/10.1063/1.437734} {\bibfield  {journal} {\bibinfo  {journal} {J. Chem. Phys.}\ }\textbf {\bibinfo {volume} {70}},\ \bibinfo {pages} {2284} (\bibinfo {year} {1979})}\BibitemShut {NoStop}%
\bibitem [{\citenamefont {Yuan}\ \emph {et~al.}(2018)\citenamefont {Yuan}, \citenamefont {Guan}, \citenamefont {Chen}, \citenamefont {Zhao}, \citenamefont {Yu}, \citenamefont {Luo}, \citenamefont {Tan}, \citenamefont {Xie}, \citenamefont {Wang}, \citenamefont {Sun}, \citenamefont {Zhang},\ and\ \citenamefont {Yang}}]{Yuan2018}%
  \BibitemOpen
  \bibfield  {author} {\bibinfo {author} {\bibfnamefont {D.}~\bibnamefont {Yuan}}, \bibinfo {author} {\bibfnamefont {Y.}~\bibnamefont {Guan}}, \bibinfo {author} {\bibfnamefont {W.}~\bibnamefont {Chen}}, \bibinfo {author} {\bibfnamefont {H.}~\bibnamefont {Zhao}}, \bibinfo {author} {\bibfnamefont {S.}~\bibnamefont {Yu}}, \bibinfo {author} {\bibfnamefont {C.}~\bibnamefont {Luo}}, \bibinfo {author} {\bibfnamefont {Y.}~\bibnamefont {Tan}}, \bibinfo {author} {\bibfnamefont {T.}~\bibnamefont {Xie}}, \bibinfo {author} {\bibfnamefont {X.}~\bibnamefont {Wang}}, \bibinfo {author} {\bibfnamefont {Z.}~\bibnamefont {Sun}}, \bibinfo {author} {\bibfnamefont {D.~H.}\ \bibnamefont {Zhang}},\ and\ \bibinfo {author} {\bibfnamefont {X.}~\bibnamefont {Yang}},\ }\bibfield  {title} {\bibinfo {title} {{Observation of the geometric phase effect in the H + HD → H$_2$ + D reaction}},\ }\href {https://doi.org/10.1126/science.aav1356} {\bibfield  {journal} {\bibinfo  {journal} {Science}\ }\textbf {\bibinfo {volume} {362}},\ \bibinfo
  {pages} {1289} (\bibinfo {year} {2018})}\BibitemShut {NoStop}%
\bibitem [{\citenamefont {Hyllus}\ \emph {et~al.}(2012)\citenamefont {Hyllus}, \citenamefont {Laskowski}, \citenamefont {Krischek}, \citenamefont {Schwemmer}, \citenamefont {Wieczorek}, \citenamefont {Weinfurter}, \citenamefont {Pezz{\'{e}}},\ and\ \citenamefont {Smerzi}}]{Hyllus2012}%
  \BibitemOpen
  \bibfield  {author} {\bibinfo {author} {\bibfnamefont {P.}~\bibnamefont {Hyllus}}, \bibinfo {author} {\bibfnamefont {W.}~\bibnamefont {Laskowski}}, \bibinfo {author} {\bibfnamefont {R.}~\bibnamefont {Krischek}}, \bibinfo {author} {\bibfnamefont {C.}~\bibnamefont {Schwemmer}}, \bibinfo {author} {\bibfnamefont {W.}~\bibnamefont {Wieczorek}}, \bibinfo {author} {\bibfnamefont {H.}~\bibnamefont {Weinfurter}}, \bibinfo {author} {\bibfnamefont {L.}~\bibnamefont {Pezz{\'{e}}}},\ and\ \bibinfo {author} {\bibfnamefont {A.}~\bibnamefont {Smerzi}},\ }\bibfield  {title} {\bibinfo {title} {{Fisher information and multiparticle entanglement}},\ }\href {https://doi.org/10.1103/PhysRevA.85.022321} {\bibfield  {journal} {\bibinfo  {journal} {Phys. Rev. A}\ }\textbf {\bibinfo {volume} {85}},\ \bibinfo {pages} {022321} (\bibinfo {year} {2012})}\BibitemShut {NoStop}%
\bibitem [{\citenamefont {T{\'{o}}th}(2012)}]{Toth2012}%
  \BibitemOpen
  \bibfield  {author} {\bibinfo {author} {\bibfnamefont {G.}~\bibnamefont {T{\'{o}}th}},\ }\bibfield  {title} {\bibinfo {title} {{Multipartite entanglement and high-precision metrology}},\ }\href {https://doi.org/10.1103/PhysRevA.85.022322} {\bibfield  {journal} {\bibinfo  {journal} {Phys. Rev. A}\ }\textbf {\bibinfo {volume} {85}},\ \bibinfo {pages} {022322} (\bibinfo {year} {2012})}\BibitemShut {NoStop}%
\bibitem [{\citenamefont {Kang}\ \emph {et~al.}(2025)\citenamefont {Kang}, \citenamefont {Kim}, \citenamefont {Qian}, \citenamefont {Neves}, \citenamefont {Ye}, \citenamefont {Jung}, \citenamefont {Puntel}, \citenamefont {Mazzola}, \citenamefont {Fang}, \citenamefont {Jozwiak}, \citenamefont {Bostwick}, \citenamefont {Rotenberg}, \citenamefont {Fuji}, \citenamefont {Vobornik}, \citenamefont {Park}, \citenamefont {Checkelsky}, \citenamefont {Yang},\ and\ \citenamefont {Comin}}]{Kang2025}%
  \BibitemOpen
  \bibfield  {author} {\bibinfo {author} {\bibfnamefont {M.}~\bibnamefont {Kang}}, \bibinfo {author} {\bibfnamefont {S.}~\bibnamefont {Kim}}, \bibinfo {author} {\bibfnamefont {Y.}~\bibnamefont {Qian}}, \bibinfo {author} {\bibfnamefont {P.~M.}\ \bibnamefont {Neves}}, \bibinfo {author} {\bibfnamefont {L.}~\bibnamefont {Ye}}, \bibinfo {author} {\bibfnamefont {J.}~\bibnamefont {Jung}}, \bibinfo {author} {\bibfnamefont {D.}~\bibnamefont {Puntel}}, \bibinfo {author} {\bibfnamefont {F.}~\bibnamefont {Mazzola}}, \bibinfo {author} {\bibfnamefont {S.}~\bibnamefont {Fang}}, \bibinfo {author} {\bibfnamefont {C.}~\bibnamefont {Jozwiak}}, \bibinfo {author} {\bibfnamefont {A.}~\bibnamefont {Bostwick}}, \bibinfo {author} {\bibfnamefont {E.}~\bibnamefont {Rotenberg}}, \bibinfo {author} {\bibfnamefont {J.}~\bibnamefont {Fuji}}, \bibinfo {author} {\bibfnamefont {I.}~\bibnamefont {Vobornik}}, \bibinfo {author} {\bibfnamefont {J.-H.}\ \bibnamefont {Park}}, \bibinfo {author} {\bibfnamefont {J.~G.}\ \bibnamefont {Checkelsky}},
  \bibinfo {author} {\bibfnamefont {B.-J.}\ \bibnamefont {Yang}},\ and\ \bibinfo {author} {\bibfnamefont {R.}~\bibnamefont {Comin}},\ }\bibfield  {title} {\bibinfo {title} {{Measurements of the quantum geometric tensor in solids}},\ }\href {https://doi.org/10.1038/s41567-024-02678-8} {\bibfield  {journal} {\bibinfo  {journal} {Nat. Phys.}\ }\textbf {\bibinfo {volume} {21}},\ \bibinfo {pages} {110} (\bibinfo {year} {2025})}\BibitemShut {NoStop}%
\bibitem [{\citenamefont {Tan}\ \emph {et~al.}(2019)\citenamefont {Tan}, \citenamefont {Zhang}, \citenamefont {Yang}, \citenamefont {Chu}, \citenamefont {Zhu}, \citenamefont {Li}, \citenamefont {Yang}, \citenamefont {Song}, \citenamefont {Han}, \citenamefont {Li}, \citenamefont {Dong}, \citenamefont {Yu}, \citenamefont {Yan}, \citenamefont {Zhu},\ and\ \citenamefont {Yu}}]{Tan2019a}%
  \BibitemOpen
  \bibfield  {author} {\bibinfo {author} {\bibfnamefont {X.}~\bibnamefont {Tan}}, \bibinfo {author} {\bibfnamefont {D.-W.}\ \bibnamefont {Zhang}}, \bibinfo {author} {\bibfnamefont {Z.}~\bibnamefont {Yang}}, \bibinfo {author} {\bibfnamefont {J.}~\bibnamefont {Chu}}, \bibinfo {author} {\bibfnamefont {Y.-Q.}\ \bibnamefont {Zhu}}, \bibinfo {author} {\bibfnamefont {D.}~\bibnamefont {Li}}, \bibinfo {author} {\bibfnamefont {X.}~\bibnamefont {Yang}}, \bibinfo {author} {\bibfnamefont {S.}~\bibnamefont {Song}}, \bibinfo {author} {\bibfnamefont {Z.}~\bibnamefont {Han}}, \bibinfo {author} {\bibfnamefont {Z.}~\bibnamefont {Li}}, \bibinfo {author} {\bibfnamefont {Y.}~\bibnamefont {Dong}}, \bibinfo {author} {\bibfnamefont {H.-F.}\ \bibnamefont {Yu}}, \bibinfo {author} {\bibfnamefont {H.}~\bibnamefont {Yan}}, \bibinfo {author} {\bibfnamefont {S.-L.}\ \bibnamefont {Zhu}},\ and\ \bibinfo {author} {\bibfnamefont {Y.}~\bibnamefont {Yu}},\ }\bibfield  {title} {\bibinfo {title} {{Experimental Measurement of the Quantum Metric
  Tensor and Related Topological Phase Transition with a Superconducting Qubit}},\ }\href {https://doi.org/10.1103/PhysRevLett.122.210401} {\bibfield  {journal} {\bibinfo  {journal} {Phys. Rev. Lett.}\ }\textbf {\bibinfo {volume} {122}},\ \bibinfo {pages} {210401} (\bibinfo {year} {2019})}\BibitemShut {NoStop}%
\bibitem [{\citenamefont {Hackl}\ \emph {et~al.}(2020)\citenamefont {Hackl}, \citenamefont {Guaita}, \citenamefont {Shi}, \citenamefont {Haegeman}, \citenamefont {Demler},\ and\ \citenamefont {Cirac}}]{Hackl2020}%
  \BibitemOpen
  \bibfield  {author} {\bibinfo {author} {\bibfnamefont {L.}~\bibnamefont {Hackl}}, \bibinfo {author} {\bibfnamefont {T.}~\bibnamefont {Guaita}}, \bibinfo {author} {\bibfnamefont {T.}~\bibnamefont {Shi}}, \bibinfo {author} {\bibfnamefont {J.}~\bibnamefont {Haegeman}}, \bibinfo {author} {\bibfnamefont {E.}~\bibnamefont {Demler}},\ and\ \bibinfo {author} {\bibfnamefont {I.}~\bibnamefont {Cirac}},\ }\bibfield  {title} {\bibinfo {title} {{Geometry of variational methods: dynamics of closed quantum systems}},\ }\href {https://doi.org/10.21468/SciPostPhys.9.4.048} {\bibfield  {journal} {\bibinfo  {journal} {SciPost Phys.}\ }\textbf {\bibinfo {volume} {9}},\ \bibinfo {pages} {048} (\bibinfo {year} {2020})}\BibitemShut {NoStop}%
\bibitem [{\citenamefont {Coelho}\ \emph {et~al.}(2016)\citenamefont {Coelho}, \citenamefont {Costanzo}, \citenamefont {Zavatta}, \citenamefont {Hughes}, \citenamefont {Kim},\ and\ \citenamefont {Bellini}}]{Coelho2016}%
  \BibitemOpen
  \bibfield  {author} {\bibinfo {author} {\bibfnamefont {A.~S.}\ \bibnamefont {Coelho}}, \bibinfo {author} {\bibfnamefont {L.~S.}\ \bibnamefont {Costanzo}}, \bibinfo {author} {\bibfnamefont {A.}~\bibnamefont {Zavatta}}, \bibinfo {author} {\bibfnamefont {C.}~\bibnamefont {Hughes}}, \bibinfo {author} {\bibfnamefont {M.~S.}\ \bibnamefont {Kim}},\ and\ \bibinfo {author} {\bibfnamefont {M.}~\bibnamefont {Bellini}},\ }\bibfield  {title} {\bibinfo {title} {{Universal Continuous-Variable State Orthogonalizer and Qubit Generator}},\ }\href {https://doi.org/10.1103/PhysRevLett.116.110501} {\bibfield  {journal} {\bibinfo  {journal} {Phys. Rev. Lett.}\ }\textbf {\bibinfo {volume} {116}},\ \bibinfo {pages} {110501} (\bibinfo {year} {2016})}\BibitemShut {NoStop}%
\bibitem [{\citenamefont {Lenhard}\ \emph {et~al.}(2015)\citenamefont {Lenhard}, \citenamefont {Bock}, \citenamefont {Becher}, \citenamefont {Kucera}, \citenamefont {Brito}, \citenamefont {Eich}, \citenamefont {M{\"{u}}ller},\ and\ \citenamefont {Eschner}}]{Lenhard2015}%
  \BibitemOpen
  \bibfield  {author} {\bibinfo {author} {\bibfnamefont {A.}~\bibnamefont {Lenhard}}, \bibinfo {author} {\bibfnamefont {M.}~\bibnamefont {Bock}}, \bibinfo {author} {\bibfnamefont {C.}~\bibnamefont {Becher}}, \bibinfo {author} {\bibfnamefont {S.}~\bibnamefont {Kucera}}, \bibinfo {author} {\bibfnamefont {J.}~\bibnamefont {Brito}}, \bibinfo {author} {\bibfnamefont {P.}~\bibnamefont {Eich}}, \bibinfo {author} {\bibfnamefont {P.}~\bibnamefont {M{\"{u}}ller}},\ and\ \bibinfo {author} {\bibfnamefont {J.}~\bibnamefont {Eschner}},\ }\bibfield  {title} {\bibinfo {title} {{Telecom-heralded single-photon absorption by a single atom}},\ }\href {https://doi.org/10.1103/PhysRevA.92.063827} {\bibfield  {journal} {\bibinfo  {journal} {Phys. Rev. A}\ }\textbf {\bibinfo {volume} {92}},\ \bibinfo {pages} {063827} (\bibinfo {year} {2015})}\BibitemShut {NoStop}%
\bibitem [{\citenamefont {Baragiola}\ \emph {et~al.}(2012)\citenamefont {Baragiola}, \citenamefont {Cook}, \citenamefont {Bra{\'{n}}czyk},\ and\ \citenamefont {Combes}}]{Baragiola2012}%
  \BibitemOpen
  \bibfield  {author} {\bibinfo {author} {\bibfnamefont {B.~Q.}\ \bibnamefont {Baragiola}}, \bibinfo {author} {\bibfnamefont {R.~L.}\ \bibnamefont {Cook}}, \bibinfo {author} {\bibfnamefont {A.~M.}\ \bibnamefont {Bra{\'{n}}czyk}},\ and\ \bibinfo {author} {\bibfnamefont {J.}~\bibnamefont {Combes}},\ }\bibfield  {title} {\bibinfo {title} {{$N$-photon wave packets interacting with an arbitrary quantum system}},\ }\href {https://doi.org/10.1103/PhysRevA.86.013811} {\bibfield  {journal} {\bibinfo  {journal} {Phys. Rev. A}\ }\textbf {\bibinfo {volume} {86}},\ \bibinfo {pages} {013811} (\bibinfo {year} {2012})}\BibitemShut {NoStop}%
\bibitem [{\citenamefont {Knill}\ \emph {et~al.}(2001)\citenamefont {Knill}, \citenamefont {Laflamme},\ and\ \citenamefont {Milburn}}]{Knill2001}%
  \BibitemOpen
  \bibfield  {author} {\bibinfo {author} {\bibfnamefont {E.}~\bibnamefont {Knill}}, \bibinfo {author} {\bibfnamefont {R.}~\bibnamefont {Laflamme}},\ and\ \bibinfo {author} {\bibfnamefont {G.~J.}\ \bibnamefont {Milburn}},\ }\bibfield  {title} {\bibinfo {title} {{A scheme for efficient quantum computation with linear optics}},\ }\href {https://doi.org/10.1038/35051009} {\bibfield  {journal} {\bibinfo  {journal} {Nature}\ }\textbf {\bibinfo {volume} {409}},\ \bibinfo {pages} {46} (\bibinfo {year} {2001})}\BibitemShut {NoStop}%
\bibitem [{\citenamefont {Jaynes}\ and\ \citenamefont {Cummings}(1963)}]{Jaynes1963}%
  \BibitemOpen
  \bibfield  {author} {\bibinfo {author} {\bibfnamefont {E.~T.}\ \bibnamefont {Jaynes}}\ and\ \bibinfo {author} {\bibfnamefont {F.~W.}\ \bibnamefont {Cummings}},\ }\bibfield  {title} {\bibinfo {title} {{Comparison of quantum and semiclassical radiation theories with application to the beam maser}},\ }\href {https://doi.org/10.1109/PROC.1963.1664} {\bibfield  {journal} {\bibinfo  {journal} {Proc. IEEE}\ }\textbf {\bibinfo {volume} {51}},\ \bibinfo {pages} {89} (\bibinfo {year} {1963})}\BibitemShut {NoStop}%
\bibitem [{\citenamefont {{Campos Venuti}}\ and\ \citenamefont {Zanardi}(2007)}]{CamposVenuti2007}%
  \BibitemOpen
  \bibfield  {author} {\bibinfo {author} {\bibfnamefont {L.}~\bibnamefont {{Campos Venuti}}}\ and\ \bibinfo {author} {\bibfnamefont {P.}~\bibnamefont {Zanardi}},\ }\bibfield  {title} {\bibinfo {title} {{Quantum Critical Scaling of the Geometric Tensors}},\ }\href {https://doi.org/10.1103/PhysRevLett.99.095701} {\bibfield  {journal} {\bibinfo  {journal} {Phys. Rev. Lett.}\ }\textbf {\bibinfo {volume} {99}},\ \bibinfo {pages} {095701} (\bibinfo {year} {2007})}\BibitemShut {NoStop}%
\bibitem [{\citenamefont {Callen}\ and\ \citenamefont {Welton}(1951)}]{Callen1951}%
  \BibitemOpen
  \bibfield  {author} {\bibinfo {author} {\bibfnamefont {H.~B.}\ \bibnamefont {Callen}}\ and\ \bibinfo {author} {\bibfnamefont {T.~A.}\ \bibnamefont {Welton}},\ }\bibfield  {title} {\bibinfo {title} {{Irreversibility and Generalized Noise}},\ }\href {https://doi.org/10.1103/PhysRev.83.34} {\bibfield  {journal} {\bibinfo  {journal} {Phys. Rev.}\ }\textbf {\bibinfo {volume} {83}},\ \bibinfo {pages} {34} (\bibinfo {year} {1951})}\BibitemShut {NoStop}%
\bibitem [{\citenamefont {Kubo}(1966)}]{Kubo1966}%
  \BibitemOpen
  \bibfield  {author} {\bibinfo {author} {\bibfnamefont {R.}~\bibnamefont {Kubo}},\ }\bibfield  {title} {\bibinfo {title} {{The fluctuation-dissipation theorem}},\ }\href {https://doi.org/10.1088/0034-4885/29/1/306} {\bibfield  {journal} {\bibinfo  {journal} {Reports Prog. Phys.}\ }\textbf {\bibinfo {volume} {29}},\ \bibinfo {pages} {255} (\bibinfo {year} {1966})}\BibitemShut {NoStop}%
\bibitem [{\citenamefont {Harada}\ and\ \citenamefont {Sasa}(2005)}]{Harada2005}%
  \BibitemOpen
  \bibfield  {author} {\bibinfo {author} {\bibfnamefont {T.}~\bibnamefont {Harada}}\ and\ \bibinfo {author} {\bibfnamefont {S.-i.}\ \bibnamefont {Sasa}},\ }\bibfield  {title} {\bibinfo {title} {{Equality Connecting Energy Dissipation with a Violation of the Fluctuation-Response Relation}},\ }\href {https://doi.org/10.1103/PhysRevLett.95.130602} {\bibfield  {journal} {\bibinfo  {journal} {Phys. Rev. Lett.}\ }\textbf {\bibinfo {volume} {95}},\ \bibinfo {pages} {130602} (\bibinfo {year} {2005})}\BibitemShut {NoStop}%
\bibitem [{\citenamefont {Banaszek}\ and\ \citenamefont {W{\'{o}}dkiewicz}(1996)}]{Banaszek1996}%
  \BibitemOpen
  \bibfield  {author} {\bibinfo {author} {\bibfnamefont {K.}~\bibnamefont {Banaszek}}\ and\ \bibinfo {author} {\bibfnamefont {K.}~\bibnamefont {W{\'{o}}dkiewicz}},\ }\bibfield  {title} {\bibinfo {title} {{Direct Probing of Quantum Phase Space by Photon Counting}},\ }\href {https://doi.org/10.1103/PhysRevLett.76.4344} {\bibfield  {journal} {\bibinfo  {journal} {Phys. Rev. Lett.}\ }\textbf {\bibinfo {volume} {76}},\ \bibinfo {pages} {4344} (\bibinfo {year} {1996})}\BibitemShut {NoStop}%
\bibitem [{\citenamefont {M{\"{u}}ller}\ \emph {et~al.}(2016)\citenamefont {M{\"{u}}ller}, \citenamefont {Peuntinger}, \citenamefont {Dirmeier}, \citenamefont {Khan}, \citenamefont {Vogl}, \citenamefont {Marquardt}, \citenamefont {Leuchs}, \citenamefont {S{\'{a}}nchez-Soto}, \citenamefont {Teo}, \citenamefont {Hradil},\ and\ \citenamefont {^^c5^^98eh{\'{a}}{\v{c}}ek}}]{Muller2016}%
  \BibitemOpen
  \bibfield  {author} {\bibinfo {author} {\bibfnamefont {C.~R.}\ \bibnamefont {M{\"{u}}ller}}, \bibinfo {author} {\bibfnamefont {C.}~\bibnamefont {Peuntinger}}, \bibinfo {author} {\bibfnamefont {T.}~\bibnamefont {Dirmeier}}, \bibinfo {author} {\bibfnamefont {I.}~\bibnamefont {Khan}}, \bibinfo {author} {\bibfnamefont {U.}~\bibnamefont {Vogl}}, \bibinfo {author} {\bibfnamefont {C.}~\bibnamefont {Marquardt}}, \bibinfo {author} {\bibfnamefont {G.}~\bibnamefont {Leuchs}}, \bibinfo {author} {\bibfnamefont {L.~L.}\ \bibnamefont {S{\'{a}}nchez-Soto}}, \bibinfo {author} {\bibfnamefont {Y.~S.}\ \bibnamefont {Teo}}, \bibinfo {author} {\bibfnamefont {Z.}~\bibnamefont {Hradil}},\ and\ \bibinfo {author} {\bibfnamefont {J.}~\bibnamefont {^^c5^^98eh{\'{a}}{\v{c}}ek}},\ }\bibfield  {title} {\bibinfo {title} {{Evading Vacuum Noise: Wigner Projections or Husimi Samples?}},\ }\href {https://doi.org/10.1103/PhysRevLett.117.070801} {\bibfield  {journal} {\bibinfo  {journal} {Phys. Rev. Lett.}\ }\textbf {\bibinfo {volume} {117}},\
  \bibinfo {pages} {070801} (\bibinfo {year} {2016})}\BibitemShut {NoStop}%
\bibitem [{\citenamefont {L{\"{u}}ders}\ \emph {et~al.}(2021)\citenamefont {L{\"{u}}ders}, \citenamefont {Pukrop}, \citenamefont {Rozas}, \citenamefont {Schneider}, \citenamefont {H{\"{o}}fling}, \citenamefont {Sperling}, \citenamefont {Schumacher},\ and\ \citenamefont {A{\ss}mann}}]{Luders2021}%
  \BibitemOpen
  \bibfield  {author} {\bibinfo {author} {\bibfnamefont {C.}~\bibnamefont {L{\"{u}}ders}}, \bibinfo {author} {\bibfnamefont {M.}~\bibnamefont {Pukrop}}, \bibinfo {author} {\bibfnamefont {E.}~\bibnamefont {Rozas}}, \bibinfo {author} {\bibfnamefont {C.}~\bibnamefont {Schneider}}, \bibinfo {author} {\bibfnamefont {S.}~\bibnamefont {H{\"{o}}fling}}, \bibinfo {author} {\bibfnamefont {J.}~\bibnamefont {Sperling}}, \bibinfo {author} {\bibfnamefont {S.}~\bibnamefont {Schumacher}},\ and\ \bibinfo {author} {\bibfnamefont {M.}~\bibnamefont {A{\ss}mann}},\ }\bibfield  {title} {\bibinfo {title} {{Quantifying Quantum Coherence in Polariton Condensates}},\ }\href {https://doi.org/10.1103/PRXQuantum.2.030320} {\bibfield  {journal} {\bibinfo  {journal} {PRX Quantum}\ }\textbf {\bibinfo {volume} {2}},\ \bibinfo {pages} {030320} (\bibinfo {year} {2021})}\BibitemShut {NoStop}%
\bibitem [{\citenamefont {Lvovsky}\ and\ \citenamefont {Raymer}(2009)}]{Lvovsky2009}%
  \BibitemOpen
  \bibfield  {author} {\bibinfo {author} {\bibfnamefont {A.~I.}\ \bibnamefont {Lvovsky}}\ and\ \bibinfo {author} {\bibfnamefont {M.~G.}\ \bibnamefont {Raymer}},\ }\bibfield  {title} {\bibinfo {title} {{Continuous-variable optical quantum-state tomography}},\ }\href {https://doi.org/10.1103/RevModPhys.81.299} {\bibfield  {journal} {\bibinfo  {journal} {Rev. Mod. Phys.}\ }\textbf {\bibinfo {volume} {81}},\ \bibinfo {pages} {299} (\bibinfo {year} {2009})}\BibitemShut {NoStop}%
\bibitem [{\citenamefont {Smithey}\ \emph {et~al.}(1993)\citenamefont {Smithey}, \citenamefont {Beck}, \citenamefont {Raymer},\ and\ \citenamefont {Faridani}}]{Smithey1993}%
  \BibitemOpen
  \bibfield  {author} {\bibinfo {author} {\bibfnamefont {D.~T.}\ \bibnamefont {Smithey}}, \bibinfo {author} {\bibfnamefont {M.}~\bibnamefont {Beck}}, \bibinfo {author} {\bibfnamefont {M.~G.}\ \bibnamefont {Raymer}},\ and\ \bibinfo {author} {\bibfnamefont {A.}~\bibnamefont {Faridani}},\ }\bibfield  {title} {\bibinfo {title} {{Measurement of the Wigner distribution and the density matrix of a light mode using optical homodyne tomography: Application to squeezed states and the vacuum}},\ }\href {https://doi.org/10.1103/PhysRevLett.70.1244} {\bibfield  {journal} {\bibinfo  {journal} {Phys. Rev. Lett.}\ }\textbf {\bibinfo {volume} {70}},\ \bibinfo {pages} {1244} (\bibinfo {year} {1993})}\BibitemShut {NoStop}%
\bibitem [{\citenamefont {Yoon}\ \emph {et~al.}(2026)\citenamefont {Yoon}, \citenamefont {Roh}, \citenamefont {Gwak},\ and\ \citenamefont {Ra}}]{Yoon2026}%
  \BibitemOpen
  \bibfield  {author} {\bibinfo {author} {\bibfnamefont {Y.-D.}\ \bibnamefont {Yoon}}, \bibinfo {author} {\bibfnamefont {C.}~\bibnamefont {Roh}}, \bibinfo {author} {\bibfnamefont {G.}~\bibnamefont {Gwak}},\ and\ \bibinfo {author} {\bibfnamefont {Y.-s.}\ \bibnamefont {Ra}},\ }\bibfield  {title} {\bibinfo {title} {{Efficient ultrafast homodyne detection of quantum light}},\ }\href {https://doi.org/10.1038/s42005-026-02887-y} {\bibfield  {journal} {\bibinfo  {journal} {Commun. Phys.}\ } (\bibinfo {year} {2026})}\BibitemShut {NoStop}%
\bibitem [{\citenamefont {Gaharwar}\ \emph {et~al.}()\citenamefont {Gaharwar}, \citenamefont {Borrego-Varillas}, \citenamefont {Ciappina}, \citenamefont {Ciriolo}, \citenamefont {Rivera-Dean}, \citenamefont {Stammer}, \citenamefont {Tzallas}, \citenamefont {Pisanty},\ and\ \citenamefont {Lewenstein}}]{Gaharwar2026}%
  \BibitemOpen
  \bibfield  {author} {\bibinfo {author} {\bibfnamefont {A.}~\bibnamefont {Gaharwar}}, \bibinfo {author} {\bibfnamefont {R.}~\bibnamefont {Borrego-Varillas}}, \bibinfo {author} {\bibfnamefont {M.~F.}\ \bibnamefont {Ciappina}}, \bibinfo {author} {\bibfnamefont {A.~G.}\ \bibnamefont {Ciriolo}}, \bibinfo {author} {\bibfnamefont {J.}~\bibnamefont {Rivera-Dean}}, \bibinfo {author} {\bibfnamefont {P.}~\bibnamefont {Stammer}}, \bibinfo {author} {\bibfnamefont {P.}~\bibnamefont {Tzallas}}, \bibinfo {author} {\bibfnamefont {E.}~\bibnamefont {Pisanty}},\ and\ \bibinfo {author} {\bibfnamefont {M.}~\bibnamefont {Lewenstein}},\ }\bibfield  {title} {\bibinfo {title} {{Geometric Control of Cat States in High Harmonic Generation}},\ }\Eprint {https://arxiv.org/abs/2608.20119} {arXiv:2608.20119} \BibitemShut {NoStop}%
\bibitem [{\citenamefont {Lipkin}\ \emph {et~al.}(1965)\citenamefont {Lipkin}, \citenamefont {Meshkov},\ and\ \citenamefont {Glick}}]{Lipkin1965a}%
  \BibitemOpen
  \bibfield  {author} {\bibinfo {author} {\bibfnamefont {H.}~\bibnamefont {Lipkin}}, \bibinfo {author} {\bibfnamefont {N.}~\bibnamefont {Meshkov}},\ and\ \bibinfo {author} {\bibfnamefont {A.}~\bibnamefont {Glick}},\ }\bibfield  {title} {\bibinfo {title} {{Validity of many-body approximation methods for a solvable model}},\ }\href {https://doi.org/10.1016/0029-5582(65)90862-X} {\bibfield  {journal} {\bibinfo  {journal} {Nucl. Phys.}\ }\textbf {\bibinfo {volume} {62}},\ \bibinfo {pages} {188} (\bibinfo {year} {1965})}\BibitemShut {NoStop}%
\bibitem [{\citenamefont {Dusuel}\ and\ \citenamefont {Vidal}(2004)}]{Dusuel2004}%
  \BibitemOpen
  \bibfield  {author} {\bibinfo {author} {\bibfnamefont {S.}~\bibnamefont {Dusuel}}\ and\ \bibinfo {author} {\bibfnamefont {J.}~\bibnamefont {Vidal}},\ }\bibfield  {title} {\bibinfo {title} {{Finite-Size Scaling Exponents of the Lipkin-Meshkov-Glick Model}},\ }\href {https://doi.org/10.1103/PhysRevLett.93.237204} {\bibfield  {journal} {\bibinfo  {journal} {Phys. Rev. Lett.}\ }\textbf {\bibinfo {volume} {93}},\ \bibinfo {pages} {237204} (\bibinfo {year} {2004})}\BibitemShut {NoStop}%
\bibitem [{\citenamefont {Kamenev}\ and\ \citenamefont {Levchenko}(2009)}]{Kamenev2009a}%
  \BibitemOpen
  \bibfield  {author} {\bibinfo {author} {\bibfnamefont {A.}~\bibnamefont {Kamenev}}\ and\ \bibinfo {author} {\bibfnamefont {A.}~\bibnamefont {Levchenko}},\ }\bibfield  {title} {\bibinfo {title} {{Keldysh technique and non-linear $\sigma$-model: basic principles and applications}},\ }\href {https://doi.org/10.1080/00018730902850504} {\bibfield  {journal} {\bibinfo  {journal} {Adv. Phys.}\ }\textbf {\bibinfo {volume} {58}},\ \bibinfo {pages} {197} (\bibinfo {year} {2009})}\BibitemShut {NoStop}%
\bibitem [{\citenamefont {Feynman}\ and\ \citenamefont {Vernon}(1963)}]{Feynman1963}%
  \BibitemOpen
  \bibfield  {author} {\bibinfo {author} {\bibfnamefont {R.}~\bibnamefont {Feynman}}\ and\ \bibinfo {author} {\bibfnamefont {F.}~\bibnamefont {Vernon}},\ }\bibfield  {title} {\bibinfo {title} {{The theory of a general quantum system interacting with a linear dissipative system}},\ }\href {https://doi.org/10.1016/0003-4916(63)90068-X} {\bibfield  {journal} {\bibinfo  {journal} {Ann. Phys. (N. Y).}\ }\textbf {\bibinfo {volume} {24}},\ \bibinfo {pages} {118} (\bibinfo {year} {1963})}\BibitemShut {NoStop}%
\bibitem [{\citenamefont {Brion}\ \emph {et~al.}(2007)\citenamefont {Brion}, \citenamefont {Pedersen},\ and\ \citenamefont {M{\o}lmer}}]{Brion2007}%
  \BibitemOpen
  \bibfield  {author} {\bibinfo {author} {\bibfnamefont {E.}~\bibnamefont {Brion}}, \bibinfo {author} {\bibfnamefont {L.~H.}\ \bibnamefont {Pedersen}},\ and\ \bibinfo {author} {\bibfnamefont {K.}~\bibnamefont {M{\o}lmer}},\ }\bibfield  {title} {\bibinfo {title} {{Adiabatic elimination in a lambda system}},\ }\href {https://doi.org/10.1088/1751-8113/40/5/011} {\bibfield  {journal} {\bibinfo  {journal} {J. Phys. A Math. Theor.}\ }\textbf {\bibinfo {volume} {40}},\ \bibinfo {pages} {1033} (\bibinfo {year} {2007})}\BibitemShut {NoStop}%
\bibitem [{\citenamefont {Hammerer}\ \emph {et~al.}(2010)\citenamefont {Hammerer}, \citenamefont {S{\o}rensen},\ and\ \citenamefont {Polzik}}]{Hammerer2010}%
  \BibitemOpen
  \bibfield  {author} {\bibinfo {author} {\bibfnamefont {K.}~\bibnamefont {Hammerer}}, \bibinfo {author} {\bibfnamefont {A.~S.}\ \bibnamefont {S{\o}rensen}},\ and\ \bibinfo {author} {\bibfnamefont {E.~S.}\ \bibnamefont {Polzik}},\ }\bibfield  {title} {\bibinfo {title} {{Quantum interface between light and atomic ensembles}},\ }\href {https://doi.org/10.1103/RevModPhys.82.1041} {\bibfield  {journal} {\bibinfo  {journal} {Rev. Mod. Phys.}\ }\textbf {\bibinfo {volume} {82}},\ \bibinfo {pages} {1041} (\bibinfo {year} {2010})}\BibitemShut {NoStop}%
\bibitem [{\citenamefont {Kiesel}\ \emph {et~al.}(2008)\citenamefont {Kiesel}, \citenamefont {Vogel}, \citenamefont {Parigi}, \citenamefont {Zavatta},\ and\ \citenamefont {Bellini}}]{Kiesel2008}%
  \BibitemOpen
  \bibfield  {author} {\bibinfo {author} {\bibfnamefont {T.}~\bibnamefont {Kiesel}}, \bibinfo {author} {\bibfnamefont {W.}~\bibnamefont {Vogel}}, \bibinfo {author} {\bibfnamefont {V.}~\bibnamefont {Parigi}}, \bibinfo {author} {\bibfnamefont {A.}~\bibnamefont {Zavatta}},\ and\ \bibinfo {author} {\bibfnamefont {M.}~\bibnamefont {Bellini}},\ }\bibfield  {title} {\bibinfo {title} {{Experimental determination of a nonclassical Glauber-Sudarshan $P$ function}},\ }\href {https://doi.org/10.1103/PhysRevA.78.021804} {\bibfield  {journal} {\bibinfo  {journal} {Phys. Rev. A}\ }\textbf {\bibinfo {volume} {78}},\ \bibinfo {pages} {021804} (\bibinfo {year} {2008})}\BibitemShut {NoStop}%
\bibitem [{\citenamefont {Ekert}\ \emph {et~al.}(2002)\citenamefont {Ekert}, \citenamefont {Alves}, \citenamefont {Oi}, \citenamefont {Horodecki}, \citenamefont {Horodecki},\ and\ \citenamefont {Kwek}}]{Ekert2002}%
  \BibitemOpen
  \bibfield  {author} {\bibinfo {author} {\bibfnamefont {A.~K.}\ \bibnamefont {Ekert}}, \bibinfo {author} {\bibfnamefont {C.~M.}\ \bibnamefont {Alves}}, \bibinfo {author} {\bibfnamefont {D.~K.~L.}\ \bibnamefont {Oi}}, \bibinfo {author} {\bibfnamefont {M.}~\bibnamefont {Horodecki}}, \bibinfo {author} {\bibfnamefont {P.}~\bibnamefont {Horodecki}},\ and\ \bibinfo {author} {\bibfnamefont {L.~C.}\ \bibnamefont {Kwek}},\ }\bibfield  {title} {\bibinfo {title} {{Direct Estimations of Linear and Nonlinear Functionals of a Quantum State}},\ }\href {https://doi.org/10.1103/PhysRevLett.88.217901} {\bibfield  {journal} {\bibinfo  {journal} {Phys. Rev. Lett.}\ }\textbf {\bibinfo {volume} {88}},\ \bibinfo {pages} {217901} (\bibinfo {year} {2002})}\BibitemShut {NoStop}%
\bibitem [{\citenamefont {Griffet}\ \emph {et~al.}(2023)\citenamefont {Griffet}, \citenamefont {Arnhem}, \citenamefont {{De Bi{\`{e}}vre}},\ and\ \citenamefont {Cerf}}]{Griffet2023}%
  \BibitemOpen
  \bibfield  {author} {\bibinfo {author} {\bibfnamefont {C.}~\bibnamefont {Griffet}}, \bibinfo {author} {\bibfnamefont {M.}~\bibnamefont {Arnhem}}, \bibinfo {author} {\bibfnamefont {S.}~\bibnamefont {{De Bi{\`{e}}vre}}},\ and\ \bibinfo {author} {\bibfnamefont {N.~J.}\ \bibnamefont {Cerf}},\ }\bibfield  {title} {\bibinfo {title} {{Interferometric measurement of the quadrature coherence scale using two replicas of a quantum optical state}},\ }\href {https://doi.org/10.1103/PhysRevA.108.023730} {\bibfield  {journal} {\bibinfo  {journal} {Phys. Rev. A}\ }\textbf {\bibinfo {volume} {108}},\ \bibinfo {pages} {023730} (\bibinfo {year} {2023})}\BibitemShut {NoStop}%
\bibitem [{\citenamefont {Adesso}\ \emph {et~al.}(2009)\citenamefont {Adesso}, \citenamefont {Dell'Anno}, \citenamefont {{De Siena}}, \citenamefont {Illuminati},\ and\ \citenamefont {Souza}}]{Adesso2009}%
  \BibitemOpen
  \bibfield  {author} {\bibinfo {author} {\bibfnamefont {G.}~\bibnamefont {Adesso}}, \bibinfo {author} {\bibfnamefont {F.}~\bibnamefont {Dell'Anno}}, \bibinfo {author} {\bibfnamefont {S.}~\bibnamefont {{De Siena}}}, \bibinfo {author} {\bibfnamefont {F.}~\bibnamefont {Illuminati}},\ and\ \bibinfo {author} {\bibfnamefont {L.~A.~M.}\ \bibnamefont {Souza}},\ }\bibfield  {title} {\bibinfo {title} {{Optimal estimation of losses at the ultimate quantum limit with non-Gaussian states}},\ }\href {https://doi.org/10.1103/PhysRevA.79.040305} {\bibfield  {journal} {\bibinfo  {journal} {Phys. Rev. A}\ }\textbf {\bibinfo {volume} {79}},\ \bibinfo {pages} {040305} (\bibinfo {year} {2009})}\BibitemShut {NoStop}%
\bibitem [{\citenamefont {Casacio}\ \emph {et~al.}(2021)\citenamefont {Casacio}, \citenamefont {Madsen}, \citenamefont {Terrasson}, \citenamefont {Waleed}, \citenamefont {Barnscheidt}, \citenamefont {Hage}, \citenamefont {Taylor},\ and\ \citenamefont {Bowen}}]{Casacio2021}%
  \BibitemOpen
  \bibfield  {author} {\bibinfo {author} {\bibfnamefont {C.~A.}\ \bibnamefont {Casacio}}, \bibinfo {author} {\bibfnamefont {L.~S.}\ \bibnamefont {Madsen}}, \bibinfo {author} {\bibfnamefont {A.}~\bibnamefont {Terrasson}}, \bibinfo {author} {\bibfnamefont {M.}~\bibnamefont {Waleed}}, \bibinfo {author} {\bibfnamefont {K.}~\bibnamefont {Barnscheidt}}, \bibinfo {author} {\bibfnamefont {B.}~\bibnamefont {Hage}}, \bibinfo {author} {\bibfnamefont {M.~A.}\ \bibnamefont {Taylor}},\ and\ \bibinfo {author} {\bibfnamefont {W.~P.}\ \bibnamefont {Bowen}},\ }\bibfield  {title} {\bibinfo {title} {{Quantum-enhanced nonlinear microscopy}},\ }\href {https://doi.org/10.1038/s41586-021-03528-w} {\bibfield  {journal} {\bibinfo  {journal} {Nature}\ }\textbf {\bibinfo {volume} {594}},\ \bibinfo {pages} {201} (\bibinfo {year} {2021})}\BibitemShut {NoStop}%
\bibitem [{\citenamefont {Lamprou}\ \emph {et~al.}(2025)\citenamefont {Lamprou}, \citenamefont {Rivera-Dean}, \citenamefont {Stammer}, \citenamefont {Lewenstein},\ and\ \citenamefont {Tzallas}}]{Lamprou2025}%
  \BibitemOpen
  \bibfield  {author} {\bibinfo {author} {\bibfnamefont {T.}~\bibnamefont {Lamprou}}, \bibinfo {author} {\bibfnamefont {J.}~\bibnamefont {Rivera-Dean}}, \bibinfo {author} {\bibfnamefont {P.}~\bibnamefont {Stammer}}, \bibinfo {author} {\bibfnamefont {M.}~\bibnamefont {Lewenstein}},\ and\ \bibinfo {author} {\bibfnamefont {P.}~\bibnamefont {Tzallas}},\ }\bibfield  {title} {\bibinfo {title} {{Nonlinear Optics Using Intense Optical Coherent State Superpositions}},\ }\href {https://doi.org/10.1103/PhysRevLett.134.013601} {\bibfield  {journal} {\bibinfo  {journal} {Phys. Rev. Lett.}\ }\textbf {\bibinfo {volume} {134}},\ \bibinfo {pages} {013601} (\bibinfo {year} {2025})}\BibitemShut {NoStop}%
\bibitem [{\citenamefont {Imai}(2026{\natexlab{a}})}]{Imai2026}%
  \BibitemOpen
  \bibfield  {author} {\bibinfo {author} {\bibfnamefont {S.}~\bibnamefont {Imai}},\ }\bibfield  {title} {\bibinfo {title} {{Macroscopic Schr{\"{o}}dinger-cat states of nonequilibrium electrons induced by cat-state optical driving and projective measurements on the light field}},\ }\href {https://doi.org/10.1103/kzr9-fxx3} {\bibfield  {journal} {\bibinfo  {journal} {Phys. Rev. A}\ }\textbf {\bibinfo {volume} {113}},\ \bibinfo {pages} {023702} (\bibinfo {year} {2026}{\natexlab{a}})}\BibitemShut {NoStop}%
\bibitem [{\citenamefont {Imai}\ \emph {et~al.}(2026)\citenamefont {Imai}, \citenamefont {Ono},\ and\ \citenamefont {Tsuji}}]{Imai2026b}%
  \BibitemOpen
  \bibfield  {author} {\bibinfo {author} {\bibfnamefont {S.}~\bibnamefont {Imai}}, \bibinfo {author} {\bibfnamefont {A.}~\bibnamefont {Ono}},\ and\ \bibinfo {author} {\bibfnamefont {N.}~\bibnamefont {Tsuji}},\ }\bibfield  {title} {\bibinfo {title} {{Electron dynamics induced by quantum cat-state light}},\ }\href {https://doi.org/10.1103/7vll-vh3l} {\bibfield  {journal} {\bibinfo  {journal} {Phys. Rev. A}\ }\textbf {\bibinfo {volume} {114}},\ \bibinfo {pages} {L011102} (\bibinfo {year} {2026})}\BibitemShut {NoStop}%
\bibitem [{\citenamefont {Rasputnyi}\ \emph {et~al.}(2024)\citenamefont {Rasputnyi}, \citenamefont {Chen}, \citenamefont {Birk}, \citenamefont {Cohen}, \citenamefont {Kaminer}, \citenamefont {Kr{\"{u}}ger}, \citenamefont {Seletskiy}, \citenamefont {Chekhova},\ and\ \citenamefont {Tani}}]{Rasputnyi2024}%
  \BibitemOpen
  \bibfield  {author} {\bibinfo {author} {\bibfnamefont {A.}~\bibnamefont {Rasputnyi}}, \bibinfo {author} {\bibfnamefont {Z.}~\bibnamefont {Chen}}, \bibinfo {author} {\bibfnamefont {M.}~\bibnamefont {Birk}}, \bibinfo {author} {\bibfnamefont {O.}~\bibnamefont {Cohen}}, \bibinfo {author} {\bibfnamefont {I.}~\bibnamefont {Kaminer}}, \bibinfo {author} {\bibfnamefont {M.}~\bibnamefont {Kr{\"{u}}ger}}, \bibinfo {author} {\bibfnamefont {D.}~\bibnamefont {Seletskiy}}, \bibinfo {author} {\bibfnamefont {M.}~\bibnamefont {Chekhova}},\ and\ \bibinfo {author} {\bibfnamefont {F.}~\bibnamefont {Tani}},\ }\bibfield  {title} {\bibinfo {title} {{High-harmonic generation by a bright squeezed vacuum}},\ }\href {https://doi.org/10.1038/s41567-024-02659-x} {\bibfield  {journal} {\bibinfo  {journal} {Nat. Phys.}\ }\textbf {\bibinfo {volume} {20}},\ \bibinfo {pages} {1960} (\bibinfo {year} {2024})}\BibitemShut {NoStop}%
\bibitem [{\citenamefont {Eto}\ \emph {et~al.}(2025)\citenamefont {Eto}, \citenamefont {Nuriya},\ and\ \citenamefont {Kano}}]{Eto2025}%
  \BibitemOpen
  \bibfield  {author} {\bibinfo {author} {\bibfnamefont {Y.}~\bibnamefont {Eto}}, \bibinfo {author} {\bibfnamefont {M.}~\bibnamefont {Nuriya}},\ and\ \bibinfo {author} {\bibfnamefont {H.}~\bibnamefont {Kano}},\ }\bibfield  {title} {\bibinfo {title} {{Spectral-focusing coherent anti-Stokes Raman scattering driven by bright correlated twin beams}},\ }\href {https://doi.org/10.1103/7pvj-3dqt} {\bibfield  {journal} {\bibinfo  {journal} {Phys. Rev. Appl.}\ }\textbf {\bibinfo {volume} {23}},\ \bibinfo {pages} {064015} (\bibinfo {year} {2025})}\BibitemShut {NoStop}%
\bibitem [{\citenamefont {Imai}(2026{\natexlab{b}})}]{Imai2026a}%
  \BibitemOpen
  \bibfield  {author} {\bibinfo {author} {\bibfnamefont {S.}~\bibnamefont {Imai}},\ }\bibfield  {title} {\bibinfo {title} {{Heralded ultrafast generation of macroscopic quantum states in matter with bright squeezed vacuum light}},\ }\href {https://doi.org/10.1088/1367-2630/ae868e} {\bibfield  {journal} {\bibinfo  {journal} {New J. Phys.}\ }\textbf {\bibinfo {volume} {28}},\ \bibinfo {pages} {074506} (\bibinfo {year} {2026}{\natexlab{b}})}\BibitemShut {NoStop}%
\bibitem [{\citenamefont {Bures}(1969)}]{Bures1969}%
  \BibitemOpen
  \bibfield  {author} {\bibinfo {author} {\bibfnamefont {D.}~\bibnamefont {Bures}},\ }\bibfield  {title} {\bibinfo {title} {{An extension of Kakutani's theorem on infinite product measures to the tensor product of semifinite {$w^*$}-algebras}},\ }\href {https://doi.org/10.1090/S0002-9947-1969-0236719-2} {\bibfield  {journal} {\bibinfo  {journal} {Trans. Am. Math. Soc.}\ }\textbf {\bibinfo {volume} {135}},\ \bibinfo {pages} {199} (\bibinfo {year} {1969})}\BibitemShut {NoStop}%
\end{thebibliography}%
\end{document}